\documentclass{article}

\usepackage[enableskew]{youngtab}

\newcommand{\be}{\begin{equation}}
\newcommand{\ee}{\end{equation}}
\newcommand{\ba}{\begin{eqnarray}}
\newcommand{\ea}{\end{eqnarray}}
\newcommand{\bas}{\begin{eqnarray*}}
\newcommand{\eas}{\end{eqnarray*}}

\newcommand{\zero}{{0}}
\newcommand{\one}{{1}}
\newcommand{\two}{{2}}
\newcommand{\three}{{3}}

\title{%
Exact Dynamics of the $SU(K)$ Haldane-Shastry Model
}

\bigskip

\author{%
Takashi Yamamoto$^1$, Yasuhiro Saiga$^2$,
Mitsuhiro Arikawa$^3$, and Yoshio Kuramoto$^3$ \\
$^1$Max-Planck-Institut f\"{u}r Physik komplexer Systeme,\\
N\"{o}thnizer Str. 38, D-01187 Dresden, Germany\\
$^2$Institute for Solid State Physics, University of Tokyo,\\
Roppongi 7-22-1, Tokyo 106-8666, Japan\\
$^3$Department of Physics, Tohoku University,\\
Sendai 980-8578, Japan}
\begin{document}

\maketitle

\bigskip

\begin{abstract}
The dynamical structure factor $S(q,\omega)$ of
the $SU(K)$ ($K=2,3,4$) Haldane-Shastry model
is derived exactly at zero temperature for arbitrary size of the system.
The result is interpreted in terms of free quasi-particles
which are generalization of spinons in the $SU(2)$ case;
the excited states relevant to $S(q,\omega)$ consist of
$K$ quasi-particles each of
which is characterized by a set of $K-1$ quantum numbers.
Near the boundaries of the region where $S(q,\omega)$ is nonzero,
$S(q,\omega)$ shows the power-law singularity.
It is found that the divergent singularity occurs only
in the lowest edges starting from $(q,\omega) = (0,0)$
toward positive and negative $q$.
The analytic result is checked numerically for finite systems
via exact diagonalization and recursion methods.
\end{abstract}

\bigskip





\section{Introduction}
\label{Introduction}

In recent years, there has been a renewal of interest in study
of the spin chains which describe the systems with the orbital degeneracy.
In the case of two-fold orbital degeneracy, the total degeneracy per
site becomes $4 \ (=2\times 2)$, and
the simplest model to realize this situation in one dimension is
a chain with the $SU(4)$ symmetry ($SU(4)$ spin-orbital model)
\cite{LMSZ,PSK,YSU1,YSU2,FMT,MFDT,AGLN,ABL,LL,IQA}.
The static property of the $SU(4)$ spin-orbital model has been studied
mainly by numerical methods.
It has been reported that the spin correlation has a period of four unit
cells, and that the asymptotic decay
has a power-law exponent different from unity \cite{YSU1,FMT}.
Such exponent has also been derived by use of
conformal field theory \cite{Affleck}.

In addition to these static properties,
dynamical information of the system has become more and more important
partly because experimental investigation for dynamical properties
has been performed with increasing accuracy.
In particular experiments of some orbitally degenerate
quasi-one-dimensional magnetic compounds
such as NaV$_2$O$_5$ \cite{Isobe}
are intensively performed in the last few years.
Hence it is useful to clarify the difference from systems without
orbital degeneracy not only for static properties,
but also for dynamic ones.

Although several one-dimensional quantum spin models
have been exactly solved,
much less is known about their dynamical properties over
the whole energy range.
An exception is the XX model \cite{xx},
which can be mapped to a spinless fermion model.
For other models dynamics has been discussed mainly
by bosonization method
which works only in the low-frequency region.
Recently, some progress has been achieved in obtaining exact results
for a wider energy range
for the antiferromagnetic Heisenberg model \cite{xxx1,xxx2} and
the XXZ model in antiferromagnetically ordered phase \cite{xxz}.
In their articles \cite{xxx1,xxx2,xxz}, Bougourzi et al. derived
two-spinon contribution to dynamical structure factors
$S(q,\omega)$ for the  Heisenberg and XXZ
models in the thermodynamic limit
by using the results of mathematical work \cite{JM}.
For the  Heisenberg model, they obtained that
the two-spinon excitations account for 72.89\% of the total intensity.
Also they showed that the exact result
and the M\"{u}ller ansatz for $S(q,\omega)$ \cite{Muller}
have the same singularity at the lower spectral boundary, 
which is called the des Cloizeaux-Pearson mode.

On the other hand, in Ref. \cite{HZ},
Haldane and Zirnbauer have
obtained exact expression of $S(q,\omega)$ at
zero temperature for the Haldane-Shastry (HS) model,
which is a Heisenberg type antiferromagnetic chain with
inverse-square interaction \cite{HS1,HS2}.
They used the supermatrix method \cite{Efetov}
which relies on close correspondence
between the HS model and
spinless Calogero-Sutherland (CS) model \cite{Cal,Suth}
with special coupling parameter.
The most remarkable feature of their results is that
only two-spinon excitations contribute to $S(q,\omega)$.
Also, the form of $S(q,\omega)$
is regarded as a variant of the M\"{u}ller ansatz form
with quadratic dispersion of spinons.
Since the HS model can be regarded
as the model of free spinon gas \cite{Haldane}
obeying the fractional exclusion statistics \cite{fes},
their results provide clear physical interpretation
of the M\"{u}ller ansatz.

With growing interest in orbitally degenerate systems
as well as progress in exact dynamical theory,
one can naturally ask for example how the dynamical
property depends on the number of
internal degrees of freedom.
This problem has been left as a theoretical challenge.
In this paper, we report on exact dynamics of
the $SU(K)$ HS model $(K\geq 3)$
\cite{Kawakami,HaHaldane} in the whole energy and momentum region.
We derive exact formulae for the
dynamical structure factor
for arbitrary size of the system
including the thermodynamic limit.
We can construct the quasi-particle picture of the spin dynamics
at zero temperature.
In Ref. \cite{YSAK}, we have briefly reported the results.
We present here the details of the derivations.
Our exact result is inconsistent with a conjecture
proposed several years ago \cite{conjecture}.

The $SU(K)$ HS model is described by the
following Heisenberg-type Hamiltonian with
the periodic boundary condition \cite{Kawakami,HaHaldane}:
\be
\label{HS}
H_{\mbox{{\scriptsize HS}}}
:=\frac{1}{2}\sum_{1\leq i<j\leq N}J_{ij}P_{ij}.
\ee
Here the exchange interaction $J_{ij}$ is given by
the following inverse-square type:
\be
\label{exchange-int}
J_{ij}
:=
J\left[\frac{N}{\pi}\sin\frac{\pi(i-j)}{N}\right]^{-2},
\ee
where $N$ is the number of
lattice sites with unit spacing and $J>0$.
The Hilbert space of this model is
$\Omega_N
:=\underbrace{%
\mbox{{\bf C}}^K\otimes\cdots\otimes\mbox{{\bf C}}^K
}_{N\mbox{-{\scriptsize times}}}$,
and the linear operator $P_{ij}$ is the color exchange operator
which exchanges the $i$-th element and $j$-th element of vectors
in $\Omega_N$.
In the particular case of $SU(2)$, $P_{ij}$ is reduced to the spin
exchange $2\vec S_i\cdot\vec S_j +1/2$.
In this paper, we assume that $N\equiv 0$ mod $K$.
Under this condition, there is the non-degenerate
$SU(K)$ singlet groundstate of the system.
The operator $P_{ij}$ can be written in the form:
\be
\label{color-exchange-op}
P_{ij}
=\sum_{\gamma,\delta=1}^{K}X_i^{\delta\gamma}X_j^{\gamma\delta},
\ee
where $X_i^{\gamma\delta}$
with $\gamma,\delta\in\{1,2,\cdots,K\}=: \Sigma_K$
changes the color $\delta$ to $\gamma$ at site $i$.
The operators $X_i^{\gamma\delta}$ form a basis for the
Lie algebra $u(K)$ of the unitary group $U(K)$
and satisfy the following relations:
\ba
\label{gl(K)-relations}
[X_j^{\gamma\delta},X_k^{\rho\sigma}]
=
\delta_{jk}
(\delta_{\delta\rho}X_j^{\gamma\sigma}
-\delta_{\gamma\sigma}X_j^{\delta\rho}).
\ea
For later use, we introduce the following operators:
\be
\label{sl(K)-elements}
\tilde{X}_j^{\gamma\delta}
:=
{X}_j^{\gamma\delta}
-\delta_{\gamma\delta}
\frac{1}{K}\sum_{\rho=1}^K{X}_j^{\rho\rho}.
\ee
They satisfy the same commutation relations as in (\ref{gl(K)-relations})
and also satisfy the conditions that
\be
\label{sum-zero}
\sum_{\gamma=1}^K\tilde{X}_j^{\gamma\gamma}=0.
\ee

We will analyze the following quantity
of the $SU(K)$ HS model
at zero temperature:
\ba
\label{def-DSF-lattice}
S_N^{(\delta\gamma)(\rho\sigma)}(q,\omega)
&:=&
\frac{1}{N}
\sum_{r=1}^N\sum_{s=0}^{N-1}e^{-iqs}
\int_{-\infty}^\infty\frac{dt}{2\pi}e^{i\omega t}
\langle 0|
\tilde{X}_{r}^{\delta\gamma}(t)\tilde{X}_{r+s}^{\rho\sigma}(0)
|0\rangle
\nonumber
\\
&=&
\sum_{\alpha}
\langle 0|\tilde{X}_{-q}^{\delta\gamma}|\alpha\rangle
\langle\alpha|\tilde{X}_q^{\rho\sigma}|0\rangle
\delta(\omega-E_\alpha+E_0),
\ea
where $\delta,\gamma,\rho,\sigma\in\Sigma_K$,
$q=2\pi n/N$ with $n\in{\mbox{{\bf Z}}}$,
$\tilde{X}_{r}^{\gamma\delta}(t)$ is the Heisenberg representation
of the operator $\tilde{X}_{r}^{\gamma\delta}$, and
\be
\tilde{X}_q^{\gamma\delta}
:=\frac{1}{\sqrt{N}}\sum_{l=1}^N \tilde{X}_l^{\gamma\delta}e^{iql}.
\label{tilde-X}
\ee
Here
$\{|\alpha\rangle\}$ is the normalized complete basis
of the Hamiltonian (\ref{HS}) with
eigenvalues $\{E_\alpha\}$,
and $|0\rangle$ is the groundstate. From the $SU(K)$ symmetry,
the following formula holds:
\be
\label{DSF-finite-lattice-general}
S_N^{(\delta\gamma)(\rho\sigma)}(q,\omega)
=
\left(\delta_{\delta\sigma}\delta_{\gamma\rho}
-\frac{1}{K}
\delta_{\delta\gamma}\delta_{\rho\sigma}\right)
S_N(q,\omega),
\ee
where
\be
S_N(q,\omega):=S_N^{(\delta\gamma)(\gamma\delta)}(q,\omega),
\quad (\gamma\ne\delta).
\ee
Therefore it is sufficient to calculate the components
$S_N^{(\delta\gamma)(\gamma\delta)}(q,\omega)$
with $\gamma\ne\delta$.

For deriving the exact formula for
$S_N^{(\delta\gamma)(\gamma\delta)}(q,\omega)$
with $\delta\ne\gamma$,
we adopt the freezing trick
introduced in Ref. \cite{Polychronakos}.
This method is based on the fact that the $SU(K)$ HS model
is obtained by the strong coupling limit of
the $U(K)$ spin CS model \cite{HaHaldane}.
The latter continuous model is more tractable than the $SU(K)$ HS model,
because the eigenfunctions of the model have been explicitly
constructed \cite{TU,Uglov1,Uglov2}.
If we have the exact dynamical results for
the $U(K)$ spin CS model, then we can obtain
$S_N^{(\delta\gamma)(\gamma\delta)}(q,\omega)$ of the $SU(K)$ HS model
by applying the freezing trick.

Fortunately, there exists an excellent work by Uglov \cite{Uglov1}.
In the case of $K=2$,
he derived the exact expression
of the dynamical spin-density correlation function
of the spin CS model with a finite number of particles \cite{Uglov1}.
We extend his result to the case of $K\geq 3$,
and then take the strong coupling limit.

The content of the paper is as follows:
In section \ref{freezing-trick}
we introduce the spin CS model
and then explain the freezing trick.
In section \ref{corr-spinCS}
we extend Uglov's results
on the correlation functions
to the case of $K\geq 3$.
Also, following Ref. \cite{YA},
we simplify the conditions on the excited states
of the dynamical correlation functions.
In section \ref{DSF-finite}
the dynamical structure factor
for a finite system is exactly derived.
We give the quasi-particle picture of
the spin dynamics at zero temperature.
We rewrite the resultant formula of
the dynamical structure factor
in terms of the momenta of quasi-particles.
The relation
between the momenta of quasi-particles
and motif \cite{HHTBP,BGHP}
is briefly discussed.
In section \ref{DSF-infinit}
we take the thermodynamic limit of
the dynamical structure factor.
We determine the support of
the dynamical structure factor
and analyze its behavior near
the dispersion lines of elementary excitations.
Summary and discussions are given in section \ref{summary}.
In Appendix A,
for proving the statement in section \ref{freezing-trick},
we analyze the strong coupling limit of
the dynamical density correlation function
of the $U(K)$ spin CS model.
Appendices B, C, D and E
contain the data and
proofs which are used in the paper.
In Appendix F,
we derive the low energy asymptotic form of
the dynamical correlation function
and analyze the singularities of
the dynamical structure factor.


\section{Spin Calogero-Sutherland Model and Freezing Trick}
\label{freezing-trick}


In this section,
we introduce
the $U(K)$ spin CS model \cite{HaHaldane}
and explain the freezing trick \cite{Polychronakos}.
The $U(K)$ spin CS model describes $N$ particles
with coordinates $x_1,\cdots,x_N$ moving along a circle of length $L$
and interacting with inverse-square type interactions.
Moreover, each particle carries a color with $K$ possible values.
The Hamiltonian of this model is given by \cite{HaHaldane}
\ba
\label{spinCS}
H_{\mbox{{\scriptsize spinCS}}}
:=
-\frac{1}{2}\sum_{i=1}^N\frac{\partial^2}{\partial x_i^2}
+
\left(\frac{\pi}{L}\right)^2\sum_{1\leq i<j\leq N}
\frac{\beta(\beta+P_{ij})}{\sin^2\frac{\pi}{L}(x_i-x_j)},
\ea
where $\beta > 0$ is the coupling parameter and $L$ is the size of
system.
The Hilbert space of the system is
$\mbox{{\bf C}}^{\infty}(x_1,\cdots,x_N)\otimes\Omega_N$.
Technically, we assume that $N\equiv 0 \mbox{ mod }K$ and
$N/K\equiv 1 \mbox{ mod } 2.$\footnote{%
The results of the paper on the dynamical structure factor
of the $SU(K)$ HS model
hold for the case
of $N/K\equiv 0 \mbox{ mod } 2$.}
In this case, the system has
the non-degenerate $U(K)$ singlet groundstate.
The spinless CS model is given by (\ref{spinCS})
where $P_{ij}$ has been set to 1, i.e., $K=1$.

It is convenient to consider the following transformed Hamiltonian:
\be
\label{trans-Hamil}
H_\beta
:=
W^{-\beta}H_{\mbox{{\scriptsize spinCS}}}W^\beta,
\ee
where $W:=\prod_{1\leq j<k\leq N}\sin\pi(x_j-x_k)/L$.
The Hilbert space of the system with the above Hamiltonian is
\be
\label{Hilbert-trans-Hamil}
{\cal H}
:=
\left(
\mbox{{\bf C}}[z^{\pm 1}_1,\cdots,z^{\pm 1}_N]\otimes\Omega_N
\right)_{\mbox{{\scriptsize antisymm}}},
\ee
where $z_j:=\exp(2\pi i x_j/L)$, and ``antisymm'' means
total antisymmetrization.
The inner product
$(\bullet,\bullet)_\beta$ of this Hilbert space
which depends on the coupling parameter $\beta$
is appropriately defined \cite{TU,Uglov1,Uglov2}.

Next we explain the freezing trick \cite{Polychronakos}.
Let us consider the strong coupling limit $\beta\rightarrow\infty$ of
the $U(K)$ spin CS model.
In this case particles
crystallize with the lattice parameter $L/N$ which is taken as the unit
of length. Then we are left with the
center of mass motion, the lattice vibration and the dynamics of
the color.
The color dynamics is equivalent to the dynamics of the $SU(K)$ HS
model.
The freezing trick described above was firstly introduced by
Polychronakos \cite{Polychronakos}, and
has been applied to thermodynamics of
lattice models \cite{SS,KK1,KK2}.
The present work
is the first application of the freezing trick to
dynamical quantities.

In the $U(2)$ spin CS model, Uglov has derived the exact formula
of the dynamical spin-density correlation function with a finite number
of particles \cite{Uglov1}.
In the next section
we extend his result to the case of $K\geq 3$, and
then, in section \ref{DSF-finite}, take the strong coupling limit.
In doing so we have to make correspondence between physical quantities
defined in the continuum and
discrete spaces.
In analogy to the operator $\tilde{X}_q^{\gamma\delta}$ defined by
Eq. (\ref{tilde-X}) for the lattice model,
let us define the following operator in the continuum space:
\be
\label{X_q}
\bar{X}_q^{\gamma\delta}
:=
\frac{1}{\sqrt L}\sum_{j=1}^N
{X}_{j}^{\gamma\delta}e^{-iqx_j},
\ee
where the momentum $q$ takes values $2\pi n/L$ with $n$ an arbitrary
integer.
We first derive the dynamical structure factor in the continuum model
defined by
\be
\label{def-DSF-continuum}
S_N^{(\delta\gamma)(\rho\sigma)}(q,\omega;\beta)
:=
\sum_{\nu}
(0|\bar{X}_{-q}^{\delta\gamma}|\nu)
(\nu|\bar{X}_q^{\rho\sigma}|0)
\delta(\omega-\bar{E}_\nu+\bar{E}_0),
\ee
where $\{|\nu)\}$ is the normalized complete basis of the
Hamiltonian (\ref{trans-Hamil}) with
eigenvalues $\{\bar{E}_\nu\}$,
and $|0)$ is the groundstate.

In the strong coupling limit the
coordinate $x_j$ in Eq. (\ref{X_q}) is written as
$x_j=R_j+u_j$ where  $R_j =j $ is a lattice point, and $u_j$
describes the lattice vibration.
Except for the uniform motion of the lattice
described by $u_j = const$,
we may regard $u_j$ as a small quantity.
In fact the density response can be shown to be smaller than the spin
response by ${\cal O}(\beta^{-1})$.
This fact is shown in Appendix A.
Therefore the dynamical structure factor
of the $SU(K)$ HS model is given
simply by the strong coupling limit of
Eq. (\ref{def-DSF-continuum})
provided that one restricts $q$ in the range of the
first Brillouin zone: $|q| \le \pi$.

\section{Dynamical Correlation Functions
of the Spin Calogero-Sutherland Model}
\label{corr-spinCS}

In Ref. \cite{Uglov1},
Uglov introduced the new formulation for the eigenvalue problem
of the $U(K)$ spin CS model
and derived the dynamical correlation functions
in the case of $K=2$.
In this section we extend Uglov's result
on the dynamical correlation functions
to the case of $K\geq 3$.
First we recall his formulation briefly
to establish the necessary terminology and notations.
For details, see Ref. \cite{Uglov1}.
Then we give the dynamical correlation functions
of the $U(K)$ spin CS model.
We also simplify the conditions on the excited states
relevant to  the dynamical correlation functions.

\subsection{Notations and
review of Uglov's results}
\label{Uglov}

In Ref. \cite{TU}, Takemura and Uglov introduced
the (un-normalized) complete orthogonal bases
$\{X_{\sigma+k^{(0)}}^{(\beta,K)}\}$ of ${\cal H}$
with respect to the inner product $(\bullet,\bullet)_\beta$
which is called the Yangian Gelfand-Zetlin
basis and is parametrized by the decreasing sequence of
integers $\sigma=(\sigma_1,\cdots,\sigma_N)$.
Here $k^{(0)}:=(M,M-1,\cdots,M-N+1)$
with $M:=(N+K)/2\in\mbox{{\bf Z}}_{>0}$
is the parameter which corresponds to
the groundstate of the system.
(Notice that $N\equiv 0 \mbox{ mod }K$ and
$N/K\equiv 1 \mbox{ mod } 2$.)
The label $\sigma+k^{(0)}$ of the state
contains the information
both for the momenta and for the colors
of the particles \cite{Uglov1}.
That is, if we write $\sigma+k^{(0)}=\underbar{\mbox{$k$}}+K\bar{k}$
with
$\underbar{\mbox{$k$}}
=(\underbar{\mbox{$k$}}_1,\cdots,\underbar{\mbox{$k$}}_N)
\in(\Sigma_K)^N$ and
$\bar{k}=(\bar{k}_1,\cdots,\bar{k}_N)\in\mbox{{\bf Z}}^N$,
then $\underbar{\mbox{$k$}}$ and $\bar{k}$ represent colors
and momenta of particles, respectively.\footnote{%
In Ref. \cite{Uglov1},
Uglov used the different convention:
$\sigma+k^{(0)}=\underbar{\mbox{$k$}}-K\bar{k}$.}

Let us consider the vector space
${\cal H}'
:=(\mbox{{\bf C}}[y^{\pm 1}_1,\cdots,y^{\pm 1}_N])_{\mbox{{\tiny symm}}}$
of the symmetric Laurent polynomials in $y=(y_1,\cdots,y_N)$.
In Ref. \cite{Uglov1},
Uglov defined the (un-normalized) complete orthogonal bases
$\{J_\sigma^{(K\beta+1,K)}\}$
of ${\cal H}'$ with respect to a certain $\beta$-dependent inner product
$(\bullet,\bullet)'_\beta$.
The function $J_\sigma^{(K\beta+1,K)}(y)$ is
defined by a certain degeneration of
the Macdonald symmetric function \cite{Macd}
and is called the $gl_K$-Jack symmetric function. From this definition,
several properties of the $gl_K$-Jack function
can be derived from the corresponding properties
of the Macdonald function.
Moreover Uglov defined the isomorphism $\Theta$
of Hilbert spaces
$({\cal H},(\bullet,\bullet)_\beta)$
and $({\cal H}',(\bullet,\bullet)'_\beta)$
such that
$\Theta(X_{\sigma+k^{(0)}}^{(\beta,K)}(z))=J_\sigma^{(K\beta+1,K)}(y)$
with the normalization
$\Theta(X_{k^{(0)}}^{(\beta,K)}(z))=1$
($z=(z_1,\cdots,z_N)$).
This fact enables one to calculate the correlation functions
of the $U(K)$ spin CS model
by using the $gl_K$-Jack function.
Actually, in the case of $K=2$,
Uglov exactly derived the dynamical density and spin-density
correlation functions
of the spin CS model.
In the next subsection we extend his result
on the correlation functions
to the case of $K\geq 3$.

We recall here the norm formula
of the Yangian Gelfand-Zetlin basis \cite{Uglov1}.
We also recall the spectral decomposition of the power sum
symmetric function
$p_m(y):=\sum_{i=1}^Ny_i^m$ with
$m\in\mbox{{\bf Z}}_{>0}$
in terms of the $gl_K$-Jack polynomials \cite{Uglov1}
which is used in the next subsection.
For this purpose,
we fix some notations for partitions \cite{Uglov1,Uglov2,Macd}.
For a non-negative integer $N$, let $\Lambda_N
:=\{\lambda=(\lambda_1,\lambda_2,\cdots,\lambda_N)
\in(\mbox{{\bf Z}}_{\geq 0})^N\,
|\,\lambda_1\geq\lambda_2\geq\cdots\geq\lambda_N\}$
be the set of all partitions with length less or equal to $N$.
A partition can be represented by a planar diagram,
i.e., by the so-called Young diagram.
When there is a box in the $i$-th row and $j$-th
column of $\lambda$, we write $(i,j)\in\lambda$.
The conjugate of a partition
$\lambda=(\lambda_1,\lambda_2,\cdots,\lambda_N)\in\Lambda_N$
is the partition
$\lambda'
=(\lambda'_1,\lambda'_2,\cdots,\lambda'_{\lambda_1})\in\Lambda_{\lambda_1}$
whose diagram is the transpose of the diagram $\lambda$.
For instance, if $\lambda=(4,3,1)$, then $\lambda'=(3,2,2,1)$.
For a box
$s=(i,j)\in\lambda$, the numbers
$a_\lambda(s)=\lambda_i-j$, $a_\lambda'(s)=j-1$,
$l_\lambda(s)=\lambda'_j-i$ and $l_\lambda'(s)=i-1$
are called
the arm-length, arm-colength,
leg-length and leg-colength
of box $s$, respectively.
We recall a coloring scheme of diagrams.
In considering the $U(K)$ symmetry, we need $K$ colors.
For a partition $\lambda$, we define subsets of $\lambda$ by
$C_K^{(a)}(\lambda)
:=
\{s\in\lambda\,|\,a_\lambda'(s)-l_\lambda'(s)\equiv a\ \mbox{mod}\, K\}$.
We call the color of box $s\in\lambda$ $a$
if $s\in C_K^{(a)}(\lambda)$.
Notice that $(1,1)\in C_K^{(0)}(\lambda)$ (if $\lambda\ne\O$).
For example, in the case of $K=3$, if $\lambda=(4,3,1)$ then
we have the colored diagram
{\scriptsize
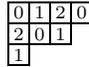
\begin{eqnarray*}
\Yvcentermath1
\young(
\zero\one\two\zero,%
\two\zero\one,%
\one
)
\end{eqnarray*}
}
We also define the subsets of $\lambda$ by
$H_K^{(a)}(\lambda)
:=\{s\in\lambda\,|\,a_\lambda(s)+l_\lambda(s)+1
\equiv a\ \mbox{mod}\, K\}$.
The quantity $a_\lambda(s)+l_\lambda(s)+1$ for
the box $s\in\lambda$ is called the
hook-length of box $s$.
Therefore the set $H_K^{(a)}(\lambda)$
is the set which consists of the boxes in $\lambda$
with the hook-length $a$ modulo $K$.
For instance, in the $K=3$ case, the modulo $K$ hook-length of
each box in the partition $\lambda=(4,3,1)$
is given by
{\scriptsize
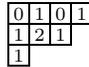
\begin{eqnarray*}
\Yvcentermath1
\young(
\zero\one\zero\one,%
\one\two\one,%
\one
)
\end{eqnarray*}
}
For any subset $\nu\subset\lambda$ the order $|\nu|$ is defined as
the number of boxes in $\nu$,
i.e.,
for example, $|C_K^{(a)}(\lambda)|$
is the number of boxes with color $a$ in $\lambda$.

The norm of the Yangian Gelfand-Zetlin
basis
$X_{\lambda+k^{(0)}}^{(\beta,K)}(z)$
for the partition $\lambda$ relative to
the norm of groundstate
is given by
\ba
\label{norm}
N_\lambda(\beta)
&:=&
\frac{(X_{\lambda+k^{(0)}}^{(\beta,K)},X_{\lambda+k^{(0)}}^{(\beta,K)})_\beta}
     {(X_{k^{(0)}}^{(\beta,K)},X_{k^{(0)}}^{(\beta,K)})_\beta}
\left(
=
\frac{(J_\lambda^{(K\beta+1,K)},J_\lambda^{(K\beta+1,K)})'_\beta}
     {(1,1)'_\beta}
\right)
\nonumber
\\
&=&
\prod_{s\in C_K^{(0)}(\lambda)}
\frac{a'_\lambda(s)+(K\beta+1)(N-l'_\lambda(s))}
     {a'_\lambda(s)+1+(K\beta+1)(N-l'_\lambda(s)-1)}
\nonumber
\\
&\times&
\prod_{s\in H_K^{(0)}(\lambda)}
\frac{a_\lambda(s)+1+(K\beta+1)l_\lambda(s)}
     {a_\lambda(s)+(K\beta+1)(l_\lambda(s)+1)}.
\ea
The spectral decomposition of the power sum $p_m(y)$ ($m>0$)
in terms of the $gl_K$-Jack polynomials
is given by
\be
\label{powe-sum2Jack}
p_m(y)
=
\sum_{{\lambda\in\Lambda_N}
      \atop
      {|\lambda|=m}}
c_\lambda(\beta)
J_\lambda^{(K\beta+1,K)}(y),
\ee
where
$c_\lambda(\beta)$
is an expansion coefficient.
To give this we introduce the notation:
$\hat{\delta}(A)=1$ if a statement $A$ is true,
$\hat{\delta}(A)=0$ otherwise.
Furthermore we define
$\omega_K:=e^{i2\pi/K}$, and
\ba
\label{c-h}
c\mbox{-}h^{(a)}_\lambda
&:=&|C_K^{(a)}(\lambda)|-|H_K^{(a)}(\lambda)|,
\\
\label{sum-l'}
n(\lambda)&:=&\sum_{s\in\lambda}l_\lambda'(s).
\ea
Then we obtain
$c_\lambda(\beta)
=
A_\lambda
|\lambda|^{\hat{\delta}(|\lambda|\equiv 0\mbox{ {\scriptsize mod} }K)}
\chi_\lambda(\beta)
$
where
\ba
\label{matrix-element}
\chi_\lambda(\beta)
:=
\frac{\prod_{s\in C_K^{(0)}(\lambda)\setminus\{(1,1)\}}
      (a'_\lambda(s)-(K\beta+1)l'_\lambda(s))}
     {\prod_{s\in H_K^{(0)}(\lambda)}
      (a_\lambda(s)+1+(K\beta+1)l_\lambda(s))},
\ea
and the $\beta$-independent constant
$A_\lambda$ is given by
\ba
\label{coeff-matrix-element}
A_\lambda
:=
\left\{
\begin{array}{l}
\omega_K^{n(\lambda)}
\prod_{a=1}^{K-1}(1-\omega_K^a)^{c\mbox{-}h_\lambda^{(a)}},
\,\mbox{for } |\lambda|\equiv 0 \mbox{ mod } K,\, c\mbox{-}h_\lambda^{(0)}=0,
\\
\omega_K^{n(\lambda)}
\prod_{a=1}^{K-1}(1-\omega_K^a)^{c\mbox{-}h_\lambda^{(a)}+\delta{{\tiny ab}}},
\,\mbox{for } |\lambda|\equiv b \mbox{ mod } K,\, c\mbox{-}h_\lambda^{(0)}=1,
\,(b=1,\cdots,K-1),
\\
0,
\,\mbox{otherwise}.
\end{array}
\right.
\ea

\subsection{Dynamical correlation functions of
the spin Calogero-Sutherland model}
\label{DCF-spinCS}

In this subsection, we derive
the following function
of the $U(K)$ spin CS model:
\be
\label{def-DCCF-continuum}
\eta_N^{(\delta\gamma)(\rho\sigma)}(x,t;\beta)
:=
(0|\bar{X}^{\delta\gamma}(x,t)\bar{X}^{\rho\sigma}(0,0)|0),
\ee
where the ``color-density" operator is defined by
\be
\label{color-density}
\bar{X}^{\gamma\delta}(x)
:=\sum_{j=1}^N\delta(x-x_j){X}_j^{\gamma\delta}
=
\frac{1}{L}
\sum_{j=1}^N
\sum_{n\in\mbox{{\scriptsize\bf Z}}}
e^{i2\pi nx/L}z^{-n}_j{X}_j^{\gamma\delta}
=
\frac{1}{\sqrt{L}}
\sum_{q}e^{iqx}\bar{X}_{q}^{\gamma\delta},
\ee
with $q=2\pi n/L$.
The function $\eta_N^{(\delta\gamma)(\rho\sigma)}(x,t;\beta)$ is the Fourier
transform of the dynamical structure factor (\ref{def-DSF-continuum}), 
and is called the dynamical color correlation function (DCCF)
in the following.

First we should determine the $SU(K)$ spin
selection rule of the excited states
of the correlation functions.
Here the $SU(K)$ spin means
the $K-1$ eigenvalues $(s_1,\cdots,s_{K-1})$
of a set of operators $(h^1,\cdots,h^{K-1})$ where
$h^a$ is defined by
\be
\label{SU(K)-spin-op}
h^a
:=\sum_{i=1}^N(X_i^{aa}-X_i^{a+1,a+1})/2
\ee
for $a=1,\cdots,K-1$.
For the $SU(2)$ case, this definition gives the ordinary spin because
$h^1=\sum_{i=1}^N \sigma_i^z/2$ where $\sigma_i^z=\mbox{diag}(1,-1)$
is the $z$-component of the Pauli matrices.
The set $\{2h^a\}_{a=1}^{K-1}$ is the standard basis
of the Cartan subalgebra of the Lie algebra $sl_K$.
Since the Cartan subalgebra is the maximal commuting subalgebra,
the eigenvalues $(s_1,\cdots,s_{K-1})$
are good quantum numbers of the system with the $SU(K)$ symmetry.
We denote the $SU(K)$ spin of the state
$|\nu)$ which satisfies $(\nu|\bar{X}^{\gamma\delta}(0)|0)\ne 0$
by $s(\gamma,\delta)=(s_1(\gamma,\delta),\cdots,s_{K-1}(\gamma,\delta))$.
We see that 
the above state $|\nu)$ 
transforms as one of the weight vectors for 
the adjoint representation $Ad$ of $SU(K)$.
For example, in the $SU(3)$ case,
we have $s(1,3)=(1/2,1/2)$.
The complete lists of $s(\gamma,\delta)$ for $K=2,3,4$
are given in Appendix B.

We cannot
directly calculate the DCCF (\ref{def-DCCF-continuum}),
since the action of
the operator $\Theta\bar{X}^{\gamma\delta}(0)\Theta^{-1}$
on ${\cal H}'$ does not have simple expression.
The key step for deriving the DCCF (\ref{def-DCCF-continuum})
is to choose appropriate local operators on ${\cal H}$.
We define local operators
$J_a(z)
=
\sum_{n\in\mbox{{\scriptsize\bf Z}}}
J_{a,n}(z)$ ($a=1,\cdots,K-1$) on ${\cal H}$
by
\be
\label{local-op}
J_{a,n}(z)
:=
\sum_{i=1}^N z_i^{-n-1}\sum_{b=1}^a X_i^{b,K-a+b}
+
\sum_{i=1}^N z_i^{-n}\sum_{b=1}^{K-a} X_i^{a+b,b}.
\ee
The DCCF $\eta_N^{(\delta\gamma)(\rho\sigma)}(x,t;\beta)$
with $\delta\ne\gamma,\rho\ne\sigma$
can be obtained from the dynamical correlation function
of the operator $J_a(z)$.
This is based on the following formula:
if $f$ and $g$ are eigenfunctions of the spin operators $h^b$
with eigenvalues $s_b(f)$ and
$s_b(g)$, respectively, then we have
\be
\label{color2local-op}
(f,\bar{X}^{\gamma\delta}(0)g)_\beta
=
\prod_{b=1}^{K-1}
\hat{\delta}(s_b(f)-s_b(g)=s_b(\gamma,\delta))
(f,(1/L)J_a(z)g)_\beta,
\ee
where $\gamma-\delta\equiv a\mbox{ mod }K$.
Moreover, we have
\be
\label{color-densityinH'}
\Theta J_{a,n}(z)\Theta^{-1}=p_{nK+a}(y),
\ee
i.e., the operator $\Theta J_{a,n}(z)\Theta^{-1}$ on ${\cal H}'$
is the multiplication of the power sum.
Then we have
\be
\label{color2-power-sum}
(f,\bar{X}^{\gamma\delta}(0)g)_\beta
=
\prod_{b=1}^{K-1}
\hat{\delta}(s_b(f)-s_b(g)=s_b(\gamma,\delta))
(\Theta(f),
(1/L)\sum_{n\in\mbox{{\scriptsize\bf Z}}}p_{nK+a}(y)\Theta(g))'_\beta.
\ee
If we know
the norm of the Yangian Gelfand-Zetlin basis
$X_{\sigma+k^{(0)}}^{(\beta,K)}(z)$
(or $gl_K$-Jack function $J_\sigma^{(K\beta+1,K)}(y)$) and
the matrix element of the power sum
with respect to the bases $\{J_\sigma^{(K\beta+1,K)}\}$,
we can calculate the DCCF
$\eta_N^{(\gamma\delta)(\rho\sigma)}(x,t;\beta)$
with $\gamma\ne\delta,\rho\ne\sigma$.
We already know these quantities explicitly.

Now we can give the formula for
${\eta}_N^{(\delta\gamma)(\gamma\delta)}(x,t;\beta)$
and
$S_N^{(\delta\gamma)(\gamma\delta)}(q,\omega;\beta)$
of the $U(K)$ spin CS model.
The momentum ${\cal P}_\lambda$,
excited energy ${\cal E}_\lambda$
and
$SU(K)$ spins $S^a_\lambda$
for the Yangian Gelfand-Zetlin basis
$X_{\lambda+k^{(0)}}^{(\beta,K)}(z)$ with
partition $\lambda$
are respectively given by
\ba
\label{momentum-finite-continuum}
{\cal P}_\lambda
&=&
\frac{2\pi}{L}|C_K^{(0)}(\lambda)|,
\\
\label{ex-energy-finite-continuum}
{\cal E}_\lambda
&=&
\frac{1}{2K}\left(\frac{2\pi}{L}\right)^2
\left[
2n_K(\lambda')-2(K\beta+1)n_K(\lambda)
+((N-1)(K\beta+1)+1)|C_K^{(0)}(\lambda)|
\right],
\\
\label{spin}
S^a_\lambda
&=&
\frac{1}{2}\left(
|C_K^{(a-1)}(\lambda)|-2|C_K^{(a)}(\lambda)|+|C_K^{(a+1)}(\lambda)|
\right),
\quad (a=1,\cdots,K-1),
\ea
where
\ba
\label{sum-l0'}
n_K(\lambda):=\sum_{s\in C_K^{(0)}(\lambda)}l_\lambda'(s),
\\
\label{sum-a0'}
n_K(\lambda'):=\sum_{s\in C_K^{(0)}(\lambda)}a_\lambda'(s).
\ea
The following relations hold for these quantities:
\ba
\label{momentum-negative}
{\cal P}_{\lambda^*}&=&-{\cal P}_{\lambda},
\\
\label{ex-energy-negative}
{\cal E}_{\lambda^*}&=&-{\cal E}_{\lambda},
\\
\label{spin-negative}
S^a_{\lambda^*}&=&S^a_{\lambda},
\ea
where $\lambda=(\lambda_1,\cdots,\lambda_N)$ and
$\lambda^*:=(-\lambda_N,\cdots,-\lambda_1)$.
We use above relations for deriving the DCCF.
Then, for $\gamma,\delta\in\Sigma_K$
with $\gamma\ne\delta$, we have
\ba
\label{DCCF-finite-continuum}
{\eta}_N^{(\delta\gamma)(\gamma\delta)}(x,t;\beta)
&=&
\frac{1}{L}
\sum_{\lambda\in\Lambda_N}
\hspace{-1.1mm}{}^{'}\hspace{1mm}
|{F}_\lambda(\beta)|^2
e^{-it{\cal E}_\lambda+ix{\cal P}_\lambda},
\\
\label{DSF-finite-continuum}
S_N^{(\delta\gamma)(\gamma\delta)}(q,\omega;\beta)
&=&
\sum_{\lambda\in\Lambda_N}
\hspace{-1.1mm}{}^{'}\hspace{1mm}
|{F}_\lambda(\beta)|^2
\delta_{n,|C_K^{(0)}(\lambda)|}\delta(\omega-{\cal E}_\lambda),
\ea
where $q=2\pi n/L$, and
$\delta_{n,|C_K^{(0)}(\lambda)|}$ represents
the momentum conservation $q={\cal P}_\lambda$.
The form factor ${F}_\lambda(\beta)$ is given by
\be
\label{ff-finite-continuum}
|{F}_\lambda(\beta)|^2
:=\frac{1}{L}|A_\lambda \chi_\lambda(\beta)|^2 N_\lambda(\beta).
\ee
The primed summation in the formulae
(\ref{DCCF-finite-continuum}) and (\ref{DSF-finite-continuum})
is taken over partitions $\lambda\in\Lambda_N$
which satisfy the following conditions:
\ba
\label{conditions-DSF}
\left\{
\begin{array}{l}
|\lambda|\equiv \gamma-\delta \mbox{ mod }K,
\\
S^a_\lambda=s_a(\gamma,\delta),\,a=1,\cdots,K-1,
\\
c\mbox{-}h^{(0)}_\lambda=1.
\end{array}
\right.
\ea

The non-zero condition of the form factor
is equivalent to that of
the matrix element $\chi_\lambda(\beta)$
of the local operator.
Notice that, if $\beta=r/s$ with $r,s$ coprime,
then
\be
\label{nonzero-continuum}
\chi_\lambda(\beta)\ne 0
\Leftrightarrow
(s+1,rK+s+1)\notin\lambda.
\ee
Under the above condition on $\beta$,
we denote the subset $\Lambda_N^{<K\beta+1>}$ of $\Lambda_N$
by
$\{\lambda\in \Lambda_N\,|\,
(s+1,rK+s+1)\notin \lambda\}$.
Then we can replace the set $\Lambda_N$ in the sum
of the formulae
(\ref{DCCF-finite-continuum}) and
(\ref{DSF-finite-continuum})
by $\Lambda_N^{<K\beta+1>}$.

\subsection{Classification of excited states}
\label{type}

The conditions on the excited states of
the formulae (\ref{DCCF-finite-continuum})
and (\ref{DSF-finite-continuum})
are rather complicated.
In this subsection, following Ref. \cite{YA},
we simplify these conditions.
We define the concept {\it type} of a partition.
We introduce a reductive transformation $\tau$
on the set of all partitions as follows
(see Ref. \cite{YA} for the case of $K=2$):\\
(i) If there exist $K$ rows or $K$ columns which have same
number of boxes in a partition,  remove those rows or columns;\\
(ii) Apply the reduction (i) repeatedly until the newly generated
partition is no longer reducible.\\
It is important to note that
this transformation has the following properties:
\ba
\label{properties}
\left\{
\begin{array}{l}
|\lambda|\equiv a \mbox{ mod } K
\Leftrightarrow
|\tau(\lambda)|\equiv a \mbox{ mod } K,
\\
S^{b}_{\lambda}
=
S^{b}_{\tau(\lambda)},
\\
c\mbox{-}h^{(c)}_{\lambda}
=
c\mbox{-}h^{(c)}_{\tau(\lambda)}.
\end{array}
\right.
\ea
We can determine the subset ${\cal A}_N$ of the set $\Lambda_N$
as the image of $\tau$, i.e., ${\cal A}_N:=\tau(\Lambda_N)$.
Notice that ${\cal A}_N$ is the finite set.
Then, for a partition $\lambda\in\Lambda_N$,
we say that the {\it type} of $\lambda$ is $\nu\in{\cal A}_N$ if
$\nu=\tau(\lambda)$.
For example, in the case of $K=3$,
the {\it type} of the partition $(7,4,4,2,2,2,1)$ is $(1)$:
{\scriptsize
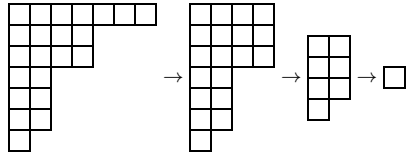
\begin{eqnarray*}
\Yvcentermath1
\yng(7,4,4,2,2,2,1)
\rightarrow
\yng(4,4,4,2,2,2,1)
\rightarrow
\yng(2,2,2,1)
\rightarrow
\yng(1)
\end{eqnarray*}
}

Let us consider the subset
${\cal A}^{<K\beta+1>}_N:=\tau(\Lambda^{<K\beta+1>}_N)$
of ${\cal A}_N$.
We have the decomposition
${\cal A}^{<K\beta+1>}_N
=
\sqcup_{\gamma,\delta=1,\gamma\ne\delta}^K
{\cal A}^{<K\beta+1;\gamma\delta>}_N
\sqcup{\cal D}_N,
$
where
\be
\label{A-continuum}
{\cal A}^{<K\beta+1;\gamma\delta>}_N
:=
\{\lambda\in{\cal A}^{<K\beta+1>}_N\,|\,
\lambda \mbox{ satisfies conditions (\ref{conditions-DSF})}
\}
\ee
and ${\cal D}_N$ is a certain set.
Then, from the properties
(\ref{properties}) and conditions (\ref{conditions-DSF}),
we can rewrite the formulae (\ref{DCCF-finite-continuum})
and (\ref{DSF-finite-continuum}).
For example, we have
\ba
\label{DCCF-finite-continuum'}
{\eta}_N^{(\delta\gamma)(\gamma\delta)}(x,t;\beta)
=
\frac{1}{L}
\sum_{\lambda\in\Lambda_N^{<K\beta+1>}}
\hspace{-5mm}{}^{''}\hspace{1.3mm}
|{F}_\lambda(\beta)|^2
e^{-it{\cal E}_\lambda+ix{\cal P}_\lambda},
\ea
where the double-primed summation is taken over
partitions $\lambda\in\Lambda_N^{<K\beta+1>}$
such that $\tau (\lambda)=\nu$ for some
$\nu\in {\cal A}_N^{<K\beta+1;\gamma\delta>}$.
The important point is that
the conditions (\ref{conditions-DSF}) on the excited states
which mainly come from $SU(K)$ spin selection rule
are conveniently implemented by introducing
a finite set ${\cal A}_N^{<K\beta+1;\gamma\delta>}$,
and by decomposing the summation over $\lambda$ by each {\it type}
$\nu = \tau (\lambda)$ such that
$\nu\in {\cal A}_N^{<K\beta+1;\gamma\delta>}$.
We do not give the explicit forms of the sets
${\cal A}_N^{<K\beta+1;\gamma\delta>}$, since
they are not needed in the following discussion.


\section{Dynamical Structure Factor
for Finite Systems}
\label{DSF-finite}


\subsection{Strong coupling limit}
\label{SU(K)}

{}From the discussion in section \ref{freezing-trick},
the DCCF
$\eta_N^{(\delta\gamma)(\rho\sigma)}(r,t)
:=
\langle 0|
\tilde{X}_{r+s}^{\delta\gamma}(t)
\tilde{X}_s^{\rho\sigma}|0\rangle$
and dynamical structure factor
$S_N^{(\delta\gamma)(\rho\sigma)}(q,\omega)$
of the $SU(K)$ HS model with
$\delta\ne\gamma,\,\rho\ne\sigma$
are simply given by
$\eta_N^{(\delta\gamma)(\rho\sigma)}(r,t;\infty)$
and $S_N^{(\delta\gamma)(\rho\sigma)}(q,\omega;\infty)$,
respectively. Following subsection \ref{type},
we first discuss the conditions on the excited states for
$\eta_N^{(\delta\gamma)(\gamma\delta)}(r,t)$
and $S_N^{(\delta\gamma)(\gamma\delta)}(q,\omega)$.
Due to the conditions $c\mbox{-}h_\lambda^{(0)}=1$
and $|\lambda|\equiv\gamma-\delta(\ne 0)\mbox{ mod }K$,
the quantity $\chi_\lambda(\beta)$ is regular
in the limit $\beta\rightarrow\infty$.
In this limit, the simplification occurs
because of the following fact:
\be
\label{nonzero-lattice}
\chi_\lambda(\infty)\ne 0
\Leftrightarrow
(1,K+1)\notin \lambda.
\ee
This means the excited state $\lambda$
consists of $K$ columns, i.e.,
$\lambda'$ has the form $\lambda'=(\lambda'_1,\cdots,\lambda'_K)$.
We define the subset $\Lambda_N^{(K)}$ of $\Lambda_N$
by $\Lambda_N^{(K)}
:=\{\lambda=(\lambda_1,\cdots\lambda_M)\,|\,\lambda_1\leq K\}$.
This is the finite set of partitions whose largest entry is less than $K+1$.
We then determine a subset ${\cal A}^{(K)}_N$ of $\Lambda^{(K)}_N$
as the image of $\tau$, i.e., ${\cal A}^{(K)}_N:=\tau(\Lambda^{(K)}_N)$.
We can explicitly determine this finite set ${\cal A}^{(K)}_N$.
The total number of elements in the set ${\cal A}^{(K)}_N$
increases from 3 in the case of $SU(2)$
to 25 in $SU(3)$, and
to 252 in $SU(4)$.
We define the subsets of ${\cal A}^{(K)}_N$ by
${\cal A}^{(K;\gamma\delta)}_N
:=
\{\nu\in{\cal A}^{(K)}_N\,|\,
S^{a}_{\nu}=s_a(\gamma,\delta),\,a=1,\cdots,K-1\}.
$
For example, in the case of $K=3$, we have
${\cal A}_N^{(3;21)}=\{(2,1,1),(3,3,1),(3,3,2,2)\}$ and
${\cal A}_N^{(3;13)}=\{(1),(2,2),(3,2,1,1)\}$.
In Appendix B, we give these sets explicitly
for $K=2,3,4$.
We have checked that, for $K=2,3,4$,
the element $\nu\in{\cal A}^{(K;\gamma\delta)}_N$
satisfies the conditions $|\nu|\equiv \gamma-\delta\mbox{ mod }K$,
$c\mbox{-}h^{(0)}_\nu=1$ and $|A_\nu|^2=1$.
For the case of $K=3$, we give these data in Appendix B.
Noticing the relation $A_\lambda=A_{\tau(\lambda)}$,
we have $|A_\lambda|^2=1$ for any partition
$\lambda\in\Lambda_N^{(K)}$
with {\it type} $\nu\in{\cal A}^{(K;\gamma\delta)}_N$.

We give our results on the DCCF and
dynamical structure factor
of the $SU(K)$ HS model with the finite number of lattice sites.
The momentum and excited energy
are respectively given by
\ba
\label{momentum-finite-lattice}
P_\lambda
&:=&
\frac{2\pi}{N}|C_K^{(0)}(\lambda)|,
\\
\label{ex-energy-finite-lattice}
E_\lambda
&:=& \frac{J}{4}\left(\frac{2\pi}{N}\right)^2
\Big[(N-1)|C^{(0)}_K(\lambda)|-2n_K(\lambda)\Big].
\ea
Then, for
$\gamma,\delta\in\Sigma_K$ with
$\gamma\ne\delta$,
we have
\ba
\label{DCCF-finite-lattice}
\eta_N^{(\delta\gamma)(\gamma\delta)}(r,t)
&=&
\frac{1}{N}
\sum_{\lambda\in\Lambda_N^{(K)}}
|F_\lambda^{(K)}|^2 e^{-itE_\lambda+irP_\lambda},
\\
\label{DSF-finite-lattice}
S_N^{(\delta\gamma)(\gamma\delta)}(q,\omega)
&=&
\sum_{\lambda\in\Lambda_N^{(K)}}
|F_\lambda^{(K)}|^2
\delta_{n,|C_K^{(0)}(\lambda)|}\delta(\omega-E_\lambda),
\ea
where $q=2\pi n/N$, and the summation is taken over
partitions $\lambda\in\Lambda_N^{(K)}$
such that $\tau (\lambda)=\nu$ for some
$\nu\in {\cal A}_N^{(K;\gamma\delta)}$.
The squared form factor is given by
\ba
\label{ff-finite}
|F_\lambda^{(K)}|^2
&:=&
\frac{1}{N}
|\chi_\lambda(\infty)|^2 N_\lambda(\infty)
\nonumber
\\
&=&
\frac{1}{N}
\frac{\prod_{s\in C^{(0)}_K(\lambda)\setminus\{(1,1)\}}l_\lambda'(s)^2}
     {\prod_{s\in H^{(0)}_K(\lambda)}l_\lambda(s)(l_\lambda(s)+1)}
\prod_{s\in
C^{(0)}_K(\lambda)}\frac{N-l_\lambda'(s)}{N-l_\lambda'(s)-1}.
\ea
Due to the limit $\beta\rightarrow\infty$,
the above quantities depend only on
$l_\lambda(s)$ and $l'_\lambda(s)$.
In the particular case of $K=2$,
our formulae (\ref{DCCF-finite-lattice})
and (\ref{DSF-finite-lattice})
generalize the known ones \cite{HZ}
to arbitrary size of the system.
We will see that our formulae have
the advantage in clarifying the role of
the internal degrees of freedom
of quasi-particles.

Since we consider the case of zero external magnetic fields,
the $SU(K)$ symmetry demands
that
$\eta_N^{(\delta\gamma)(\gamma\delta)}(q,\omega)$
and
$S_N^{(\delta\gamma)(\gamma\delta)}(q,\omega)$
are actually independent of
$(\gamma,\delta)$ as long as $\gamma\ne\delta$.
Therefore we put $\eta_N(r,t)=
\eta_N^{(\delta\gamma)(\gamma\delta)}(r,t)$
and
$S_N(q,\omega)=
S_N^{(\delta\gamma)(\gamma\delta)}(q,\omega)$
for $\gamma\ne\delta$.
We can prove this independence by using
other expressions
of the DCCF and dynamical structure factor which are introduced in
the next subsection.
See Appendix D for the proof. From the formula
(\ref{DSF-finite-lattice-general}),
other non-zero components of
the dynamical structure factors are given by
\be
\label{z-z}
S_N^{(\gamma\gamma)(\delta\delta)}(q,\omega)
=
\left(\delta_{\gamma\delta}-\frac{1}{K}\right)
S_N(q,\omega).
\ee

We have to determine the sets
$C_K^{(0)}(\lambda)$ and $H_K^{(0)}(\lambda)$
explicitly to derive the dynamical structure factor.
The {\it type} of given partition
has sufficient information for determining these sets.
For example, in the case of $K=3$,
we have the following explicit form of the sets
$C_3^{(0)}(\lambda)$ and $H_3^{(0)}(\lambda)$
for $\lambda\in\Lambda_N^{(K)}$ with {\it type} $(1)$:
\ba
\label{C_3}
C_3^{(0)}(\lambda)
&=&
\left\{
(1,1),(4,1),\cdots,(\lambda'_1,1),
\right.
\nonumber
\\
& &
\,\,\,(2,2),(5,2),\cdots,(\lambda'_2-1,2),
\nonumber
\\
& &
\left.
\,\,(3,3),(6,3),\cdots,(\lambda'_3,3)
\right\},
\\
\label{H_3}
H_3^{(0)}(\lambda)
&=&
\left\{
(1,1),(4,1),\cdots,(\lambda'_3-2,1),\right.
\nonumber
\\
& &
\,\,\,(\lambda'_3+3,1),(\lambda'_3+6,1),\cdots,(\lambda'_2,1),
\nonumber
\\
& &
\,\,\,(\lambda'_2+2,1),(\lambda'_2+5,1),\cdots,(\lambda'_1-2,1),
\nonumber
\\
& &
\,\,\,(2,2),(5,2),\cdots,(\lambda'_3-1,2),
\nonumber
\\
& &
\,\,\,(\lambda'_3+1,2),(\lambda'_3+4,2),\cdots,(\lambda'_2-2,2),
\nonumber
\\
& &
\left.
\,\,(1,3),(4,3),\cdots,(\lambda'_2,3)
\right\}.
\ea
We have determined these sets for all {\it types}
in the case of $K=2,3,4$.
Therefore
$\eta_N(q,\omega)$ and
$S_N(q,\omega)$
can be explicitly calculated.
In general, $S_N(q,\omega)$
is the rational function of $N$.
In Appendix C we give examples.

We have checked the validity of Eq.
(\ref{DSF-finite-lattice}) with $K=2$ and $3$
by comparing with the numerical result for $N\leq 24$ and $N\leq 15$,
respectively.
The numerical result
is obtained via exact diagonalization and the recursion
method.
We truncated the continued fraction at 100 iterations
and took the Lorentzian width ${\cal O}(10^{-5}J)$.
The agreement is excellent in both cases of $K=2$ and $3$.
The comparison between the exact result and numerical data
is given in Table \ref{table-eAn}
for the case of $K=3$ and $N=15$.
\bigskip
\begin{center}
{\bf \large Table \ref{table-eAn}}
\end{center}
\bigskip
In Fig. \ref{fig1}, we present the exact results of
$S_N(q,\omega)$
for $(K,N)=(3,18)$ and $(4,16)$.
The dynamical structure factors $S_N(q,\omega)$ in Fig. \ref{fig1} have
rather complicated structure.
The origin of this structure
will be clarified in section \ref{DSF-infinit}.
\bigskip
\begin{center}
{\bf \large Fig. \ref{fig1}}
\end{center}
\bigskip
The static structure factor
$S_N(q):=S_N^{(\delta\gamma)(\gamma\delta)}(q)$ with
$\gamma\ne\delta$ is
obtained by
integrating over
the energy in
the formula (\ref{DSF-finite-lattice}).
We give the results of
$S_N(q)$ for $(K,N)=(3,18)$ and $(4,16)$
in Fig. \ref{fig2}. From Fig. \ref{fig2}, we see that
$S_N(q)$ has
the cusp structure at $q=2\pi/3$ for $K=3$
and at $q=\pi/2,\pi$ for $K=4$.
In section \ref{DSF-infinit}
we will give the detailed discussion
of this point.
\bigskip
\begin{center}
{\bf \large Fig. \ref{fig2}}
\end{center}
\bigskip
%
%

\subsection{Quasi-particle interpretation}
\label{quasi-particle-interpretation}

As an effect of $\beta\rightarrow\infty$,
the summation in the right hand side of
the formulae (\ref{DCCF-finite-lattice}) and
(\ref{DSF-finite-lattice})
is restricted to the finite set $\Lambda_N^{(K)}$.
The conditions on the {\it type} of excited states
come from the selection rules for the $SU(K)$ spin.
We consider the quasi-particle interpretations of these conditions.
For labelling the excited states relevant to
$S_N(q,\omega)$,
it is more convenient to use the conjugate partition
$\lambda'=(\lambda'_1,\cdots,\lambda'_K)\in\Lambda_K^{(N)}$
instead of $\lambda\in\Lambda_N^{(K)}$.
Each $\lambda'_i$ has the information on the momentum and $SU(K)$ spin
of a quasi-particle.
We call this quasi-particle a spinon following the $SU(2)$ case.
The spinon is considered to be an object possessing
an $SU(K)$ spin. An $SU(K)$ spin of the spinon
with $\lambda'_i$ is specified by
a certain condition on the pair $(i,\lambda'_i)$.
Since the condition for general $K$
is rather complicated,
we give examples in the case of $K=2,3,4$:
\begin{eqnarray}
\label{SU(2)spin-spinon}
\mbox{$SU(2)$ spin of spinon with }\lambda'_i
:=
\left\{
\begin{array}{ll}
1/2,
\,& \mbox{ if }(i,\lambda'_i)\equiv
(0,0), (1,1), \mbox{ mod } 2,\\
-1/2,
\,& \mbox{ if }(i,\lambda'_i)\equiv
(0,1), (1,0), \mbox{ mod } 2.
\end{array}
\right.
\end{eqnarray}
\begin{eqnarray}
\label{SU(3)spin-spinon}
\mbox{$SU(3)$ spin of spinon with }\lambda'_i
:=
\left\{
\begin{array}{ll}
(0,1/2),
\,& \mbox{ if }(i,\lambda'_i)\equiv
(0,0), (1,1), (2,2) \mbox{ mod } 3,\\
(1/2,-1/2),
\,& \mbox{ if }(i,\lambda'_i)\equiv
(0,1), (1,2), (2,0) \mbox{ mod } 3,\\
(-1/2,0),
\,& \mbox{ if }(i,\lambda'_i)\equiv
(0,2), (1,0), (2,1) \mbox{ mod } 3.
\end{array}
\right.
\end{eqnarray}
\begin{eqnarray}
\label{SU(4)spin-spinon}
\mbox{$SU(4)$ spin of spinon with }\lambda'_i
:=
\left\{
\begin{array}{ll}
(0,0,1/2),
\,& \mbox{ if }(i,\lambda'_i)\equiv
(0,0), (1,1), (2,2), (3,3) \mbox{ mod } 4,\\
(0,1/2,-1/2),
\,& \mbox{ if }(i,\lambda'_i)\equiv
(0,1), (1,2), (2,3), (3,0) \mbox{ mod } 4,\\
(1/2,-1/2,0),
\,& \mbox{ if }(i,\lambda'_i)\equiv
(0,2), (1,3), (2,0), (3,1) \mbox{ mod } 4,\\
(-1/2,0,0),
\,& \mbox{ if }(i,\lambda'_i)\equiv
(0,3), (1,0), (2,1), (3,2) \mbox{ mod } 4.
\end{array}
\right.
\end{eqnarray}
It is important to note that
a spinon transforms as a weight vector of 
the fundamental representation $\bar{K}$ of $SU(K)$.

{}From the formulae
(\ref{DCCF-finite-lattice}) and
(\ref{DSF-finite-lattice}),
we can conclude that relevant excited states of
$S_N(q,\omega)$
consist of $K$ spinons.
Moreover the conditions on the {\it type} of excited states
lead to an important consequence:
$K$-spinon excited states have $K-1$ different $SU(K)$ spins.
That is, the excited states contain $K-1$ species of quasi-particles.
For instance, in the case of $K=3$
excited states relevant to $S_N^{(31)(13)}(q,\omega)$,
which have the $SU(3)$ spin $(1/2,1/2)$,
consist of three spinons with $SU(3)$ spins $(0,1/2)$, $(0,1/2)$
and $(1/2,-1/2)$.
For the $SU(2)$ case, we recover the well-known fact
that only two spinons with the same spin contribute
to $S_N(q,\omega)$ \cite{HZ}.
The complete lists of the $SU(K)$ spins of
$K$ spinons are given in Appendix B for $K=2,3,4$.

The relevant states of
$S_N^{(\gamma\gamma)(\delta\delta)}(q,\omega)$
have the $SU(K)$ spin $(s_1,\cdots,s_{K-1})=(0,\cdots,0)$.
Then we can show that, in the spinon picture,
the relevant states of
$S_N^{(\gamma\gamma)(\delta\delta)}(q,\omega)$
consist of $K$-spinon excited states with $K$ different $SU(K)$ spins.
Namely, in contrast to the case of $S_N(q,\omega)$,
the excited states contain $K$ species of quasi-particles.
See Appendix B for lists of $SU(K)$ spins of $K$-spinon
excited states for
$S_N^{(\gamma\gamma)(\delta\delta)}(q,\omega)$
in the case of $K=2,3,4$.
The dynamical structure factor
$S_N^{(\gamma\gamma)(\delta\delta)}(q,\omega)$
is related to
$S_N(q,\omega)$ through the formula (\ref{z-z}).

The $K$-spinon excitation belongs to
the tensor representation $\bar{K}^{\otimes K}$ of $SU(K)$.
This representation
contains the adjoint representation $Ad$ as
an irreducible component. From
the condition stated above for $K$-spinon excitation,
we see that the $K$-spinon excitation transforms
as one of the weight vectors for $Ad$.
This is consistent with the condition for $SU(K)$ spin
of the excited states which are relevant to
$S_N^{(\delta\gamma)(\gamma\delta)}(q,\omega)$.

\subsection{$U(1)$ expression}
\label{U(1)}

Although the parametrization
$\lambda'=(\lambda'_1,\cdots,\lambda'_K)\in\Lambda_K^{(N)}$
has the advantage of
giving the spinon interpretation,
we introduce another parametrization which is
more convenient for taking the thermodynamic limit.
The new parametrization
$c(\lambda')=(c_1(\lambda'),\cdots,c_K(\lambda'))\in\Lambda_K^{(N/K)}$
for the excited states of
the DCCF (\ref{DCCF-finite-lattice}) and
dynamical structure factor (\ref{DSF-finite-lattice})
is defined by
\ba
\label{spinon-momentum}
c_i(\lambda')
&:=&
\mbox{the number of box } s\in\lambda'_i
\mbox{ such that }s\in C_K^{(0)}(\lambda)
\nonumber
\\
&=&
\left\{
\begin{array}{ll}
(\lambda'_i-1)/K,
\quad & \lambda'_i\equiv 1\mbox{ mod }K,\\
\qquad\cdot &\\
\qquad\cdot &\\
(\lambda'_i-l+1)/K,
\quad & \lambda'_i\equiv l-1\mbox{ mod }K,\\
(\lambda'_i+K-l)/K,
\quad & \lambda'_i\equiv l\mbox{ mod }K,\\
\qquad\cdot &\\
\qquad\cdot &\\
(\lambda'_i+1)/K,
\quad & \lambda'_i\equiv K-1\mbox{ mod }K,\\
\lambda'_i/K,
\quad & \lambda'_i\equiv 0\mbox{ mod }K,
\end{array}
\right.
\ea
where, in the second explicit expression, $i\equiv l\mbox{ mod }K$
$(l=0,1,\cdots,K-1)$.
This parametrization is nothing but
representing
the momenta of spinons.
It is important to note that
we can not recover the information on the $SU(K)$ spins of spinons from
this new parametrization $c(\lambda')$,
since $c(\lambda')$ is characterized by
the information on the color $0$ only.
We rewrite the formulae
(\ref{DCCF-finite-lattice}) and (\ref{DSF-finite-lattice})
by using $c(\lambda')$.
The momentum and excited energy are
respectively expressed by
\ba
\label{momentum-u1}
P_{c(\lambda')}&:=&
\frac{2\pi}{N}\sum_{i=1}^{\lambda_1}c_i(\lambda')
=
\frac{2\pi}{N}\sum_{i=1}^Kc_i(\lambda'),
\\
\label{ex-energy-u1}
E_{c(\lambda')}
&:=&
\frac{J}{4}\left(\frac{2\pi}{N}\right)^2
\Big[
\sum_{i=1}^{\lambda_1}
\left(
(N-1)c_i(\lambda')-Kc_i(\lambda')^2
\right)
\nonumber
\\
&+&K\sum_{i\equiv 1\mbox{ {\scriptsize mod} }K}c_i(\lambda')
+(K-2)\sum_{i\equiv 2\mbox{ {\scriptsize mod} }K}c_i(\lambda')
+\cdots
-(K-2)\sum_{i\equiv 0\mbox{ {\scriptsize mod} }K}c_i(\lambda'))
\Big]
\nonumber
\\
&=&\frac{J}{4}\left(\frac{2\pi}{N}\right)^2\sum_{i=1}^K
c_i(\lambda')(N+K+1-2i-Kc_i(\lambda')).
\ea
The expression of the form factor (\ref{ff-finite}) depends
on the {\it type} of state $\lambda$.
In general, we have
\ba
\label{ff-suK2u1}
|F_\lambda^{(K)}|^2/R_{c(\lambda')}^{(K)}
&=&
\frac{\Gamma(c_{ab}(\lambda')+\chi_{ab}+(K-1)/K)}
     {\Gamma(c_{ab}(\lambda')+\chi_{ab})}
\frac{\Gamma(c_{ab}(\lambda')+\chi'_{ab}+(K-1)/K)}
     {\Gamma(c_{ab}(\lambda')+\chi'_{ab})}
\nonumber
\\
&\times&
\prod_{1\leq i<j\leq K,\,(i,j)\ne(a,b)}
\frac{\Gamma(c_{ij}(\lambda')+\rho_{ij}-2/K)}
     {\Gamma(c_{ij}(\lambda')+\rho_{ij})},
\ea
where $\Gamma(z)$ is the gamma function,
$c_{ij}(\lambda'):=c_{i}(\lambda')-c_{j}(\lambda')$ and
\ba
\label{u1-ff-factor}
R_{c(\lambda')}^{(K)}
&:=&
\frac{1}{K}\prod_{i=1}^K\frac{\Gamma((K-1)/K)}{\Gamma(i/K)^2}
\nonumber
\\
&\times&
\prod_{j=1}^K
\frac{\Gamma(c_j(\lambda')+j/K-1/K)}
     {\Gamma(c_j(\lambda')+j/K)}
\frac{\Gamma((N+K-j+1)/K-c_j(\lambda')-1/K)}
     {\Gamma((N+K-j+1)/K-c_j(\lambda'))}.
\ea
In the above formula,
the pair of indeces $(a,b)$ with $a<b$
and rational numbers
$\chi_{ab}$, $\chi'_{ab}$ and $\rho_{ij}$ depend
on the {\it type} of $\lambda$.
We note that the order of $\chi_{ab}$, $\chi'_{ab}$ and $\rho_{ij}$
is ${\cal O}(N^0)$.
In Appendix D,
we derive the formulae (\ref{momentum-u1}),
(\ref{ex-energy-u1}) and (\ref{ff-suK2u1}),
and give explicit formulae for form factors.

The summation in (\ref{DSF-finite-lattice})
is reduced to the summation
$\sum_{N/K\geq c_1(\lambda')\geq \cdots\geq c_K(\lambda')\geq 0}$
under the condition that the cases of
$c_1(\lambda')=c_2(\lambda')=\cdots=c_K(\lambda')$ are excluded.
Physically, this condition means the extraction of the
density fluctuation.
The conditions on {\it type} in (\ref{DSF-finite-lattice})
can be combined into the redefinition
of the form factor.
In Appendix F, we give
the redefined form factor $F_{c(\lambda')}^{(K)}$
for $K=2,3,4$.
Then finally, we have
\be
\label{DSF-U(1)}
S_N(q,\omega)
=
\sum_{N/K\geq c_1\geq \cdots\geq c_K\geq 0}
|F_{c}^{(K)}|^2\,
\delta_{n,\sum_{i=1}^Kc_i}\,
\delta(\omega-E_{c}),
\ee
where
$q=2\pi n/N$, and
$\delta_{n,\sum_{i}c_i}$ represents
the momentum conservation $q=P_c$.
Here the summation is taken over partitions
$c=(c_1,\cdots,c_K)\in\Lambda_K^{(N/K)}$
with the conditions given above.
In the formula (\ref{DSF-U(1)}), we have omitted
the dependence on the original parameter $\lambda'$,
because
we regard the formula (\ref{DSF-U(1)})
as the expression of $S_N(q,\omega)$ by using only the momenta
$c=(c_1,\cdots,c_K)\in\Lambda_K^{(N/K)}$
of the spinons.
We call the formula (\ref{DSF-U(1)}) the $U(1)$ expression
of the dynamical structure factor.

\subsection{Motif expression}
\label{relation2motif}

In this subsection,
we discuss the relation between
the momenta of spinons and motif and then
express $S_N(q,\omega)$ in terms of motifs.
We recall definition of the motif \cite{HHTBP,BGHP}.
Eigenstates of the Hamiltonian
$H_{\mbox{{\scriptsize HS}}}$ (\ref{HS})
are characterized via motifs, which are sequences
$d=d_0d_1d_2\cdots d_{N-1}d_N$
of $N+1$ digits `0' or `1',
beginning and ending with `0', i.e., $d_0=d_N=0$,
and containing at most $K-1$
consecutive `1'.
For example, in the case of $N\equiv 0 \mbox{ mod } K$,
the unique singlet groundstate is represented by the motif
$d^{(0)}:=0\underbrace{1\cdots 1}_{K-1}0\underbrace{1\cdots 1}_{K-1}0
\cdots 0\underbrace{1\cdots 1}_{K-1}0$.
The crystal momentum $P_d$ and
energy $E_d$ of the state for the motif $d$
are respectively given by
\ba
\label{momentum-motif}
P_d
&\equiv&
\frac{2\pi}{N}\sum_{k=1}^{N-1}kd_k
\mbox{ mod } 2\pi,
\\
\label{energy-motif}
E_d
&:=&
\frac{J}{4}\left(\frac{2\pi}{N}\right)^2
\Big[
-\sum_{k=1}^{N-1}kd_k(N-kd_k)
+\frac{1}{6}N(N^2-1)
\Big].
\ea
The groundstate energy, i.e.,
the energy for the groundstate motif $d^{(0)}$,
is explicitly given by
\be
\label{groundstate-energy-motif}
E_{d^{(0)}}
=
\frac{J}{4}\left(\frac{2\pi}{N}\right)^2
\frac{1}{6}\frac{N}{K}(N^2-K^2).
\ee

Now we consider the relation between
the parametrization $c(\lambda')$ with
$\lambda\in\Lambda_N^{(K)}$
and the motif.
For this purpose, it is convenient to use
another expression for the motif \cite{HHTBP,BGHP,BS1}.
For a given motif, replacing every intermediate `0' by $`)(`$,
first `0' by `$($' and last `0' by `$)$',
one can obtain a string of ``elementary" motifs
$(\underbrace{1\cdots 1}_{a})$ ($a=0,1,\cdots,K-1$).
We denote a motif with this expression by
$\cdots e_2 e_1 e_0$ where $e_i$ is one of the elementary motifs.
Namely we enumerate the elementary motifs from
right to left.
For example, for the motif $0110010=(11)(\,)(1)$, we have
$e_0=(1)$, $e_1=(\,)$ and $e_2=(11)$.
The groundstate motif $d^{(0)}$ is represented by
$e_{N/K-1}\cdots e_1 e_0$ with $e_j=(\underbrace{1\cdots 1}_{K-1})$
for all $j$.

The groundstate motif has $(N-N/K)$ `1'.
We construct the set of motifs by
changing one `1' by `0' in the groundstate motif
and rearranging those $(N-N/K-1)$ `1' possible ways.
We call these motifs the $K$-spinon motifs
because they satisfy the following constraint
which leads to the $K$-spinon excitation:
\be
\label{K-spinon-motif}
\sum_{a=0}^{K-2}
(K-a-1)\times
(\mbox{the number of elementary motif }
(\underbrace{1\cdots 1}_{a}) \mbox{ in the motif})
=
K.
\ee
The $K$-spinon motif consists of $(N/K+1)$ elementary motifs, i.e.,
we can write its as $e_{N/K}\cdots e_{1}e_{0}$.

The correspondence between the parametrization
$c(\lambda')=(c_1(\lambda'),\cdots,c_K(\lambda'))$
which is the relevant state for $S_N(q,\omega)$
and the $K$-spinon motif
$d=e_{N/K}\cdots e_{1}e_{0}$ is explained as follows.
We have the following correspondence:
\be
\label{motif2spinon}
e_i=(\underbrace{1\cdots 1}_{a})
\longleftrightarrow
(K-a-1)\mbox{ spinons with the same momentum } i.
\ee
For example, in the case of $K=3$ and $N=12$,
the motif $0110101100110=(11)(1)(11)(\,)(11)$
is regarded as three spinons with
$c_1(\lambda')=3,c_2(\lambda')=c_3(\lambda')=1$.
As shown above, the groundstate motif contains no spinon
and consists of $N/K$ elementary motifs.

The above simple rule for the correspondence between momenta of
spinons and motif is valid only
for the $K$-spinon motifs and the groundstate motif.
For generalizing this rule to arbitrary motifs,
we have to use the correspondence between
states of the $SU(K)$ HS model and semistandard tableaux
given in Refs. \cite{KKN,Uglov2}.
In Appendix E, we recall the results of Refs. \cite{KKN,Uglov2}
and then generalize the correspondence between momenta of
spinons and $K$-spinon motifs to arbitrary motifs.

{}From the above discussion, we can rewrite
the formula (\ref{DSF-U(1)}) in terms of the motif
as follows:
\be
\label{DSF-motif}
S_N(q,\omega)
=
\sum_{d:\mbox{{\scriptsize $K$-spinon motif}}}
|F_d^{(K)}|^2
\,
\hat{\delta}
\Big(q-\frac{2\pi}{N}\sum_{k=1}^{N-1}kd_k\equiv 0 \mbox{ mod }2\pi\Big)
\,
\delta(\omega-E_d+E_{d^{(0)}}),
\ee
where $\hat{\delta}$ represents the
momentum conservation, and
$F^{(K)}_d=F_{c}^{(K)}$ through the above identification
of $K$-spinon motif $d$ and the momenta of spinons $c$.
We call the formula (\ref{DSF-motif}) the motif
(or domain-wall) expression
of the dynamical structure factor.

\section{Dynamical Structure Factor
in the Thermodynamic Limit}
\label{DSF-infinit}

\subsection{Thermodynamic limit}
\label{thermodynamic-limit}

Next our task is to take the thermodynamic limit of
the formulae (\ref{DCCF-finite-lattice}) and
(\ref{DSF-finite-lattice}).
Performing a procedure similar to that in Refs. \cite{Ha,LPS},
for $K=2,3$ and $4$,
we can take this limit.
We introduce the momentum of the spinon $k_i$ in the thermodynamic limit
by $2K(c_i(\lambda')-N)/N\rightarrow k_i$ as $N\rightarrow\infty$.
In this limit, the momentum and excited energy
are respectively given by
\ba
\label{momentum-thermo}
p(k)&:=&k_F\sum_{i=1}^K k_i,
\\
\label{ex-energy-thermo}
\epsilon(k)&:=&\sum_{i=1}^K \epsilon_s(k_i),
\ea
where $k_F:=\pi/K$,
$k=(k_1,\cdots,k_K)$, and
the spinon dispersion is given by
\be
\label{spinon-dispersion}
\epsilon_s(u):=\frac{\pi v_s}{2K}(1-u^2)
\ee
with the velocity $v_s:=J\pi/2$.
We have adopted the above normalization of $k_i$ so that
the Fermi points coincide with $\{\pm 1\}$.
Then, using the formula
$\Gamma(p+z)/\Gamma(p)
\rightarrow p^z$ as $|p|\rightarrow\infty$
and changing the summations to integrals
in the formula (\ref{DSF-U(1)}) and
similar formula for the DCCF,
we have the following formulae for
the DCCF and
dynamical structure factor of the $SU(K)$ HS model
in the thermodynamic limit:
\ba
\label{DCCF-thermo}
\eta^{(\delta\gamma)(\gamma\delta)}(r,t)
&=&
(-1)^r \frac{A_K}{2\pi} \sum_{1\leq a<b\leq K}
\prod_{j=1}^K\int_{-1}^{1}dk_j
|F_{ab}^{(K)}(k)|^2 e^{-it\epsilon(k)+irp(k)},
\\
S^{(\delta\gamma)(\gamma\delta)}(q,\omega)
&=&
A_K \sum_{1\leq a<b\leq K}
\prod_{j=1}^K\int_{-1}^{1}dk_j
|F_{ab}^{(K)}(k)|^2
\delta(q-\pi-p(k))\delta(\omega-\epsilon(k)),
\label{DSF-thermo}
\ea
where a normalization constant is given by
\ba
\label{coeff-thermo}
A_K
&:=&
\frac{2^{K}\pi}{K^3(K-1)}
\prod_{j=1}^K\frac{\Gamma((K-1)/K)}{\Gamma(j/K)^2}.
\ea
The form factor is given by
the following Jastrow form in the momentum space:
\ba
\label{ff-thermo}
F_{ab}^{(K)}(k)
:=
\frac{|k_a-k_b|^{g_K}
      \prod_{1\leq i<j\leq K,(i,j)\ne(a,b)}|k_i-k_j|^{g'_K}}
     {\prod_{i=1}^K(1-k_i^2)^{(1-g_K)/2}}
\ea
with $g_K:=(K-1)/K$ and $g'_K:=-1/K$.
Since
$\eta^{(\delta\gamma)(\gamma\delta)}(r,t)$
and $S^{(\delta\gamma)(\gamma\delta)}(q,\omega)$
with $\gamma\ne\delta$
do not depend on the pair $(\gamma,\delta)$,
we denote them by $\eta(r,t)$
and $S(q,\omega)$, respectively.
Other non-zero components of
dynamical structure factors are given by
\be
\label{DSF-thermo-general}
S^{(\gamma\gamma)(\delta\delta)}(q,\omega)
=
\left(
\delta_{\gamma\delta}g_K
+
(1-\delta_{\gamma\delta})g'_K
\right)
S(q,\omega).
\ee

Since
$p(k)$, $\epsilon(k)$ and $\sum_{a<b}|F_{ab}^{(K)}(k)|^2$
are symmetric in $k_1,\cdots,k_K$,
the summation in the formulae (\ref{DCCF-thermo}) and
(\ref{DSF-thermo}) can be replaced
by $K(K-1)/2$ if we appropriately change the subscripts
of $k_i$'s.
Performing the integrals twice,
we have the following $(K-2)$-fold integral expression
of $S(q,\omega)$:
\ba
\label{DSF-thermo-(K-2)-int}
&&
S(q,\omega)
=
\frac{2^K}{\pi v_s}
\prod_{i=1}^K\frac{\Gamma((K-1)/K)}{\Gamma(i/K)^2}
\prod_{i=1}^{K-2}\int_{-1}^{1}
\frac{dk_i}{(1-k_i^2)^{1-g_K}}\hat{\delta}(\Omega)
\nonumber
\\
&&
\times
\frac{(a^2-4b)^{(g_K+g_K')/2}
      \prod_{i=1}^{K-2}(k_i^2-ak_i+b)^{2g_K'}
      \prod_{1\leq i<j\leq K-2}(k_i-k_j)^{2g_K'}}
     {(1+a+b)^{1-g_K}(1-a+b)^{1-g_K}},
\ea
where
\ba
\label{a}
a
&:=&
\tilde{q}/k_F-\sum_{i=1}^{K-2}k_i,
\\
\label{b}
b
&:=&
a^2/2+K\tilde{\omega}/(\pi v_s)+\sum_{i=1}^{K-2}k_i^2/2,
\ea
with $\tilde{q}:=q-\pi$ and $\tilde{\omega}:=\omega-\pi v_s/2$.
The region $\Omega$ is determined by the following conditions:
\ba
\label{1}
&&
1-a+b\geq 0,
\\
\label{2}
&&
1+a+b\geq 0,
\\
\label{3}
&&
4b-a^2 \leq 0,
\\
\label{4}
&&
-2\leq a\leq 2.
\ea

For $K=2$, the formulae
(\ref{DCCF-thermo}) and (\ref{DSF-thermo}) reproduce the result
of Haldane-Zirnbauer \cite{HZ}
which was obtained by a completely different method.
Notice that the second product in the numerator
of Eq. (\ref{ff-thermo}) is absent in the $SU(2)$ case.

In the low energy limit we recover the results of Ref. \cite{Kawakami}
which are obtained by conformal field theory.
Performing the same procedure as in Ref. \cite{Ha},
we obtain
the asymptotic behavior of the DCCF (\ref{DCCF-thermo}):
\begin{equation}
\label{asympt-DCCF}
\eta(r,t)
\sim
A_0(\xi_L^{-2}+\xi_R^{-2})
+
\sum_{l=1}^{K-1}
A_l (\xi_L \xi_R )^{-\alpha_l/2}
\cos (2lk_Fr),
\end{equation}
where $\xi_L := r + v_s t$, $\xi_R := r - v_s t$, and
$A_l$ ($l=0,1,\cdots,K-1$) are constants.
The critical exponents are given by
\be
\label{exponent-DCCF}
\alpha_i
:=
2e_i^tC^{-1}e_i
=
2i(1-i/K),
\ee
where $e_i^t:=(0,\cdots,0,1,0,\cdots,0)$ for $i$-th entry
($i=1,\cdots,K-1$) and $(K-1)\times(K-1)$ matrix
$C=(C_{ij}):=(2\delta_{ij}-\delta_{i-1j}-\delta_{i+1j})$
is called the Cartan matrix of the Lie algebra $sl_K$.
Since spinons do not interact in the $SU(K)$ HS model
in contrast to the corresponding nearest-neighbor model,
we expect absence of the logarithmic correction
in the DCCF of the $SU(K)$ HS model.
The derivation of Eq. (\ref{asympt-DCCF}) is given in Appendix F.
These $K-1$ singularities
with exponents $\alpha_i$
correspond to $K-1$ gapless bosonic modes,
i.e.,
the exponents $\alpha_i$ agree with that obtained by
conformal field theory
for the $SU(K)_1$ WZW model
\cite{Affleck,Kawakami}.
We can derive static structure factor
$S(q)$ by integrating over
$\omega$ in $S(q,\omega)$. From Eq. (\ref{asympt-DCCF}),
we can perform the Fourier
transform of $\eta(r,0)$.
Then we have the following asymptotic behavior of $S(q)$:
\begin{equation}
\label{asympt-SSF}
S(q)
\sim
B_0|q|
+
\sum_{l=1}^{K-1}
B_l |q-2lk_F|^{\alpha_l-1},
\end{equation}
where $B_l$ are non-universal constants.
The cusp structure in Fig. \ref{fig2} for $K=3,4$
is consistent with this formula.
Note that $\alpha_1=1$
in the case of $K=2$
corresponds to the logarithmic singularity in
$S(q)$ of the $SU(2)$
HS model \cite{HS1,HS2}.

The spinon interpretation of the formulae (\ref{DCCF-thermo})
and
(\ref{DSF-thermo})
goes as follows.
As in the case of finite systems,
the excited states for
$S(q,\omega)$ in the thermodynamic limit
consist of $K$ spinons with $K-1$ different $SU(K)$ spins.
This fact means, as in the low energy limit,
the dynamics of the $SU(K)$ HS model can be described by $K-1$ species
of quasi-particles. This simple structure reflects
the Yangian symmetry of the $SU(K)$ HS model
\cite{HHTBP}.
In the form factor (\ref{ff-thermo}),
the factor $|k_a-k_b|^{g_K}$
represents the statistical interactions of spinons with the same $SU(K)$
spin,
while the factor
$
|k_i-k_j|^{g'_K}
$
represents those of spinons with different $SU(K)$ spins.
We refer to Ref. \cite{KYA}
for more detailed explanation of statistical interactions.

\subsection{Support of dynamical structure factor}
\label{support-DSF}

The support of
$S^{(\gamma\delta)(\rho\sigma)}(q,\omega)$ represents the region
in the momentum-frequency plane where
$S^{(\gamma\delta)(\rho\sigma)}(q,\omega)$  takes the non-zero
value.
We see that the support of
$S(q,\omega)$ as determined from the formula
(\ref{DSF-thermo}) is compact,
i.e., there is no intensity outside of the finite area.
The support of $S(q,\omega)$
is determined as follows:
\ba
\label{upper-condition}
\omega
&\leq&
\epsilon^{(\mbox{{\tiny U}})}(q)
\quad\mbox{for }0\leq q\leq 2\pi,
\\
\label{lower-condition}
\omega
&\geq&
\epsilon_j^{(\mbox{{\tiny L}})}(q)
\quad\mbox{for } 2(j-1)k_F\leq q\leq 2jk_F
\ea
with $j=1,\ldots ,K$.
Here upper and lower boundaries are respectively given by
\ba
\label{upper-boundary}
\epsilon^{(\mbox{{\tiny U}})}(q)
&:=&
\frac{v_s}{2\pi}q\left(2\pi-q\right),
\,\mbox{ for } 0\leq q\leq 2\pi,
\\
\label{lower-boundary}
\epsilon_j^{(\mbox{{\tiny L}})}(q)
&:=&
\frac{Kv_s}{2\pi}\left(q-2(j-1)k_F\right)
                 \left(2jk_F-q\right),
\,\mbox{ for } 2(j-1)k_F\leq q\leq 2jk_F.
\ea
See Fig. \ref{fig1} (and also Fig. \ref{fig3})
for the case of $K=3,4$.

The behavior of $S(q,\omega)$ near
the dispersion lines of elementary excitations
is derived for general $K$ as follows.
We introduce the following dispersion:
\ba
\label{general-dispersion-line}
\epsilon_{m,n,f}(q)
&:=&
\frac{Kv_s}{2f\pi}\left(q-2nk_F\right)
                 \left(2(K-m)k_F-q\right),
\,\mbox{ for } 2nk_F\leq q\leq 2(K-m)k_F,
\ea
where $m,n\geq 0$ and $f\geq 1$ are integers
with $m+n+f=K$.
This dispersion represents $K$ spinons for which
$m$ spinons are near the left Fermi point $k=-1$,
$n$ spinons are near the right Fermi point $k=1$,
and $f$ spinons are moving together
with same momentum $Q/f$ with $Q\sim{\cal O}(1)$.
For example,
$\epsilon_{0,0,K}(q)=\epsilon^{(\mbox{{\tiny U}})}(q)$
and
$\epsilon_{K-j,j-1,1}(q)=\epsilon_j^{(\mbox{{\tiny L}})}(q)$.
Near the dispersion line $\omega=\epsilon_{m,n,f}(q)$,
$S(q,\omega)$ has the singularity
\be
\label{asympt-DSF}
S(q,\omega)
\sim
|\omega-\epsilon_{m,n,f}(q)|^{\eta_{m,n,f}}
\ee
with exponent
\begin{eqnarray}
\label{dispersion-line-exponent}
\eta_{m,n,f}
:=
\left\{
\begin{array}{ll}
0,
\quad & \mbox{ if } m=n=0,f=K,\\
K-2
-\frac{1}{K}
\left[
m^2+n^2+f(f-1)
\right],
\quad & \mbox{ otherwise}.
\end{array}
\right.
\end{eqnarray}
In Fig. \ref{fig3},
we give values of exponents explicitly for $K=3,4$.
\bigskip
\begin{center}
{\bf \large Fig. \ref{fig3}}
\end{center}
\bigskip
The derivation of Eq. (\ref{asympt-DSF}) is given in Appendix F.
There is a stepwise discontinuity at the upper boundary
$\omega=\epsilon^{(\mbox{{\tiny U}})}(q)$.
On the other hand, there are divergent singularities at the lower
boundaries
$\omega
= \epsilon_1^{(\mbox{{\tiny L}})}(q)$ and $\omega
= \epsilon_{K}^{(\mbox{{\tiny L}})}(q)$.
Here
$S(q,\omega)$ diverges by the power law with the exponent $-1/K$.
For $K\geq 3$,
at other lower boundaries $\omega = \epsilon_j^{(\mbox{{\tiny L}})}(q)$
with $j\neq 1, K$,
$S(q,\omega)$ has threshold singularities but no divergence.
Also, for $K\geq 3$,
$S(q,\omega)$ has cusp type singularities
at other dispersion lines in the support.
These cusp type singularities
originate from the statistical interactions for
spinons with different $SU(K)$ spins.
The complicated structures which appeared in Fig. \ref{fig1}
reflect the cusp type singularities.
(Notice that there are finite size effects in Fig. \ref{fig1}.)

\section{Summary and Discussions}
\label{summary}

In conclusion,
we have derived the exact formulae
(\ref{DSF-finite-lattice}) and (\ref{DSF-thermo})
for $S(q,\omega)$ of
the $SU(K)$ HS model for arbitrary size of the system
at zero temperature.
We have clarified the quasi-particle picture
of the spin dynamics.
The relevant excited states consist of $K$ spinons with $K-1$ different
$SU(K)$ spins.
We have discussed the relation between
momenta of spinons and motif.
We have analyzed the singularities of $S(q,\omega)$
in the thermodynamic limit.
In contrast to the $SU(2)$ case,
cusp type singularities in the support,
which originate from the
statistical interactions of different species of the spinons,
appear in the $SU(K)$ ($K\geq 3$) case.

As in the $SU(2)$ case \cite{HZ,xxx2},
we expect that the divergences at two of the
lowest
boundaries occur also in the $SU(K)$ Heisenberg model with
the nearest-neighbor exchange.
Our exact result of
$S(q,\omega)$ for $K \leq 4$ is likely to be valid
for larger $K$ as well.

As mentioned in section \ref{Introduction},
there is
a close correspondence
between
the $SU(2)$ HS model and spinless CS model
with special coupling parameter $\beta=-2$
in the normalization of Ref. \cite{BGHP}.
Thus it is expected that there is relation
between our result for $K=2$
and the dynamical density correlation function
of the spinless CS model calculated in Refs. \cite{Ha,LPS}.
In the thermodynamic limit,
we checked that the dynamical density correlation function
of the spinless CS model with $\beta=-2$
is almost same as our DCCF (\ref{DCCF-thermo}) for $K=2$.
The difference between them is the range of integration.

The $SU(K)$ HS model
and one-dimensional $SU(K|1)$
supersymmetric $1/r^2$ $t$-$J$ model \cite{KY}
have been an interesting laboratory for
study of the quantum number fractionalization
in one dimension \cite{Laughlin}.
We have applied the freezing trick to
the $SU(K|1)$ supersymmetric $1/r^2$ $t$-$J$ model
and then obtained the dynamical charge structure factor \cite{AYSK}.
It will be interesting to consider the relation between our results
and the exclusion statistics in conformal field theory
discussed in Refs. \cite{BS2,Suzuki,BCR}.

\bigskip
\bigskip

\noindent
{\bf \large Acknowledgement}

\bigskip

T.Y. wishes to thank the support of the Visitor Program of the MPI-PKS.
T.Y. and Y.S. wish to acknowledge the support of the CREST from the
Japan Science and Technology Corporation.

\bigskip
\bigskip

\section*{Appendix A:
Dynamical Density Correlation Function
in the Strong Coupling Limit}
\label{appA}

In this Appendix,
we analyze the strong coupling limit
of the dynamical density correlation function
$(0|\rho(x,t)\rho(0,0)|0)$
of the $U(K)$ spin CS model.
Here $\rho(x,t)$ is the Heisenberg representation
of the density fluctuation operator
\ba
\rho(x)
&:=&
\sum_{j=1}^N\delta(x-x_j)
\sum_{\gamma=1}^KX^{\gamma\gamma}_j-\frac{N}{L}
\nonumber
\\
&=&
\frac{1}{L}\sum_{j=1}^N\sum_{n(\ne 0)}
e^{i2\pi nx/L}z_j^{-n}
\sum_{\gamma=1}^KX^{\gamma\gamma}_j
=
\frac{1}{\sqrt{L}}\sum_{q(\ne 0)}
e^{iqx}\sum_{\gamma=1}^K\bar{X}^{\gamma\gamma}_q.
\ea
We use the same notations as in section \ref{corr-spinCS}.
Uglov derived the following formula \cite{Uglov1}:
\be
\label{densityinH'}
\Theta \rho(0)\Theta^{-1}
=
\frac{1}{L}
\sum_{n\in\mbox{{\scriptsize\bf Z}}\setminus \{0\}}
p_{nK}(y).
\ee
Thus we can derive the dynamical density correlation function
of the $U(K)$ spin CS model
in the same manner as
in Ref. \cite{Uglov1} and in section \ref{corr-spinCS}:
\ba
\label{DDCF-finite-continuum}
(0|\rho(x,t)\rho(0,0)|0)
&=&
\frac{K^2L}{4\pi^2}
\sum_{\lambda\in\Lambda_N}
|{\cal P}_\lambda|^2 |{F}_\lambda(\beta)|^2
e^{-it{\cal E}_\lambda+ix{\cal P}_\lambda},
\ea
where the summation is taken over the partitions
$\lambda\in\Lambda_N$ which satisfy the following conditions:
\ba
\label{conditions-DDCF}
\left\{
\begin{array}{l}
|\lambda|\equiv 0 \mbox{ mod }K,
\\
S^a_\lambda=0,\,a=1,\cdots,K-1,
\\
c\mbox{-}h^{(0)}_\lambda=0.
\end{array}
\right.
\ea

Careful analysis of the formula for $(0|\rho(x,t)\rho(0,0)|0)$
reveals the following behavior in the strong coupling limit:
\ba
\label{strong-lim-DDCF}
(0|\rho(x,t)\rho(0,0)|0)
\longrightarrow
\frac{1}{\beta}
\sum_{l=1}^{N-1}
\sum_{(K^l)(\in\Lambda_N^{(K)})}G_\lambda
+
{\cal O}(\beta^{-2})
\ea
as $\beta\rightarrow\infty$. Here $G_\lambda$
is the $\beta$-independent quantity.
The $SU(K)$ singlet states
$\lambda^{\mbox{{\tiny ph}};l}:=(K^l)$
($l=1,\cdots,N-1$)
satisfy the conditions (\ref{conditions-DDCF})
and
represent the phonon contributions.
Thus we can conclude that the phonon contribution
to the dynamical correlation functions appears in the order
$\beta^{-1}$.

Let us derive Eq. (\ref{strong-lim-DDCF}).
It is sufficient to consider the strong coupling limit
of the matrix element $\chi_{\lambda}$ of local operator
and the norm $N_{\lambda}$.
We can easily see that
$|C_K^{(0)}(\lambda^{\mbox{{\tiny ph}};l})|
=|H_K^{(0)}(\lambda^{\mbox{{\tiny ph}};l})|
=l$.
Notice that
$s=((\lambda^{\mbox{{\tiny ph}};l})'_1,1)
=(l,1)\in H_K^{(0)}(\lambda^{\mbox{{\tiny ph}};l})$.
For this point $s=((\lambda^{\mbox{{\tiny ph}};l})'_1,1)$, we have
$l_{\lambda^{\mbox{{\tiny ph}};l}}(s)=0$.
Therefore, from this fact and conditions
$c\mbox{-}h^{(0)}_{\lambda^{\mbox{{\tiny ph}};l}}=0$,
$|\lambda^{\mbox{{\tiny ph}};l}|\equiv 0\mbox{ mod }K$,
we see that the matrix element
$\chi_{\lambda^{\mbox{{\tiny ph}};l}}(\beta)$
is regular
for the limit $\beta\rightarrow \infty$.
On the other hand,
$N_{\lambda^{\mbox{{\tiny ph}};l}}(\beta)\rightarrow
const\times(1/(\beta+1))$
as $\beta\rightarrow \infty$.
Indeed, noticing
\ba
&&
a_{\lambda^{\mbox{{\tiny ph}};l}}(s)+1
+(K\beta+1)l_{\lambda^{\mbox{{\tiny ph}};l}}(s)=K,
\\
&&
a_{\lambda^{\mbox{{\tiny ph}};l}}(s)
+(K\beta+1)(l_{\lambda^{\mbox{{\tiny ph}};l}}(s)+1)=K(\beta+1)
\ea
for $s=(l,1)$,
we have
\ba
\label{matrix-element-cm}
&&
\chi_{\lambda^{\mbox{{\tiny ph}};l}}(\beta)
=
\frac{1}{K}
\frac{\prod_{s\in C_K^{(0)}(\lambda^{\mbox{{\tiny ph}};l})\setminus\{(1,1)\}}
      (a'_{\lambda^{\mbox{{\tiny ph}};l}}(s)
       -(K\beta+1)l'_{\lambda^{\mbox{{\tiny ph}};l}}(s))}
     {\prod_{s\in H_K^{(0)}(\lambda^{\mbox{{\tiny ph}};l})\setminus\{(l,1)\}}
      (a_{\lambda^{\mbox{{\tiny ph}};l}}(s)+1
       +(K\beta+1)l_{\lambda^{\mbox{{\tiny ph}};l}}(s))}
\\
&&
\longrightarrow
\chi_{\lambda^{\mbox{{\tiny ph}};l}}(\infty)
=
\frac{1}{K}
\frac{\prod_{s\in C_K^{(0)}(\lambda^{\mbox{{\tiny ph}};l})\setminus\{(1,1)\}}
      (-l'_{\lambda^{\mbox{{\tiny ph}};l}}(s))}
     {\prod_{s\in H_K^{(0)}(\lambda^{\mbox{{\tiny ph}};l})\setminus\{(l,1)\}}
      l_{\lambda^{\mbox{{\tiny ph}};l}}(s)},
\\
\label{norm-cm}
&&
N_{\lambda^{\mbox{{\tiny ph}};l}}(\beta)
=
\frac{1}{\beta+1}
\prod_{s\in C_K^{(0)}(\lambda^{\mbox{{\tiny ph}};l})}
\frac{a'_{\lambda^{\mbox{{\tiny ph}};l}}(s)
      +(K\beta+1)(N-l'_{\lambda^{\mbox{{\tiny ph}};l}}(s))}
     {a'_{\lambda^{\mbox{{\tiny ph}};l}}(s)+1
      +(K\beta+1)(N-l'_{\lambda^{\mbox{{\tiny ph}};l}}(s)-1)}
\nonumber
\\
&&
\qquad\qquad
\times
\prod_{s\in H_K^{(0)}(\lambda^{\mbox{{\tiny ph}};l})\setminus\{(l,1)\}}
\frac{a_{\lambda^{\mbox{{\tiny ph}};l}}(s)+1
      +(K\beta+1)l_{\lambda^{\mbox{{\tiny ph}};l}}(s)}
     {a_{\lambda^{\mbox{{\tiny ph}};l}}(s)
      +(K\beta+1)(l_{\lambda^{\mbox{{\tiny ph}};l}}(s)+1)}
\\
&&
\longrightarrow
\frac{1}{\beta+1}
\prod_{s\in C_K^{(0)}(\lambda^{\mbox{{\tiny ph}};l})}
\frac{N-l'_{\lambda^{\mbox{{\tiny ph}};l}}(s)}
     {N-l'_{\lambda^{\mbox{{\tiny ph}};l}}(s)-1}
\prod_{s\in H_K^{(0)}(\lambda^{\mbox{{\tiny ph}};l})\setminus\{(l,1)\}}
\frac{l_{\lambda^{\mbox{{\tiny ph}};l}}(s)}
     {l_{\lambda^{\mbox{{\tiny ph}};l}}(s)+1},
\ea
as $\beta\rightarrow \infty$.
Then we have Eq. (\ref{strong-lim-DDCF}).

\section*{Appendix B:
$SU(K)$ Spin Content and Enumeration of ${\cal A}_N^{(K)}$}
\label{appB}

\noindent
{\it (i) Data of $SU(K)$ spin}

The $SU(K)$ spins
of the excited states of $S_N(q,\omega)$
and those of $K$-spinon excitations
are summarised in the following Tables for $K=2,3,4$.
\bigskip
\begin{center}
{\bf \large Table \ref{selection2}}
\end{center}
\bigskip
\bigskip
\begin{center}
{\bf \large Table \ref{selection3}}
\end{center}
\bigskip
\bigskip
\begin{center}
{\bf \large Table \ref{selection4}}
\end{center}
\bigskip

\bigskip
\noindent
{\it (ii) Set ${\cal A}_N^{(K)}$}

We give the explicit form of the set
${\cal A}_N^{(K)}$.
In general, the set ${\cal A}_N^{(K)}$
has the decomposition
${\cal A}_N^{(K)}
=
\sqcup_{\gamma,\delta=1,\gamma\ne\delta}^K{\cal A}_N^{(K;\gamma\delta)}
\sqcup{\cal B}_N^{(K)}\sqcup{\cal C}_N^{(K)},
$
where ${\cal A}_N^{(K;\gamma\delta)}$ is defined
in subsection \ref{SU(K)},
${\cal B}_N^{(K)}$ and ${\cal C}_N^{(K)}$ are certain subsets
of ${\cal A}_N^{(K)}$.
The set ${\cal B}_N^{(K)}$ is related to the excited states
for $S^{(\gamma\gamma)(\delta\delta)}_N(q,\omega)$
and contains the partition $\O$.

For $K=2$ and $N\geq 2$, we have
${\cal A}_N^{(2)}
=
\sqcup_{\gamma,\delta=1,\gamma\ne\delta}^2{\cal A}_N^{(2;\gamma\delta)}
\sqcup{\cal B}_N^{(2)}
$
where
${\cal B}_N^{(2)}=\{\O\}$,
and
\ba
{\cal A}_N^{(2;12)}
&=&
\left\{%
(1)
\right\},
\\
{\cal A}_N^{(2;21)}
&=&
\left\{%
(2,1)
\right\}.
\ea
For $K=3$ and $N\geq 6$, we have
${\cal A}_N^{(3)}
=
\sqcup_{\gamma,\delta=1,\gamma\ne\delta}^3{\cal A}_N^{(3;\gamma\delta)}
\sqcup{\cal B}_N^{(3)}\sqcup{\cal C}_N^{(3)}
$
where
${\cal B}_N^{(3)}=\{\O,(2,1),(3,2,1),(3,3,2,1)\}$,
${\cal C}_N^{(3)}=\{(2,2,1,1),(3,2,2,1,1),(3,3,2,2,1,1)\}$,
and
\ba
{\cal A}_N^{(3;12)}
&=&
\left\{%
(1,1),(3,2),(3,2,2,1)
\right\},
\\
{\cal A}_N^{(3;21)}
&=&
\left\{%
(2,1,1),(3,3,1),(3,3,2,2)
\right\},
\\
{\cal A}_N^{(3;13)}
&=&
\left\{%
(1),(2,2),(3,2,1,1)
\right\},
\\
{\cal A}_N^{(3;31)}
&=&
\left\{%
(3,1,1),(3,3,2),(3,3,2,2,1)
\right\},
\\
{\cal A}_N^{(3;23)}
&=&
\left\{%
(2),(2,2,1),(3,3,1,1)
\right\},
\\
{\cal A}_N^{(3;32)}
&=&
\left\{%
(3,1),(3,2,2),(3,3,2,1,1)
\right\}.
\ea
For $K=4$ and $N\geq 10$, we have
${\cal A}_N^{(4)}
=
\sqcup_{\gamma,\delta=1,\gamma\ne\delta}^4{\cal A}_N^{(4;\gamma\delta)}
\sqcup{\cal B}_N^{(4)}\sqcup{\cal C}_N^{(4)},
$
where
\ba
{\cal A}_N^{(4;12)}
&=&
\left\{%
(1,1,1), (3,2,2), (4,2,1), (3,2,2,2,1,1),
\right.
\nonumber
\\
& &
\,\,(4,2,2,2,1), (4,4,3), (4,3,3,2,1,1,1),(4,4,3,2,1,1),
\nonumber
\\
& &
\left.
(4,4,3,3,1), (4,4,3,3,2,2,1), (4,4,3,3,3,2), (4,4,3,3,3,2,2,1,1)
\right\},
\\
{\cal A}_N^{(4;21)}
&=&
\left\{%
(2,1,1,1), (3,3,2,1), (4,3,1,1), (3,3,2,2,2,1),
\right.
\nonumber
\\
& &
\,\,(4,3,2,2,2), (4,4,4,1), (4,3,3,2,2,1,1,1),(4,4,4,2,2,1),
\nonumber
\\
& &
\left.
(4,4,4,3,2), (4,4,4,3,2,2,1,1), (4,4,4,3,3,3), (4,4,4,3,3,3,2,1,1)
\right\},
\\
{\cal A}_N^{(4;13)}
&=&
\left\{%
(1,1), (2,2,2), (4,2), (3,2,2,1,1,1),
\right.
\nonumber
\\
& &
\,\,(4,2,2,1,1), (4,3,3), (4,3,3,2,1,1),(4,4,3,1,1,,1),
\nonumber
\\
& &
\left.
(4,3,3,3,1), (4,3,3,3,2,2,1), (4,4,3,3,2,2), (4,4,3,3,2,2,2,1,1)
\right\},
\\
{\cal A}_N^{(4;31)}
&=&
\left\{%
(3,1,1,1), (3,3,2,2), (4,4,1,1), (3,3,2,2,2,1,1),
\right.
\nonumber
\\
& &
\,\,(4,4,2,2,2), (4,4,4,2), (4,4,3,2,2,1,1,1),(4,4,4,2,2,1,1),
\nonumber
\\
& &
\left.
(4,4,4,3,3), (4,4,4,3,3,2,1,1), (4,4,4,3,3,3,1), (4,4,4,3,3,3,2,2,1)
\right\},
\\
{\cal A}_N^{(4;14)}
&=&
\left\{%
(1), (2,2,1), (3,2), (3,2,2,1,1),
\right.
\nonumber
\\
& &
\,\,(4,2,1,1,1), (3,3,3), (3,3,3,2,1,1),(4,3,3,2,1),
\nonumber
\\
& &
\left.
(4,4,3,1,1), (4,3,3,2,2,2,1), (4,4,3,2,2,2), (4,4,3,3,2,2,1,1,1)
\right\},
\\
{\cal A}_N^{(4;41)}
&=&
\left\{%
(4,1,1,1), (4,3,2,2), (4,4,2,1), (4,3,2,2,2,1,1),
\right.
\nonumber
\\
& &
\,\,(4,4,2,2,2,1), (4,4,4,3), (4,4,3,3,2,1,1,1),(4,4,4,3,2,1,1),
\nonumber
\\
& &
\left.
(4,4,4,3,3,1), (4,4,4,3,3,2,2,1), (4,4,4,3,3,3,2), (4,4,4,3,3,3,2,2,1,1)
\right\},
\\
{\cal A}_N^{(4;23)}
&=&
\left\{%
(2,1), (2,2,2,1), (4,3), (3,3,2,1,1,1),
\right.
\nonumber
\\
& &
\,\,(4,3,2,1,1), (4,3,3,1), (4,3,3,2,2,1),(4,4,4,1,1,1),
\nonumber
\\
& &
\left.
(4,3,3,3,2), (4,3,3,3,2,2,1,1), (4,4,4,3,2,2), (4,4,4,3,2,2,2,1,1)
\right\},
\\
{\cal A}_N^{(4;32)}
&=&
\left\{%
(3,1,1), (3,2,2,2), (4,4,1), (3,3,2,2,1,1,1),
\right.
\nonumber
\\
& &
\,\,(4,4,2,2,1), (4,4,3,2), (4,4,3,2,2,1,1),(4,4,4,2,1,1,1),
\nonumber
\\
& &
\left.
(4,4,3,3,3), (4,4,4,3,3,3,2,1,1), (4,4,4,3,3,2,1), (4,4,4,3,3,2,2,2,1)
\right\},
\\
{\cal A}_N^{(4;24)}
&=&
\left\{%
(2), (2,2,1,1), (3,3), (3,3,2,1,1),
\right.
\nonumber
\\
& &
\,\,(4,3,1,1,1), (3,3,3,1), (3,3,3,2,2,1),(4,3,3,2,2),
\nonumber
\\
& &
\left.
(4,4,4,1,1), (4,3,3,2,2,2,1,1), (4,4,4,2,2,2), (4,4,4,3,2,2,1,1,1)
\right\},
\\
{\cal A}_N^{(4;42)}
&=&
\left\{%
(4,1,1), (4,2,2,2), (4,4,2), (4,3,2,2,1,1,1),
\right.
\nonumber
\\
& &
\,\,(4,4,2,2,1,1), (4,4,3,3), (4,4,3,3,2,1,1),(4,4,4,3,1,1,1),
\nonumber
\\
& &
\left.
(4,4,3,3,3,1), (4,4,3,3,3,2,2,1), (4,4,4,3,3,2,2), (4,4,4,3,3,2,2,2,1,1)
\right\},
\\
{\cal A}_N^{(4;34)}
&=&
\left\{%
(3), (3,2,1,1), (3,3,1), (3,3,2,2,1),
\right.
\nonumber
\\
& &
\,\,(4,4,1,1,1), (3,3,3,2), (3,3,3,2,2,1,1),(4,4,3,2,2),
\nonumber
\\
& &
\left.
(4,4,4,2,1), (4,4,3,2,2,2,1,1), (4,4,4,2,2,2,1), (4,4,4,3,3,2,2,1,1,1)
\right\},
\\
{\cal A}_N^{(4;43)}
&=&
\left\{%
(4,1), (4,2,2,1), (4,3,2), (4,3,2,2,1,1),
\right.
\nonumber
\\
& &
\,\,(4,4,2,1,1,1), (4,3,3,3), (4,3,3,3,2,1,1),(4,4,3,3,2,1),
\nonumber
\\
& &
\left.
(4,4,4,3,1,1), (4,4,3,3,2,2,2,1), (4,4,4,3,2,2,2), (4,4,4,3,3,2,2,1,1,1)
\right\}.
\ea
We do not give ${\cal B}_N^{(4)}$ and ${\cal C}_N^{(4)}$,
since they are not important for our discussion.

\bigskip
\noindent
{\it (iii) Data for the set ${\cal A}_N^{(3)}$}

We summarize data for the set ${\cal A}_N^{(3)}$
in the following Tables:
\bigskip
\begin{center}
{\bf \large Table \ref{A12-3}}
\end{center}
\bigskip
\bigskip
\begin{center}
{\bf \large Table \ref{A13-3}}
\end{center}
\bigskip
\bigskip
\begin{center}
{\bf \large Table \ref{A23-3}}
\end{center}
\bigskip
\bigskip
\begin{center}
{\bf \large Table \ref{B-3}}
\end{center}
\bigskip
\bigskip
\begin{center}
{\bf \large Table \ref{C-3}}
\end{center}
\bigskip
%
%

\section*{Appendix C:
Some Explicit Values of the Dynamical Structure Factor for Finite
$N$}
\label{appC}

We give the explicit form of $S_N(q,\omega)$
as a function of $N$.
The notation $c=(c_1,\cdots,c_K)$
for momenta of spinons is defined in
subsection \ref{U(1)}.
For $K=2$ with $c_1=m, c_2=n$
($0\leq n<m \leq N/2$), we have
\begin{eqnarray}
&&
S_N
\left(
\frac{2(m+n)\pi}{N},
\frac{J}{4}
\left(\frac{2\pi}{N}\right)^2
((m+n)N-2(m^2+n^2)+m-n)
\right)
\nonumber
\\
&&
=
\frac{2m-2n-1}{2m-1}
\prod_{i=n+1}^{m-1}\frac{2i}{2i-1}
\frac{1}{N-2n-1}
\prod_{i=n+1}^{m-1}\frac{N-2i}{N-2i-1}.
\end{eqnarray}
We can give the similar formulae for $K=3,4$.
However  these formulae are rather complicated.
We give more explicit formulae for the small momentum $q$.
For $K=3$ and $q\leq 8\pi/N$, we have
\begin{eqnarray}
S_N
\left(
\frac{2\pi}{N},
\frac{J}{4}
\left(\frac{2\pi}{N}\right)^2
(N-1)
\right)
&=&
\frac{1}{N-1},
\,(c_1=1,c_2=c_3=0),
\\
S_N
\left(
\frac{4\pi}{N},
\frac{J}{4}
\left(\frac{2\pi}{N}\right)^2
(2N-8)
\right)
&=&
\frac{3}{2}\frac{N-3}{(N-1)(N-4)},
\,(c_1=2,c_2=c_3=0),
\\
S_N
\left(
\frac{4\pi}{N},
\frac{J}{4}
\left(\frac{2\pi}{N}\right)^2
(2N-4)
\right)
&=&
\frac{1}{2}\frac{1}{N-2},
\,(c_1=c_2=1,c_3=0),
\\
S_N
\left(
\frac{6\pi}{N},
\frac{J}{4}
\left(\frac{2\pi}{N}\right)^2
(3N-21)
\right)
&=&
\frac{9}{5}\frac{(N-3)(N-6)}{(N-1)(N-4)(N-7)},
\nonumber
\\
&&(c_1=3,c_2=c_3=0),
\\
S_N
\left(
\frac{6\pi}{N},
\frac{J}{4}
\left(\frac{2\pi}{N}\right)^2
(3N-11)
\right)
&=&
\frac{6}{5}\frac{N-3}{(N-2)(N-4)},
\,(c_1=2,c_2=1,c_3=0),
\\
S_N
\left(
\frac{8\pi}{N},
\frac{J}{4}
\left(\frac{2\pi}{N}\right)^2
(4N-40)
\right)
&=&
\frac{81}{40}
\frac{(N-3)(N-6)(N-9)}
     {(N-1)(N-4)(N-7)(N-10)},
\nonumber
\\
&&(c_1=4,c_2=c_3=0),
\\
S_N
\left(
\frac{8\pi}{N},
\frac{J}{4}
\left(\frac{2\pi}{N}\right)^2
(4N-24)
\right)
&=&
\frac{9}{8}\frac{(N-3)(N-6)}{(N-2)(N-4)(N-7)},
\nonumber
\\
&&(c_1=3,c_2=1,c_3=0),
\\
S_N
\left(
\frac{8\pi}{N},
\frac{J}{4}
\left(\frac{2\pi}{N}\right)^2
(4N-20)
\right)
&=&
\frac{3}{5}\frac{N-3}{(N-2)(N-5)},
\,(c_1=c_2=2,c_3=0),
\\
S_N
\left(
\frac{8\pi}{N},
\frac{J}{4}
\left(\frac{2\pi}{N}\right)^2
(4N-16)
\right)
&=&
\frac{1}{4}\frac{1}{N-4},
\,(c_1=2,c_2=c_3=1).
\end{eqnarray}
For $K=4$ and $q\leq 8\pi/N$, we have
\begin{eqnarray}
S_N
\left(
\frac{2\pi}{N},
\frac{J}{4}
\left(\frac{2\pi}{N}\right)^2
(N-1)
\right)
&=&
\frac{1}{N-1},
\,(c_1=1,c_2=c_3=c_4=0),
\\
S_N
\left(
\frac{4\pi}{N},
\frac{J}{4}
\left(\frac{2\pi}{N}\right)^2
(2N-10)
\right)
&=&
\frac{4}3{}\frac{N-4}{(N-1)(N-5)},
\,(c_1=2,c_2=c_3=c_4=0),
\\
S_N
\left(
\frac{4\pi}{N},
\frac{J}{4}
\left(\frac{2\pi}{N}\right)^2
(2N-4)
\right)
&=&
\frac{2}{3}\frac{1}{N-2},
\,(c_1=c_2=1,c_3=c_4=0),
\\
S_N
\left(
\frac{6\pi}{N},
\frac{J}{4}
\left(\frac{2\pi}{N}\right)^2
(3N-27)
\right)
&=&
\frac{32}{21}\frac{(N-4)(N-8)}{(N-1)(N-5)(N-9)},
\nonumber
\\
&&(c_1=3,c_2=c_3=c_4=0),
\\
S_N
\left(
\frac{6\pi}{N},
\frac{J}{4}
\left(\frac{2\pi}{N}\right)^2
(3N-13)
\right)
&=&
\frac{8}{7}\frac{N-4}{(N-2)(N-5)},
\,(c_1=2,c_2=1,c_3=c_4=0),
\\
S_N
\left(
\frac{6\pi}{N},
\frac{J}{4}
\left(\frac{2\pi}{N}\right)^2
(3N-9)
\right)
&=&
\frac{1}{3}\frac{1}{N-3},
\,(c_1=c_2=c_3=1,c_4=0),
\\
S_N
\left(
\frac{8\pi}{N},
\frac{J}{4}
\left(\frac{2\pi}{N}\right)^2
(4N-52)
\right)
&=&
\frac{128}{77}
\frac{(N-4)(N-8)(N-12)}{(N-1)(N-5)(N-9)(N-13)},
\nonumber
\\
&&(c_1=4,c_2=c_3=c_4=0),
\\
S_N
\left(
\frac{8\pi}{N},
\frac{J}{4}
\left(\frac{2\pi}{N}\right)^2
(4N-30)
\right)
&=&
\frac{112}{99}\frac{(N-4)(N-8)}{(N-2)(N-5)(N-9)},
\nonumber
\\
&&(c_1=3,c_2=1,c_3=c_4=0),
\\
S_N
\left(
\frac{8\pi}{N},
\frac{J}{4}
\left(\frac{2\pi}{N}\right)^2
(4N-24)
\right)
&=&
\frac{20}{63}\frac{N-4}{(N-2)(N-6)},
\,(c_1=c_2=2,c_3=c_4=0),
\\
S_N
\left(
\frac{8\pi}{N},
\frac{J}{4}
\left(\frac{2\pi}{N}\right)^2
(4N-18)
\right)
&=&
\frac{8}{9}\frac{N-4}{(N-3)(N-5)},
\,(c_1=2,c_2=c_3=1,c_4=0).
\end{eqnarray}

\section*{Appendix D:
Some Formulae and Form Factors in the
$U(1)$ Expression}
\label{appD}

\bigskip
\noindent
{\it (i) Derivation of formulae (\ref{momentum-u1}),
(\ref{ex-energy-u1}) and (\ref{ff-suK2u1})}

We can easily derive the following formulae:
\ba
|C_K^{(0)}|
&=&
\sum_{i=1}^{\lambda_1}c_i(\lambda'),
\\
n_K(\lambda)
&=&
\frac{K}{2}\sum_{i=1}^{\lambda_1}c_i(\lambda')^2
\nonumber
\\
&-&
\frac{K}{2}
\sum_{i\equiv 1\mbox{ {\scriptsize mod} }K}c_i(\lambda')
-\frac{(K-2)}{2}
\sum_{i\equiv 2\mbox{ {\scriptsize mod} }K}c_i(\lambda')
-\cdots
+\frac{(K-2)}{2}
\sum_{i\equiv 0\mbox{ {\scriptsize mod} }K}c_i(\lambda')).
\ea
Then, from Eqs.
(\ref{momentum-finite-lattice}) and
(\ref{ex-energy-finite-lattice}), we have
formulae (\ref{momentum-u1}) and (\ref{ex-energy-u1}).

Next we derive the formula (\ref{ff-suK2u1})
for $K=3$ and $\lambda$ with {\it type} $(1)$.
In the following, we use the abbreviated notation
$c_i=c_i(\lambda')$. From the explicit forms
of the sets
$C_K^{(0)}(\lambda)$ and $H_K^{(0)}(\lambda)$
given in (\ref{C_3}) and (\ref{H_3}),
we have
\ba
\prod_{s\in C_3^{(0)}(\lambda)\setminus\{(1,1)\}}
      (l'_{\lambda}(s))^2
&=&
3^{2(c_1+c_2+c_3-1)}
\frac{\Gamma(c_1)^2\Gamma(c_2+1/3)^2\Gamma(c_3+2/3)^2}
     {\Gamma(1/3)^2\Gamma(2/3)^2},
\\
\prod_{s\in H_3^{(0)}(\lambda)}
      l_{\lambda}(s)(l_{\lambda}(s)+1)
&=&
3^{2(c_1+c_2+c_3-1)}
\nonumber
\\
&\times&
\frac{\Gamma(c_1)\Gamma(c_1+1/3)
      \Gamma(c_2+1/3)\Gamma(c_2+2/3)
      \Gamma(c_3+2/3)\Gamma(c_3+1)}
     {\Gamma(2/3)^3}
\nonumber
\\
&\times&
\frac{\Gamma(c_1-c_2)}{\Gamma(c_1-c_2-2/3)}
\frac{\Gamma(c_2-c_3+1)}{\Gamma(c_2-c_3+1/3)}
\frac{\Gamma(c_1-c_3-2/3)}{\Gamma(c_1-c_3)}
\frac{\Gamma(c_1-c_3-1/3)}{\Gamma(c_1-c_3+1/3)},
\\
\prod_{s\in C_3^{(0)}(\lambda)}
\frac{N-l'_{\lambda}(s)}
     {N-l'_{\lambda}(s)-1}
&=&
\frac{N}{3}
\frac{\Gamma(N/3-c_1+2/3)}{\Gamma(N/3-c_1+1)}
\frac{\Gamma(N/3-c_2+1/3)}{\Gamma(N/3-c_2+2/3)}
\frac{\Gamma(N/3-c_3)}{\Gamma(N/3-c_3+1/3)}.
\ea
Then we have the formula (\ref{ff-suK2u1})
for $K=3$ and $\lambda$ with {\it type} $(1)$.
Other cases can be derived in the same way.

\bigskip
\noindent
{\it (ii) $U(1)$ expression of the form factor $F_\lambda^{(K)}$}

We give the expression of the form factor $F_\lambda^{(K)}$
in terms of the momenta of spinons $c(\lambda')$.
The form factors for $S_N^{(21)(12)}(q,\omega)$
and $S_N^{(12)(21)}(q,\omega)$
have the form:
\ba
\label{su2-u1-ff}
|F_\lambda^{(2)}|^2/R_{c(\lambda')}^{(2)}
=
\frac{\Gamma(c_{12}+1/2)}
     {\Gamma(c_{12}-1/2)},
\, (c_1>c_2),\,
& \mbox{ if }\lambda:\mbox{ type }
(1)\in{\cal A}_N^{(2;12)} \mbox{ or } (2,1)\in{\cal A}_N^{(2;21)}.
\ea
The form factors for $S_N^{(21)(12)}(q,\omega)$,
$S_N^{(31)(13)}(q,\omega)$ and $S_N^{(13)(31)}(q,\omega)$
respectively have the forms:
\ba
\label{su3-u1-ff12}
|F_\lambda^{(3)}|^2/R_{c(\lambda')}^{(3)}
&=&
\left\{
\begin{array}{l}
\frac{\Gamma(c_{12}+1/3)}
     {\Gamma(c_{12}-1/3)}
\frac{\Gamma(c_{12}+2/3)}
     {\Gamma(c_{12})}
\frac{\Gamma(c_{13})}
     {\Gamma(c_{13}+2/3)}
\frac{\Gamma(c_{23}+1/3)}
     {\Gamma(c_{23}+1)}(=:f_{1}^{(3)}),
\\
(c_1>c_2\geq c_3),
\mbox{ if }\lambda:\mbox{ type } (1,1)\in{\cal A}_N^{(3;12)},
\\
\frac{\Gamma(c_{13})}
     {\Gamma(c_{13}-2/3)}
\frac{\Gamma(c_{13}+1/3)}
     {\Gamma(c_{13}-1/3)}
\frac{\Gamma(c_{12}+1/3)}
     {\Gamma(c_{12}+1)}
\frac{\Gamma(c_{23}-2/3)}
     {\Gamma(c_{23})}(=:f_{2}^{(3)}),
\\
(c_1\geq c_2>c_3),
\mbox{ if }\lambda:\mbox{ type } (3,2)\in{\cal A}_N^{(3;12)},
\\
\frac{\Gamma(c_{23}+1/3)}
     {\Gamma(c_{23}-1/3)}
\frac{\Gamma(c_{23}+2/3)}
     {\Gamma(c_{23})}
\frac{\Gamma(c_{12}-2/3)}
     {\Gamma(c_{12})}
\frac{\Gamma(c_{13}-1)}
     {\Gamma(c_{13}-1/3)}(=:f_{3}^{(3)}),
\\
(c_1>c_2>c_3),
\mbox{ if }\lambda:\mbox{ type } (3,2,2,1)\in{\cal A}_N^{(3;12)},
\end{array}
\right.
\\
\label{su3-u1-ff13}
|F_\lambda^{(3)}|^2/R_{c(\lambda')}^{(3)}
&=&
\left\{
\begin{array}{l}
\frac{\Gamma(c_{13})}
     {\Gamma(c_{13}-2/3)}
\frac{\Gamma(c_{13}+1/3)}
     {\Gamma(c_{13}-1/3)}
\frac{\Gamma(c_{12}-2/3)}
     {\Gamma(c_{12})}
\frac{\Gamma(c_{23}+1/3)}
     {\Gamma(c_{23}+1)}(=:f_{1}^{(3)}),
\\
(c_1>c_2\geq c_3),
\mbox{ if }\lambda:\mbox{ type } (1)\in{\cal A}_N^{(3;13)},
\\
\frac{\Gamma(c_{23}+1/3)}
     {\Gamma(c_{23}-1/3)}
\frac{\Gamma(c_{23}+2/3)}
     {\Gamma(c_{23})}
\frac{\Gamma(c_{12}+1/3)}
     {\Gamma(c_{12}+1)}
\frac{\Gamma(c_{13})}
     {\Gamma(c_{13}+2/3)}(=:f_{2}^{(3)}),
\\
(c_1\geq c_2>c_3),
\mbox{ if }\lambda:\mbox{ type } (2,2)\in{\cal A}_N^{(3;13)},
\\
\frac{\Gamma(c_{12}+1/3)}
     {\Gamma(c_{12}-1/3)}
\frac{\Gamma(c_{12}+2/3)}
     {\Gamma(c_{12})}
\frac{\Gamma(c_{13}-1)}
     {\Gamma(c_{13}-1/3)}
\frac{\Gamma(c_{23}-2/3)}
     {\Gamma(c_{23})}(=:f_{3}^{(3)}),
\\
(c_1>c_2>c_3),
\mbox{ if }\lambda:\mbox{ type } (3,2,1,1)\in{\cal A}_N^{(3;13)},
\end{array}
\right.
\\
\label{su3-u1-ff31}
|F_\lambda^{(3)}|^2/R_{c(\lambda')}^{(3)}
&=&
\left\{
\begin{array}{l}
\frac{\Gamma(c_{12}+1/3)}
     {\Gamma(c_{12}-1/3)}
\frac{\Gamma(c_{12}+2/3)}
     {\Gamma(c_{12})}
\frac{\Gamma(c_{13})}
     {\Gamma(c_{13}+2/3)}
\frac{\Gamma(c_{23}+1/3)}
     {\Gamma(c_{23}+1)}(=:f_{1}^{(3)}),
\\
(c_1>c_2\geq c_3),
\mbox{ if }\lambda:\mbox{ type } (3,1,1)\in{\cal A}_N^{(3;31)},
\\
\frac{\Gamma(c_{13}+1/3)}
     {\Gamma(c_{13}-1/3)}
\frac{\Gamma(c_{13})}
     {\Gamma(c_{13}-2/3)}
\frac{\Gamma(c_{12}+1/3)}
     {\Gamma(c_{12}+1)}
\frac{\Gamma(c_{23}-2/3)}
     {\Gamma(c_{23})}(=:f_{2}^{(3)}),
\\
(c_1\geq c_2>c_3),
\mbox{ if }\lambda:\mbox{ type } (3,3,2)\in{\cal A}_N^{(3;31)},
\\
\frac{\Gamma(c_{23}+1/3)}
     {\Gamma(c_{23}-1/3)}
\frac{\Gamma(c_{23}+2/3)}
     {\Gamma(c_{23})}
\frac{\Gamma(c_{12}-2/3)}
     {\Gamma(c_{12})}
\frac{\Gamma(c_{13}-1)}
     {\Gamma(c_{13}-1/3)}(=:f_{3}^{(3)}),
\\
(c_1>c_2>c_3),
\mbox{ if }\lambda:\mbox{ type } (3,3,2,2,1)\in{\cal A}_N^{(3;31)}.
\end{array}
\right.
\ea
The form factor for $S_N^{(41)(14)}(q,\omega)$
has the form:
\ba
\label{su4-u1-ff14}
|F_\lambda^{(4)}|^2/R_{c(\lambda')}^{(4)}
=
\left\{
\begin{array}{l}
\frac{\Gamma(c_{12}-2/4)}
     {\Gamma(c_{12})}
\frac{\Gamma(c_{13}-3/4)}
     {\Gamma(c_{13}-1/4)}
\frac{\Gamma(c_{23}+2/4)}
     {\Gamma(c_{23}+1)}
\frac{\Gamma(c_{24}+1/4)}
     {\Gamma(c_{24}+3/4)}
\frac{\Gamma(c_{34}+2/4)}
     {\Gamma(c_{34}+1)}
\frac{\Gamma(c_{14})}
     {\Gamma(c_{14}-3/4)}
\frac{\Gamma(c_{14}+1/4)}
     {\Gamma(c_{14}-2/4)}
\\
(=:f_{1}^{(4)}),\,
(c_1>c_2\geq c_3\geq c_4),
\mbox{ if }\lambda:\mbox{ type } (1)\in{\cal A}_N^{(4;14)},
\\
\frac{\Gamma(c_{12}+2/4)}
     {\Gamma(c_{12}+1)}
\frac{\Gamma(c_{13}+1/4)}
     {\Gamma(c_{13}+3/4)}
\frac{\Gamma(c_{14})}
     {\Gamma(c_{14}+2/4)}
\frac{\Gamma(c_{23}-2/4)}
     {\Gamma(c_{23})}
\frac{\Gamma(c_{34}+2/4)}
     {\Gamma(c_{34}+1)}
\frac{\Gamma(c_{24}+1/4)}
     {\Gamma(c_{24}-2/4)}
\frac{\Gamma(c_{24}+2/4)}
     {\Gamma(c_{24}-1/4)}
\\
(=:f_{2}^{(4)}),\,
(c_1\geq c_2>c_3\geq c_4),
\mbox{ if }\lambda:\mbox{ type } (2,2,1)\in{\cal A}_N^{(4;14)},
\\
\frac{\Gamma(c_{12}+2/4)}
     {\Gamma(c_{12}+1)}
\frac{\Gamma(c_{13}-3/4)}
     {\Gamma(c_{13}-1/4)}
\frac{\Gamma(c_{14})}
     {\Gamma(c_{14}+2/4)}
\frac{\Gamma(c_{23}-2/4)}
     {\Gamma(c_{23})}
\frac{\Gamma(c_{34}+2/4)}
     {\Gamma(c_{34}+1)}
\frac{\Gamma(c_{24}+1/4)}
     {\Gamma(c_{24}-2/4)}
\frac{\Gamma(c_{24}+2/4)}
     {\Gamma(c_{24}-1/4)}
\\
(=:f_{3}^{(4)}),\,
(c_1\geq c_2>c_3\geq c_4),
\mbox{ if }\lambda:\mbox{ type } (3,2)\in{\cal A}_N^{(4;14)},
\\
\frac{\Gamma(c_{12}-2/4)}
     {\Gamma(c_{12})}
\frac{\Gamma(c_{13}-3/4)}
     {\Gamma(c_{13}-1/4)}
\frac{\Gamma(c_{23}-2/4)}
     {\Gamma(c_{23})}
\frac{\Gamma(c_{24}+1/4)}
     {\Gamma(c_{24}+3/4)}
\frac{\Gamma(c_{34}+2/4)}
     {\Gamma(c_{34}+1)}
\frac{\Gamma(c_{14})}
     {\Gamma(c_{14}-3/4)}
\frac{\Gamma(c_{14}+1/4)}
     {\Gamma(c_{14}-2/4)}
\\
(=:f_{4}^{(4)}),\,
(c_1>c_2>c_3\geq c_4),
\mbox{ if }\lambda:\mbox{ type } (3,2,2,1,1)\in{\cal A}_N^{(4;14)},
\\
\frac{\Gamma(c_{13}-3/4)}
     {\Gamma(c_{13}-1/4)}
\frac{\Gamma(c_{14}-1)}
     {\Gamma(c_{14}-2/4)}
\frac{\Gamma(c_{23}-2/4)}
     {\Gamma(c_{23})}
\frac{\Gamma(c_{24}-3/4)}
     {\Gamma(c_{24}-1/4)}
\frac{\Gamma(c_{34}+2/4)}
     {\Gamma(c_{34}+1)}
\frac{\Gamma(c_{12}+2/4)}
     {\Gamma(c_{12}-1/4)}
\frac{\Gamma(c_{12}+3/4)}
     {\Gamma(c_{12})}
\\
(=:f_{5}^{(4)}),\,
(c_1>c_2>c_3\geq c_4),
\mbox{ if }\lambda:\mbox{ type } (4,2,1,1,1)\in{\cal A}_N^{(4;14)},
\\
\frac{\Gamma(c_{12}+2/4)}
     {\Gamma(c_{12}+1)}
\frac{\Gamma(c_{13}+1/4)}
     {\Gamma(c_{13}+3/4)}
\frac{\Gamma(c_{14})}
     {\Gamma(c_{14}+2/4)}
\frac{\Gamma(c_{23}+2/4)}
     {\Gamma(c_{23}+1)}
\frac{\Gamma(c_{24}+1/4)}
     {\Gamma(c_{24}+3/4)}
\frac{\Gamma(c_{34}+2/4)}
     {\Gamma(c_{34}-1/4)}
\frac{\Gamma(c_{34}+3/4)}
     {\Gamma(c_{34})}
\\
(=:f_{6}^{(4)}),\,
(c_1\geq c_2\geq c_3>c_4),
\mbox{ if }\lambda:\mbox{ type } (3,3,3)\in{\cal A}_N^{(4;14)},
\\
\frac{\Gamma(c_{12}-2/4)}
     {\Gamma(c_{12})}
\frac{\Gamma(c_{13}+1/4)}
     {\Gamma(c_{13}+3/4)}
\frac{\Gamma(c_{14})}
     {\Gamma(c_{14}+2/4)}
\frac{\Gamma(c_{23}+2/4)}
     {\Gamma(c_{23}+1)}
\frac{\Gamma(c_{24}+1/4)}
     {\Gamma(c_{24}+3/4)}
\frac{\Gamma(c_{34}+2/4)}
     {\Gamma(c_{34}-1/4)}
\frac{\Gamma(c_{34}+3/4)}
     {\Gamma(c_{34})}
\\
(=:f_{7}^{(4)}),\,
(c_1>c_2\geq c_3>c_4),
\mbox{ if }\lambda:\mbox{ type } (3,3,3,2,1,1)\in{\cal A}_N^{(4;14)},
\\
\frac{\Gamma(c_{12}-2/4)}
     {\Gamma(c_{12})}
\frac{\Gamma(c_{14}-1)}
     {\Gamma(c_{14}-2/4)}
\frac{\Gamma(c_{23}+2/4)}
     {\Gamma(c_{23}+1)}
\frac{\Gamma(c_{24}+1/4)}
     {\Gamma(c_{24}+3/4)}
\frac{\Gamma(c_{34}-2/4)}
     {\Gamma(c_{34})}
\frac{\Gamma(c_{13}+1/4)}
     {\Gamma(c_{13}-2/4)}
\frac{\Gamma(c_{13}+2/4)}
     {\Gamma(c_{13}-1/4)}
\\
(=:f_{8}^{(4)}),\,
(c_1>c_2\geq c_3>c_4),
\mbox{ if }\lambda:\mbox{ type } (4,3,3,2,1)\in{\cal A}_N^{(4;14)},
\\
\frac{\Gamma(c_{12}-2/4)}
     {\Gamma(c_{12})}
\frac{\Gamma(c_{14}-1)}
     {\Gamma(c_{14}-2/4)}
\frac{\Gamma(c_{23}+2/4)}
     {\Gamma(c_{23}+1)}
\frac{\Gamma(c_{24}-3/4)}
     {\Gamma(c_{24}-1/4)}
\frac{\Gamma(c_{34}-2/4)}
     {\Gamma(c_{34})}
\frac{\Gamma(c_{13}+1/4)}
     {\Gamma(c_{13}-2/4)}
\frac{\Gamma(c_{13}+2/4)}
     {\Gamma(c_{13}-1/4)}
\\
(=:f_{9}^{(4)}),\,
(c_1>c_2\geq c_3>c_4),
\mbox{ if }\lambda:\mbox{ type } (4,4,3,1,1)\in{\cal A}_N^{(4;14)},
\\
\frac{\Gamma(c_{12}+2/4)}
     {\Gamma(c_{12}+1)}
\frac{\Gamma(c_{13}+1/4)}
     {\Gamma(c_{13}+3/4)}
\frac{\Gamma(c_{14})}
     {\Gamma(c_{14}+2/4)}
\frac{\Gamma(c_{24}-3/4)}
     {\Gamma(c_{24}-1/4)}
\frac{\Gamma(c_{34}-2/4)}
     {\Gamma(c_{34})}
\frac{\Gamma(c_{23}+2/4)}
     {\Gamma(c_{23}-1/4)}
\frac{\Gamma(c_{23}+3/4)}
     {\Gamma(c_{23})}
\\(=:f_{10}^{(4)}),\,
(c_1\geq c_2>c_3>c_4),
\mbox{ if }\lambda:\mbox{ type } (4,3,3,2,2,2,1)\in{\cal A}_N^{(4;14)},
\\
\frac{\Gamma(c_{12}+2/4)}
     {\Gamma(c_{12}+1)}
\frac{\Gamma(c_{13}+1/4)}
     {\Gamma(c_{13}+3/4)}
\frac{\Gamma(c_{14}-1)}
     {\Gamma(c_{14}-2/4)}
\frac{\Gamma(c_{24}-3/4)}
     {\Gamma(c_{24}-1/4)}
\frac{\Gamma(c_{34}-2/4)}
     {\Gamma(c_{34})}
\frac{\Gamma(c_{23}+2/4)}
     {\Gamma(c_{23}-1/4)}
\frac{\Gamma(c_{23}+3/4)}
     {\Gamma(c_{23})}
\\
(=:f_{11}^{(4)}),\,
(c_1\geq c_2>c_3>c_4),
\mbox{ if }\lambda:\mbox{ type } (4,4,3,2,2,2)\in{\cal A}_N^{(4;14)},
\\
\frac{\Gamma(c_{13}-3/4)}
     {\Gamma(c_{13}-1/4)}
\frac{\Gamma(c_{14}-1)}
     {\Gamma(c_{14}-2/4)}
\frac{\Gamma(c_{23}-2/4)}
     {\Gamma(c_{23})}
\frac{\Gamma(c_{24}-3/4)}
     {\Gamma(c_{24}-1/4)}
\frac{\Gamma(c_{34}-2/4)}
     {\Gamma(c_{34})}
\frac{\Gamma(c_{12}+2/4)}
     {\Gamma(c_{12}-1/4)}
\frac{\Gamma(c_{12}+3/4)}
     {\Gamma(c_{12})}
\\
(=:f_{12}^{(4)}),\,
(c_1>c_2>c_3>c_4),
\mbox{ if }\lambda:\mbox{ type } (4,4,3,3,2,2,1,1,1)\in{\cal A}_N^{(4;14)}.
\end{array}
\right.
\ea

\bigskip
\noindent
{\it (iii) Definition of the form factor
$F_{c(\lambda')}^{(K)}$}

The form factors
$F_{c(\lambda')}^{(K)}$ for $K=2,3,4$
are defined as follows:
\ba
\label{su2-c}
F_{c(\lambda')}^{(2)}
&:=&
F_\lambda^{(2)},
\\
\label{su3-c}
|F_{c(\lambda')}^{(3)}|^2/R_{c(\lambda')}^{(3)}
&:=&
\left\{
\begin{array}{ll}
f_{1}^{(3)},
\, (c_1>c_2= c_3),
\\
f_{2}^{(3)},
\, (c_1=c_2>c_3),
\\
\sum_{i=1}^3f_{i}^{(3)},
\, (c_1>c_2>c_3),
\end{array}
\right.
\\
\label{su4-c}
|F_{c(\lambda')}^{(4)}|^2/R_{c(\lambda')}^{(4)}
&:=&
\left\{
\begin{array}{ll}
f_{1}^{(4)}
\, (c_1>c_2=c_3=c_4),
\\
f_{2}^{(4)}+f_{3}^{(4)},
\, (c_1=c_2>c_3=c_4),
\\
f_{6}^{(4)},
\, (c_1=c_2=c_3>c_4),
\\
f_{1}^{(4)}+f_{2}^{(4)}+f_{3}^{(4)}+f_{4}^{(4)}+f_{5}^{(4)},
\, (c_1>c_2>c_3=c_4),
\\
f_{1}^{(4)}+f_{6}^{(4)}+f_{7}^{(4)}+f_{8}^{(4)}+f_{9}^{(4)},
\, (c_1>c_2=c_3>c_4),
\\
f_{2}^{(4)}+f_{3}^{(4)}+f_{6}^{(4)}+f_{10}^{(4)}+f_{11}^{(4)},
\, (c_1=c_2>c_3>c_4),
\\
\sum_{i=1}^{12}f_{i}^{(4)},
\, (c_1>c_2>c_3>c_4),
\end{array}
\right.
\ea
where $f_i^{(K)}$ are defined in
(\ref{su3-u1-ff12}),
(\ref{su3-u1-ff13}),
(\ref{su3-u1-ff31}) and
(\ref{su4-u1-ff14}).

\bigskip
\noindent
{\it (iv) Proof of the independence
of
$S_N^{(\delta\gamma)(\gamma\delta)}(q,\omega)$
for the pair $(\gamma,\delta)$ with $\gamma\ne\delta$}

We prove that
$S_N^{(\delta\gamma)(\gamma\delta)}(q,\omega)$
is independent of the pair $(\gamma,\delta)$
as long as $\gamma\ne\delta$.
We introduce the contragradient partition
$\lambda^c=(\lambda^c_1,\cdots,\lambda^c_N)$
of the partition $\lambda=(\lambda_1,\cdots,\lambda_N)\in\Lambda_N$
by $\lambda^c_i=\lambda_1-\lambda_{N-i+1}$ for $i=1,\cdots,N$.
For an excited state $\lambda\in\Lambda_N^{(K)}$ of
$S_N^{(\delta\gamma)(\gamma\delta)}(q,\omega)$,
the contragradient partition
$\lambda^c$ is also
the excited state of $S_N^{(\delta\gamma)(\gamma\delta)}(q,\omega)$
and gives the intensity
at point $(2\pi-P_\lambda,E_\lambda)$ in the momentum-frequency plane.
That is, the states $\lambda$ and $\lambda^c$ give the intensity
at points which are symmetric with respect to the axis $q=\pi$.
We can easily prove this fact by using the following relations:
\be
\label{c-c}
c_i(\lambda')+c_{K-i+1}({\lambda^c}')=N/K,
\quad i=1,\cdots,K.
\ee
Notice that {\it types} of $\lambda$ and  $\lambda^c$
have the simple relation.
For example, in the case of $K=3$
and $(\gamma,\delta)=(1,3)$, $(\rho,\sigma)=(1,2)$,
we can easily show that
\ba
\begin{array}{lcl}
\lambda:\mbox{ type } (1)\in{\cal A}_N^{(3;13)}
&\Leftrightarrow&
\lambda^c:\mbox{ type } (3,2)\in{\cal A}_N^{(3;12)},
\\
\lambda:\mbox{ type } (2,2)\in{\cal A}_N^{(3;13)}
&\Leftrightarrow&
\lambda^c:\mbox{ type } (1,1)\in{\cal A}_N^{(3;12)},
\\
\lambda:\mbox{ type } (3,2,1,1)\in{\cal A}_N^{(3;13)}
&\Leftrightarrow&
\lambda^c:\mbox{ type } (3,2,2,1)\in{\cal A}_N^{(3;12)}.
\end{array}
\ea
Using the above fact and relation (\ref{c-c}),
we can show that
\be
\label{s2s'}
S_N^{(\delta\gamma)(\gamma\delta)}(q,\omega)
=S_N^{(\sigma\rho)(\rho\sigma)}(2\pi-q,\omega)
\ee
for any $(\gamma,\delta)\ne(\rho,\sigma)$.
We can also prove the following formula:
\be
\label{s2s}
S_N^{(\delta\gamma)(\gamma\delta)}(q,\omega)
=S_N^{(\delta\gamma)(\gamma\delta)}(2\pi-q,\omega).
\ee
Therefore we proved that
$S_N^{(\delta\gamma)(\gamma\delta)}(q,\omega)$
is independent of the pair $(\gamma,\delta)$
with $\gamma\ne\delta$.

\section*{Appendix E:
Correspondence
between States and Semistandard Tableaux}
\label{appE}

In this Appendix, we recall the correspondence
between the states of the $SU(K)$ HS model and
the semistandard tableaux given in Refs. \cite{KKN,Uglov2}. From
this result, we can generalize
the correspondence
between the momenta of spinons and $K$-spinon motifs
given in subsection \ref{relation2motif}
to arbitrary motifs.
In the following, we assume that
$N\equiv 0\mbox{ mod }K$ and $N/K\equiv 1\mbox{ mod }2$.

\noindent
{\it (i) $K$-regular partition}

The $K$-regular partition is the
partition $\lambda=(\lambda_1,\cdots,\lambda_N)\in\Lambda_N$
which satisfies $\lambda_i-\lambda_{i+1}<K$
for $i=1,\cdots,N$ ($\lambda_{N+1}=0$).
In the formulation in section \ref{corr-spinCS},
the state with $K$-regular partition means
the state without phonon contribution.
We denote the set of all $K$-regular partitions
by $\Lambda_{N;K}$.
The set $\Lambda_{N;K}$ is finite.
For a $K$-regular partition $\lambda\in\Lambda_{N;K}$,
we write $\lambda+k^{(0)}=k=\underbar{\mbox{$k$}}+K\bar{k}$
where
$\underbar{\mbox{$k$}}
=(\underbar{\mbox{$k$}}_1,\cdots,\underbar{\mbox{$k$}}_N)
\in(\Sigma_K)^N$ and
$\bar{k}=(\bar{k}_1,\cdots,\bar{k}_N)\in\mbox{{\bf Z}}^N$.
Here $k^{(0)}=(M,M-1,\cdots,M-N+1)$
with $M=(N+K)/2\in\mbox{{\bf Z}}_{>0}$.
We represent $\bar{k}$ as $r_1^{p_1}r_2^{p_2}\cdots$
where $r_1>r_2>\cdots$ and $p_i$ denotes the multiplicity
of $r_i$ in $\bar{k}$.
For example, in the case of $K=3$ and $N=3$,
we have
\ba
\label{exK3N3}
\begin{array}{ccc}
\lambda & \underbar{\mbox{$k$}} & \bar{k}
\\
(1,0,0)&(1,2,1)&1^1 0^2
\\
(1,1,0)&(1,3,1)&1^1 0^2
\\
(1,1,1)&(1,3,2)&1^1 0^2
\\
(2,0,0)&(2,2,1)&1^1 0^2
\\
(2,1,0)&(2,3,1)&1^1 0^2
\\
(2,1,1)&(2,3,2)&1^1 0^2
\\
(3,1,0)&(3,3,1)&1^1 0^2
\\
(3,1,1)&(3,3,2)&1^1 0^2
\end{array}
\ea
For the groundstate
$k^{(0)}=\underbar{\mbox{$k$}}^{(0)}+K\bar{k}^{(0)}$,
we have
\ba
\underbar{\mbox{$k$}}^{(0)}
&=&
(K,K-1,\cdots,1,K,K-1,\cdots,1,\cdots,K,K-1,\cdots,1),
\\
\bar{k}^{(0)}
&=&
\left(\frac{1}{2}\left(\frac{N}{K}-1\right)\right)^K
\left(\frac{1}{2}\left(\frac{N}{K}-3\right)\right)^K
\cdots
\left(-\frac{1}{2}\left(\frac{N}{K}-1\right)\right)^K.
\ea

Clearly the set $\Lambda_{N;K}$
is the subset of $\Lambda_{N}^{(N(K-1))}$.
That is, the conjugate partition $\lambda'$ of $\lambda\in\Lambda_{N;K}$
belongs to $\Lambda_{N(K-1)}^{(N)}$.
Therefore, in general, the state with $K$-regular partition
represents the $N(K-1)$-spinon excitation.
The momenta of spinons
$c=(c_1,\cdots,c_K,c_{K+1},\cdots,c_{2K},
\cdots,c_{N(K-1)-K+1},\cdots,c_{N(K-1)})$
of the state $\lambda\in\Lambda_{N;K}$
satisfies the following conditions:
\ba
\left\{
\begin{array}{l}
c_{(l-1)K+1}\geq\cdots\geq c_{lK}
\mbox{ for } l=1,\cdots,N(K-1)/K,
\\
\mbox{the cases of } c_{(l-1)K+1}=\cdots =c_{lK}
\mbox{ are prohibited},
\\
c_i\geq c_{i+K}.
\end{array}
\right.
\ea
For example, in the case of $K=4$ and $N=4$,
the state with the $4$-regular partition
$\lambda=(12,9,6,3)$ has the momenta of spinons
$c=(1,1,1,0,1,1,0,0,1,0,0,0)$:
{\scriptsize
\begin{eqnarray*}
\Yvcentermath1
\young(
\zero\one\two\three\zero\one\two\three\zero\one\two\three,%
\three\zero\one\two\three\zero\one\two\three,%
\two\three\zero\one\two\three,%
\one\two\three
)
\end{eqnarray*}
}
In Table \ref{K-reglarK3N3},
we give the data of the $K$-regular partitions
for $(K,N)=(3,3)$.

\bigskip
\begin{center}
{\bf \large Table \ref{K-reglarK3N3}}
\end{center}
\bigskip

\noindent
{\it (ii) Ribbon diagram}

We denote the set of all motifs by ${\cal M}_N^{(K)}$.
The motifs can be expressed by the rank $K$ ribbon diagrams
with length $N$ \cite{KKN}.
The rank $K$ ribbon diagram with length $N$ consists of
columns of boxes and satisfies the following conditions:

\noindent
(a) The total number of boxes is $N$.\\
(b) The diagram is connected and contains no $2\times 2$
blocks of boxes.\\
(c) The heights of all its columns do not exceed $K$.\\
We denote the ribbon diagram
of $l$ columns such that the height of $i$-th
column from the left is $p_i$
by $[p_1,\cdots,p_l]$.
For example, $[2,3,1,1,2,2,1]$
is the rank $K(\geq3)$ ribbon diagram with length $12$:
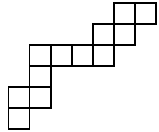
\begin{eqnarray*}
{\scriptsize
\young(%
:::::\hfil\hfil,%
::::\hfil\hfil,%
:\hfil\hfil\hfil\hfil,%
:\hfil,%
\hfil\hfil,%
\hfil
)
}
\end{eqnarray*}
We denote the set of all rank $K$ ribbon diagrams with length $N$
by ${\cal R}_N^{(K)}$.
The one-to-one
correspondence between ${\cal M}_N^{(K)}$ and ${\cal R}_N^{(K)}$
is given as follows \cite{KKN}:
for a motif $d=d_0 d_1\cdots d_N$,

\noindent
(I) Write the first box.\\
(II) Attach the second box under (resp. left to) the first box
if $d_1=1$ (resp. $d_1=0$).\\
(III) Similarly attach the $(i+1)$-th box under (resp. left to)
the $i$-th box if $d_i=1$ (resp. $d_i=0$)
for $i=2,\cdots,N-1$.\\
For example, the motif $d=0010100011010$ corresponds to the
ribbon diagram $[2,3,1,1,2,2,1]$.
The groundstate motif $d^{(0)}\in{\cal M}_N^{(K)}$ is expressed by
the ribbon diagram
$\theta^{(0)}:=[{\underbrace{K,\cdots,K}_{N/K}}]\in{\cal R}_N^{(K)}$
which contains $N/K$ columns with length $K$.
For example, in the case of $K=3$ and $N=9$,
the ribbon diagram $\theta^{(0)}=[3,3,3]$
for the groundstate motif $d^{(0)}=0110110110$
is given by
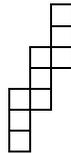
\begin{eqnarray*}
{\scriptsize
\young(%
::\hfil,%
::\hfil,%
:\hfil\hfil,%
:\hfil,%
\hfil\hfil,%
\hfil,%
\hfil
)
}
\end{eqnarray*}
In Table \ref{motif-K3N3},
we give the data of the motifs
for $(K,N)=(3,3)$.

\bigskip
\begin{center}
{\bf \large Table \ref{motif-K3N3}}
\end{center}
\bigskip

\noindent
{\it (iii) Semistandard tableau}

In each box of a given ribbon diagram $\theta$,
let us inscribe one of the numbers $1,2,\cdots, K$.
We call such an arrangement of numbers
a semistandard tableau of shape $\theta$
if it satisfies the following conditions:
let $a$ and $b$ be the inscribed numbers in any
pair of adjacent boxes, then

\noindent
(A) $a<b$ if $b$ is lower-adjacent to $a$.\\
(B) $a\geq b$ if $b$ is left-adjacent to $a$.\\
We denote the set of all
semistandard tableaux of shape $\theta$
by ${\cal SST}(\theta)$.
For example,
\ba
&&{\cal SST}
\left(
{{\tiny {\Yvcentermath1\young(\hfil,\hfil,\hfil)}}}
\right)
=
\left\{
{\scriptsize {\Yvcentermath1 \young(1,2,3)}}
\right\},
\\
&&
\label{sst}
{\cal SST}
\left(
{{\tiny {\Yvcentermath1\young(:\hfil,\hfil\hfil)}}}
\right)
=
\left\{
{\scriptsize {\Yvcentermath1 \young(:1,12)}},
{\scriptsize {\Yvcentermath1 \young(:1,13)}},
{\scriptsize {\Yvcentermath1 \young(:2,13)}},
{\scriptsize {\Yvcentermath1 \young(:1,22)}},
{\scriptsize {\Yvcentermath1 \young(:1,23)}},
{\scriptsize {\Yvcentermath1 \young(:2,23)}},
{\scriptsize {\Yvcentermath1 \young(:1,33)}},
{\scriptsize {\Yvcentermath1 \young(:2,33)}}
\right\},
\\
&&{\cal SST}
\left(
{{\tiny {\Yvcentermath1\young(\hfil\hfil,\hfil)}}}
\right)
=
\left\{
{\scriptsize {\Yvcentermath1 \young(11,2)}},
{\scriptsize {\Yvcentermath1 \young(12,2)}},
{\scriptsize {\Yvcentermath1 \young(13,2)}},
{\scriptsize {\Yvcentermath1 \young(11,3)}},
{\scriptsize {\Yvcentermath1 \young(12,3)}},
{\scriptsize {\Yvcentermath1 \young(13,3)}},
{\scriptsize {\Yvcentermath1 \young(22,3)}},
{\scriptsize {\Yvcentermath1 \young(23,3)}}
\right\},
\\
&&{\cal SST}
\left(
{{\tiny {\Yvcentermath1\young(\hfil\hfil\hfil)}}}
\right)
=
\left\{
{\scriptsize {\Yvcentermath1 \young(111)}},
{\scriptsize {\Yvcentermath1 \young(112)}},
{\scriptsize {\Yvcentermath1 \young(113)}},
{\scriptsize {\Yvcentermath1 \young(122)}},
{\scriptsize {\Yvcentermath1 \young(123)}},
\right.
\nonumber
\\
&&\qquad\qquad\quad\quad
\left.
{\scriptsize {\Yvcentermath1 \young(133)}},
{\scriptsize {\Yvcentermath1 \young(222)}},
{\scriptsize {\Yvcentermath1 \young(223)}},
{\scriptsize {\Yvcentermath1 \young(233)}},
{\scriptsize {\Yvcentermath1 \young(333)}}
\right\}.
\ea
Notice that the number of element in the set
${\cal SST}(\theta^{(0)})$ is 1.

\noindent
{\it (iv) Correspondence between $\Lambda_{N;K}$
and $\sqcup_{\theta\in{\cal R}_N^{(K)}}{\cal SST}(\theta)$
}\cite{Uglov2}

The one-to-one correspondence between $K$-regular partitions
and semistandard tableaux is given in Ref. \cite{Uglov2}.
Let us consider the $K$-regular partition $\lambda\in\Lambda_{N;K}$
with $\underbar{\mbox{$k$}}=(a_1,\cdots,a_N)$ and
$\bar{k}=r_1^{p_1}r_2^{p_2}\cdots r_l^{p_l}$.
The corresponding semistandard tableau $T$ of shape $\theta$
is assigned as follows:

\noindent
($\alpha$) The shape $\theta$ is given by $[p_1,\cdots,p_l]$.\\
($\beta$) Inscribe number $a_i$ in the $i$-th box in $\theta$
counting from the left to right and from the bottom to top. \\
For example, from Eqs. (\ref{exK3N3}) and (\ref{sst}),
we have the following correspondence:
\ba
\begin{array}{ccc}
\lambda & & T
\\
(1,0,0)&\leftrightarrow
&{\scriptsize {\Yvcentermath1 \young(:1,12)}}
\\
\\
(1,1,0)&\leftrightarrow
&{\scriptsize {\Yvcentermath1 \young(:1,13)}}
\\
\\
(1,1,1)&\leftrightarrow
&{\scriptsize {\Yvcentermath1 \young(:2,13)}}
\\
\\
(2,0,0)&\leftrightarrow
&{\scriptsize {\Yvcentermath1 \young(:1,22)}}
\\
\\
(2,1,0)&\leftrightarrow
&{\scriptsize {\Yvcentermath1 \young(:1,23)}}
\\
\\
(2,1,1)&\leftrightarrow
&{\scriptsize {\Yvcentermath1 \young(:2,23)}}
\\
\\
(3,1,0)&\leftrightarrow
&{\scriptsize {\Yvcentermath1 \young(:1,33)}}
\\
\\
(3,1,1)&\leftrightarrow
&{\scriptsize {\Yvcentermath1 \young(:2,33)}}
\end{array}
\ea
{}From the above correspondence,
we can assign the momenta of spinons to arbitrary motifs.

\section*{Appendix F:
Asymptotic Behavior}
\label{appF}

First we consider the low energy asymptotic behavior
of the DCCF given by
\ba
\label{DCCF-thermo'}
\eta(r,t)
&=&
(-1)^r \frac{A_K}{2\pi}
\prod_{j=1}^K\int_{-1}^{1}\frac{dk_j}{(1-k_j^2)^{1/K}}
\frac{\sum_{1\leq j<l\leq K}(k_j-k_l)^{2}}
     {\prod_{1\leq j<l\leq K}(k_j-k_l)^{2/K}}
e^{-it\epsilon(k)+irp(k)}.
\ea
The singularities of the DCCF
come from the momentum space region
where all $k_j$'s are near the Fermi points $\{\pm 1\}$.
These contributions are divided into $K+1$ sectors:
for $l=0,\cdots,K$,
$k_i$'s with $i=1,\cdots,l$ close to $1$
and $k_j$'s with $j=l+1,\cdots,K$ close to $-1$.
It is sufficient to consider these sectors,
because the integrand is symmetric in $k_1,\cdots,k_K$.
We put $k_i=1-u_i$ ($i=1,\cdots,l$)
and $k_i=-1+u_i$ ($i=l+1,\cdots,K$) with
$|u_i|<<1$.
We have
\be
\label{momentum-energy-condition}
rp(k)-t\epsilon(k)
\sim
2lk_Fr-\pi r
-k_F\sum_{i=1}^lu_i\xi_L
+k_F\sum_{i=l+1}^Ku_i\xi_R,
\ee
where $\xi_L=r+v_st$ and $\xi_R=r-v_st$.
Then the singularities of $\eta(r,t)$
are derived by a simple power counting.
For example, if $l\ne 0,K$, then we obtain the exponent as
\ba
-(\mbox{exponent of }\xi_L)
&=&
l
-\frac{1}{K}l
-\frac{2}{K}\frac{l(l-1)}{2},
\\
-(\mbox{exponent of }\xi_R)
&=&
K-l
-\frac{1}{K}(K-1)
-\frac{2}{K}\frac{(K-l)(K-l-1)}{2}.
\ea
In the right hand side of the above formulae,
the first terms come from the integrals,
the second terms come from $\prod(1-k_i^2)^{-1/K}$,
and the last terms come from $\prod(k_i-k_j)^{-2/K}$.
The exponents $\alpha_0$ and $\alpha_K$ can be
derived in the same way.

Next we derive the exponent $\eta_{m,n,f}$
in Eq. (\ref{asympt-DSF})
for $S(q,\omega)$ given by
\ba
\label{DSF-thermo'}
S(q,\omega)
&=&
A_K
\prod_{j=1}^K\int_{-1}^{1}\frac{dk_j}{(1-k_j^2)^{1/K}}
\frac{\sum_{1\leq j<l\leq K}(k_j-k_l)^{2}}
     {\prod_{1\leq j<l\leq K}(k_j-k_l)^{2/K}}
\delta(q-\pi-p(k))\delta(\omega-\epsilon(k)).
\label{thermo-DSF'}
\ea
We consider the
contributions of the form factor to $S(q,\omega)$
come from the momentum space region
where
$(m+n)$ spinons $k_j$'s are near the Fermi points $\{\pm 1\}$
and
$f$ spinons are moving together with same momentum $Q/f$
with $Q\sim{\cal O}(1)$,
where $m,n\geq 0$, $f\geq 1$ are integers with $m+n+f=K$.
Let us consider the sector which
$k_i$'s with $i=1,\cdots,f$ move together,
$k_i$'s with $i=f+1,\cdots,m+f$ close to $-1$
and $k_i$'s with $i=m+f+1,\cdots,K$ close to $1$.
We put
$k_i=Q/f-u_i$ ($i=1,\cdots,f$),
$k_i=-1+u_i$ ($i=f+1,\cdots,m+f$)
and $k_i=1-u_i$ ($i=m+f+1,\cdots,K$) with
$|u_i|<<1$.

The restriction on the total momentum is given by
\ba
\label{momentum-condition}
&&
q-\pi-k_Ff\frac{Q}{f}+k_Fm-k_Fn
+k_F\sum_{i=1}^{f}u_i
-k_F\sum_{i=f+1}^{f+m}u_i
+k_F\sum_{i=f+m+1}^{K}u_i
=
0
\\
&\Leftrightarrow&
Q=q/k_F+m-n-K+{\cal O}(u_i).
\ea
Then the restriction on the energy is given by
\ba
\label{energy-condition}
&&
\omega
-\frac{\pi v_s}{2K}f\Big(1-\Big(\frac{Q}{f}\Big)^2\Big)
-\frac{\pi v_s}{2K}\sum_{i=1}^{f}2\frac{Q}{f}u_i
-\frac{\pi v_s}{2K}\sum_{i=f+1}^{K}2u_i=0
\\
&\Leftrightarrow&
\omega=\epsilon_{m,n,f}(q)+{\cal O}(u_i),
\ea
where the dispersion $\epsilon_{m,n,f}(q)$ 
is defined by Eq. (\ref{general-dispersion-line}).
We obtain the exponent $\eta_{m,n,f}$
with $(m,n,f)\ne(0,0,K)$
in Eq. (\ref{asympt-DSF})
as
\be
\eta_{m,n,f}
=
K-2-\frac{1}{K}(m+n)
-\frac{2}{K}
\Big[
\frac{m(m-1)}{2}+\frac{n(n-1)}{2}+\frac{f(f-1)}{2}
\Big],
\ee
where $K-2$ comes from the number of independent integration variables,
$-(m+n)/K$ comes from $\prod (1-k_i^2)^{-1/K}$
and the last term comes from $\prod (k_i-k_j)^{-2/K}$.
The exponent $\eta_{0,0,K}$ can be derived in the same manner.



\newpage

\begin{figure}
\caption[]{%
Exact results of $S_N(q,\omega)$
in the cases of $(K,N)=(3,18)$ and $(4,16)$.
The vertical and horizontal axis represent
the rescaled energy and momentum, respectively.
The intensity is proportional to the area of the circle.
The solid lines are the dispersion lines of the elementary excitations
in the thermodynamic limit.}
\label{fig1}
\end{figure}
\begin{figure}
\caption[]{%
Exact results of the static structure factor $S_N(q)$
in the cases of $(K,N)=(3,18)$ and $(4,16)$.}
\label{fig2}
\end{figure}
\begin{figure}
\caption[]{
Critical exponents $\eta_{m,n,f}$ of $S(q,\omega)$
in the cases of $K=3$ and $4$.
}
\label{fig3}
\end{figure}

\newpage


\begin{table}[]
\caption{The comparison between the exact result
and numerical data of
$S_N(q,\omega)$ for $K=3$ and $N=15$.
The circulating decimals is denoted, for example, by
$0.a\dot{b}\dot{c}=0.abcbcbc\cdots$.}
\label{table-eAn}
\begin{center}
\begin{tabular}{cccc}
\hline
$q/\pi$
&$\omega/(\pi^2J)$
&$S_{15}(q,\omega)$: exact result
& Numerical data
\\
\hline
 $2/15$
&$14/225$
&$1/14=0.0\dot{7}\dot{1}\dot{4}\dot{2}\dot{8}\dot{5}$
&$0.0714283$
\\
\hline
 $4/15$
&$22/225$
&$9/77=0.\dot{1}\dot{1}\dot{6}\dot{8}\dot{8}\dot{3}$
&$0.1168791$
\\
\hline
 $4/15$
&$26/225$
&$1/26=0.0\dot{3}\dot{8}\dot{4}\dot{6}\dot{1}\dot{5}$
&$0.0384614$
\\
\hline
 $2/5$
&$24/225$
&$243/1540=0.15\dot{7}\dot{7}\dot{9}\dot{2}\dot{2}\dot{0}$
&$0.1577902$
\\
\hline
 $2/5$
&$34/225$
&$72/715=0.1\dot{0}\dot{0}\dot{6}\dot{9}\dot{9}\dot{3}$
&$0.1006985$
\\
\hline
 $8/15$
&$4/45$
&$6561/30800=0.2130\dot{1}\dot{9}\dot{4}\dot{8}\dot{0}\dot{5}$
&$0.2130180$
\\
\hline
 $8/15$
&$36/225$
&$243/2288=0.10620\dot{6}\dot{2}\dot{9}\dot{3}\dot{7}\dot{0}$
&$0.1062047$
\\
\hline
 $8/15$
&$8/45$
&$18/325=0.055\dot{3}\dot{8}\dot{4}\dot{6}\dot{1}\dot{5}$
&$0.0553838$
\\
\hline
 $8/15$
&$44/225$
&$1/44=0.02\dot{2}\dot{7}$
&$0.0227276$
\\
\hline
 $2/3$
&$2/45$
&$59049/169400=0.348\dot{5}\dot{7}\dot{7}\dot{3}\dot{3}\dot{1}\cdots$
&$0.3485755$
\\
\hline
 $2/3$
&$32/225$
&$39366/275275=0.14\dot{3}\dot{0}\dot{0}\dot{6}\dot{0}\dot{8}\cdots$
&$0.1430055$
\\
\hline
 $2/3$
&$42/225$
&$243/2600=0.093\dot{4}\dot{6}\dot{1}\dot{5}\dot{3}\dot{8}$
&$0.0934600$
\\
\hline
 $2/3$
&$46/225$
&$27/616=0.043\dot{8}\dot{3}\dot{1}\dot{1}\dot{6}\dot{8}$
&$0.0438308$
\\
\hline
 $2/3$
&$2/9$
&$1/50=0.02$
&$0.0200004$
\\
\hline
 $4/5$
&$22/225$
&$59049/250250=0.235\dot{9}\dot{6}\dot{0}\dot{0}\dot{3}\dot{9}$
&$0.2359586$
\\
\hline
 $4/5$
&$38/225$
&$6561/100100=0.06\dot{5}\dot{5}\dot{4}\dot{4}\dot{4}\dot{5}$
&$0.0655457$
\\
\hline
 $4/5$
&$14/75$
&$20169/143000=0.141\dot{0}\dot{4}\dot{1}\dot{9}\dot{5}\dot{8}$
&$0.1410389$
\\
\hline
 $4/5$
&$52/225$
&$54/875=0.061\dot{7}\dot{1}\dot{4}\dot{2}\dot{8}\dot{5}$
&$0.0617144$
\\
\hline
 $14/15$
&$28/225$
&$59049/625625=0.0943\dot{8}\dot{4}\dot{0}\dot{1}\dot{5}\dot{9}$
&$0.0943822$
\\
\hline
 $14/15$
&$32/225$
&$6561/57200=0.1147\dot{0}\dot{2}\dot{7}\dot{9}\dot{7}\dot{2}$
&$0.1147006$
\\
\hline
 $14/15$
&$38/225$
&$78732/625625=0.1258\dot{4}\dot{5}\dot{3}\dot{5}\dot{4}\dot{6}$
&$0.1258442$
\\
\hline
 $14/15$
&$16/75$
&$729/10000=0.729$
&$0.7289911$
\\
\hline
 $14/15$
&$52/225$
&$27/700=0.03\dot{8}\dot{5}\dot{7}\dot{1}\dot{4}\dot{2}$
&$0.0385710$
\\
\hline
 $14/15$
&$56/225$
&$1/56=0.017\dot{8}\dot{5}\dot{7}\dot{1}\dot{4}\dot{2}$
&$0.0178575$
\\
\hline
\end{tabular}
\end{center}
\end{table}


\begin{table}[]
\caption{Data of $SU(2)$ spin}
\label{selection2}
\begin{center}
\begin{tabular}{lcc}
\hline
  $S^{(\delta\gamma)(\rho\sigma)}(q,\omega)$
& $s(\gamma,\delta)$
& $SU(2)$ spins of $2$ spinons
\\
\hline
  $S^{(\gamma\gamma)(\delta\delta)}(q,\omega)$
& $0$
& $1/2,-1/2$
\\
\hline
  $S^{(21)(12)}(q,\omega)$
& $1$
& $1/2,1/2$
\\
  $S^{(12)(21)}(q,\omega)$
& $-1$
& $-1/2,-1/2$
\\
\hline
\end{tabular}
\end{center}
\end{table}


\begin{table}[]
\caption{Data of $SU(3)$ spin}
\label{selection3}
\begin{center}
\begin{tabular}{lcc}
\hline
  $S^{(\delta\gamma)(\rho\sigma)}(q,\omega)$
& $s(\gamma,\delta)$
& $SU(3)$ spins of $3$ spinons
\\
\hline
  $S^{(\gamma\gamma)(\delta\delta)}(q,\omega)$
& $(0,0)$
& $(0,1/2),(1/2,-1/2),(-1/2,0)$
\\
\hline
  $S^{(21)(12)}(q,\omega)$
& $(1,-1/2)$
& $(1/2,-1/2),(1/2,-1/2),(0,1/2)$
\\
  $S^{(12)(21)}(q,\omega)$
& $(-1,1/2)$
& $(-1/2,0),(-1/2,0),(0,1/2)$
\\
  $S^{(31)(13)}(q,\omega)$
& $(1/2,1/2)$
& $(0,1/2),(0,1/2),(1/2,-1/2)$
\\
  $S^{(13)(31)}(q,\omega)$
& $(-1/2,-1/2)$
& $(-1/2,0),(-1/2,0),(1/2,-1/2)$
\\
  $S^{(32)(23)}(q,\omega)$
& $(-1/2,1)$
& $(0,1/2),(0,1/2),(-1/2,0)$
\\
  $S^{(23)(32)}(q,\omega)$
& $(1/2,-1)$
& $(1/2,-1/2),(1/2,-1/2),(-1/2,0)$
\\
\hline
\end{tabular}
\end{center}
\end{table}

\begin{table}[]
\caption{Data of $SU(4)$ spin}
\label{selection4}
\begin{center}
\begin{tabular}{lcc}
\hline
  $S^{(\delta\gamma)(\rho\sigma)}(q,\omega)$
& $s(\gamma,\delta)$
& $SU(4)$ spins of $4$ spinons
\\
\hline
  $S^{(\gamma\gamma)(\delta\delta)}(q,\omega)$
& $(0,0,0)$
& $(0,0,1/2),(0,1/2,-1/2),(1/2,-1/2,0),(-1/2,0,0)$
\\
\hline
  $S^{(21)(12)}(q,\omega)$
& $(1,-1/2,0)$
& $(1/2,-1/2,0),(1/2,-1/2,0),(0,0,1/2),(0,1/2,-1/2)$
\\
  $S^{(12)(21)}(q,\omega)$
& $(-1,1/2,0)$
& $(-1/2,0,0),(-1/2,0,0),(0,0,1/2),(0,1/2,-1/2)$
\\
  $S^{(31)(13)}(q,\omega)$
& $(1/2,1/2,-1/2)$
& $(0,1/2,-1/2),(0,1/2,-1/2),(0,0,1/2),(1/2,-1/2,0)$
\\
  $S^{(13)(31)}(q,\omega)$
& $(-1/2,-1/2,1/2)$
& $(-1/2,0,0),(-1/2,0,0),(0,0,1/2),(1/2,-1/2,0)$
\\
  $S^{(41)(14)}(q,\omega)$
& $(1/2,0,1/2)$
& $(0,0,1/2),(0,0,1/2),(0,1/2,-1/2),(1/2,-1/2,0)$
\\
  $S^{(14)(41)}(q,\omega)$
& $(-1/2,0,-1/2)$
& $(-1/2,0,0),(-1/2,0,0),(0,1/2,-1/2),(1/2,-1/2,0)$
\\
  $S^{(32)(23)}(q,\omega)$
& $(-1/2,1,-1/2)$
& $(0,1/2,-1/2),(0,1/2,-1/2),(0,0,1/2),(-1/2,0,0)$
\\
  $S^{(23)(32)}(q,\omega)$
& $(1/2,-1,1/2)$
& $(1/2,-1/2,0),(1/2,-1/2,0),(0,0,1/2),(-1/2,0,0)$
\\
  $S^{(42)(24)}(q,\omega)$
& $(-1/2,1/2,1/2)$
& $(0,0,1/2),(0,0,1/2),(0,1/2,-1/2),(-1/2,0,0)$
\\
  $S^{(24)(42)}(q,\omega)$
& $(1/2,-1/2,-1/2)$
& $(1/2,-1/2,0),(1/2,-1/2,0),(0,1/2,-1/2),(-1/2,0,0)$
\\
  $S^{(43)(34)}(q,\omega)$
& $(0,-1/2,1)$
& $(0,0,1/2),(0,0,1/2),(1/2,-1/2,0),(-1/2,0,0)$
\\
  $S^{(34)(43)}(q,\omega)$
& $(0,1/2,-1)$
& $(0,1/2,-1/2),(0,1/2,-1/2),(1/2,-1/2,0),(-1/2,0,0)$
\\
\hline
\end{tabular}
\end{center}
\end{table}



\begin{table}[]
\caption{Data for the sets
${\cal A}_N^{(3;12)}$ and ${\cal A}_N^{(3;21)}$.}
\label{A12-3}
\begin{center}
\begin{tabular}{cllccccccc}
\hline
  $|\nu|$ (mod 3)
& $C_3^{(a)}(\nu)$
& $H_3^{(a)}(\nu)$
& $S_\nu^1$
& $S_\nu^2$
& $c\mbox{-}h_\nu^{(0)}$
& $c\mbox{-}h_\nu^{(1)}$
& $c\mbox{-}h_\nu^{(2)}$
& $n(\nu)$ (mod 3)
& $A_\nu$
\\
\hline
${\cal A}_N^{(3;12)}$&&&&&&&&&
\\
\hline
  $2\ (2)$
&
{\small $
\Yvcentermath1
\young(
\zero,%
\two
)
$}
&
{\small $
\Yvcentermath1
\young(
\two,%
\one
)
$}
& $1$
& $-1/2$
& $1$
& $-1$
& $0$
& $1\ (1)$
& $-1$
\\
\hline
  $5\ (2)$
&
{\small $
\Yvcentermath1
\young(
\zero\one\two,%
\two\zero
)
$}
&
{\small $
\Yvcentermath1
\young(
\one\zero\one,%
\two\one
)
$}
& $1$
& $-1/2$
& $1$
& $-2$
& $1$
& $2\  (2)$
& $1$
\\
\hline
  $8\ (2)$
&
{\small $
\Yvcentermath1
\young(
\zero\one\two,%
\two\zero,%
\one\two,%
\zero
)
$}
&
{\small $
\Yvcentermath1
\young(
\zero\one\one,%
\one\two,%
\zero\one,%
\one
)
$}
& $1$
& $-1/2$
& $1$
& $-3$
& $2$
& $9\ (0)$
& $-1$
\\
\hline
${\cal A}_N^{(3;21)}$&&&&&&&&&
\\
\hline
  $4\  (1)$
&
{\small $
\Yvcentermath1
\young(
\zero\one,%
\two,%
\one
)
$}
&
{\small $
\Yvcentermath1
\young(
\one\one,%
\two,%
\one
)
$}
& $-1$
& $1/2$
& $1$
& $-1$
& $0$
& $3\  (0)$
& $1$
\\
\hline
  $7\  (1)$
&
{\small $
\Yvcentermath1
\young(
\zero\one\two,%
\two\zero\one,%
\one
)
$}
&
{\small $
\Yvcentermath1
\young(
\two\zero\two,%
\one\two\one,%
\one
)
$}
& $-1$
& $1/2$
& $1$
& $0$
& $-1$
& $5\  (2)$
& $-1$
\\
\hline
  $10\  (1)$
&
{\small $
\Yvcentermath1
\young(
\zero\one\two,%
\two\zero\one,%
\one\two,%
\zero\one
)
$}
&
{\small $
\Yvcentermath1
\young(
\zero\two\two,%
\two\one\one,%
\zero\two,%
\two\one
)
$}
& $-1$
& $1/2$
& $1$
& $1$
& $-2$
& $13\  (1)$
& $1$
\\
\hline
\end{tabular}
\end{center}
\end{table}

\begin{table}[]
\caption{Data for the sets ${\cal A}_N^{(3;13)}$
and ${\cal A}_N^{(3;31)}$.}
\label{A13-3}
\begin{center}
\begin{tabular}{cllccccccc}
\hline
  $|\nu|$ (mod 3)
& $C_3^{(a)}(\nu)$
& $H_3^{(a)}(\nu)$
& $S_\nu^1$
& $S_\nu^2$
& $c\mbox{-}h_\nu^{(0)}$
& $c\mbox{-}h_\nu^{(1)}$
& $c\mbox{-}h_\nu^{(2)}$
& $n(\nu)$ (mod 3)
& $A_\nu$
\\
\hline
${\cal A}_N^{(3;13)}$&&&&&&&&&
\\
\hline
  $1\  (1)$
&
{\small $
\Yvcentermath1
\young(
\zero
)
$}
&
{\small $
\Yvcentermath1
\young(
\one
)
$}
& $1/2$
& $1/2$
& $1$
& $-1$
& $0$
& $0\  (0)$
& $1$
\\
\hline
  $4\  (1)$
&
{\small $
\Yvcentermath1
\young(
\zero\one,%
\two\zero
)
$}
&
{\small $
\Yvcentermath1
\young(
\zero\two,%
\two\one
)
$}
& $1/2$
& $1/2$
& $1$
& $0$
& $-1$
& $2\  (2)$
& $-1$
\\
\hline
  $7\  (1)$
&
{\small $
\Yvcentermath1
\young(
\zero\one\two,%
\two\zero,%
\one,%
\zero
)
$}
&
{\small $
\Yvcentermath1
\young(
\zero\zero\one,%
\one\one,%
\two,%
\one
)
$}
& $1/2$
& $1/2$
& $1$
& $-2$
& $1$
& $7\  (1)$
& $-1$
\\
\hline
${\cal A}_N^{(3;31)}$&&&&&&&&&
\\
\hline
  $5\  (2)$
&
{\small $
\Yvcentermath1
\young(
\zero\one\two,%
\two,%
\one
)
$}
&
{\small $
\Yvcentermath1
\young(
\two\two\one,%
\two,%
\one
)
$}
& $-1/2$
& $-1/2$
& $1$
& $0$
& $-1$
& $3\  (0)$
& $1$
\\
\hline
  $8\  (2)$
&
{\small $
\Yvcentermath1
\young(
\zero\one\two,%
\two\zero\one,%
\one\two
)
$}
&
{\small $
\Yvcentermath1
\young(
\two\one\two,%
\one\zero\one,%
\two\one
)
$}
& $-1/2$
& $-1/2$
& $1$
& $-1$
& $0$
& $7\  (1)$
& $-1$
\\
\hline
  $11\  (2)$
&
{\small $
\Yvcentermath1
\young(
\zero\one\two,%
\two\zero\one,%
\one\two,%
\zero\one,%
\two
)
$}
&
{\small $
\Yvcentermath1
\young(
\one\two\two,%
\zero\one\one,%
\one\two,%
\zero\one,%
\one
)
$}
& $-1/2$
& $-1/2$
& $1$
& $-2$
& $1$
& $17\  (2)$
& $1$
\\
\hline
\end{tabular}
\end{center}
\end{table}

\begin{table}[]
\caption{Data for the sets ${\cal A}_N^{(3;23)}$
and ${\cal A}_N^{(3;32)}$.}
\label{A23-3}
\begin{center}
\begin{tabular}{cllccccccc}
\hline
  $|\nu|$ (mod 3)
& $C_3^{(a)}(\nu)$
& $H_3^{(a)}(\nu)$
& $S_\nu^1$
& $S_\nu^2$
& $c\mbox{-}h_\nu^{(0)}$
& $c\mbox{-}h_\nu^{(1)}$
& $c\mbox{-}h_\nu^{(2)}$
& $n(\nu)$ (mod 3)
& $A_\nu$
\\
\hline
${\cal A}_N^{(3;23)}$&&&&&&&&&
\\
\hline
  $2\  (2)$
&
{\small $
\Yvcentermath1
\young(
\zero\one
)
$}
&
{\small $
\Yvcentermath1
\young(
\two\one
)
$}
& $-1/2$
& $1$
& $1$
& $0$
& $-1$
& $0\  (0)$
& $1$
\\
\hline
  $5\  (2)$
&
{\small $
\Yvcentermath1
\young(
\zero\one,%
\two\zero,%
\one
)
$}
&
{\small $
\Yvcentermath1
\young(
\one\two,%
\zero\one,%
\one
)
$}
& $-1/2$
& $1$
& $1$
& $-1$
& $0$
& $4\  (1)$
& $-1$
\\
\hline
  $8\  (2)$
&
{\small $
\Yvcentermath1
\young(
\zero\one\two,%
\two\zero\one,%
\one,%
\zero
)
$}
&
{\small $
\Yvcentermath1
\young(
\zero\zero\two,%
\two\two\one,%
\two,%
\one
)
$}
& $-1/2$
& $1$
& $1$
& $1$
& $-2$
& $8\  (2)$
& $-1$
\\
\hline
${\cal A}_N^{(3;32)}$&&&&&&&&&
\\
\hline
  $4\  (1)$
&
{\small $
\Yvcentermath1
\young(
\zero\one\two,%
\two
)
$}
&
{\small $
\Yvcentermath1
\young(
\one\two\one,%
\one
)
$}
& $1/2$
& $-1$
& $1$
& $-2$
& $1$
& $1\  (1)$
& $-1$
\\
\hline
  $7\  (1)$
&
{\small $
\Yvcentermath1
\young(
\zero\one\two,%
\two\zero,%
\one\two
)
$}
&
{\small $
\Yvcentermath1
\young(
\two\one\one,%
\zero\two,%
\two\one
)
$}
& $1/2$
& $-1$
& $1$
& $-1$
& $0$
& $6\  (0)$
& $1$
\\
\hline
  $10\  (1)$
&
{\small $
\Yvcentermath1
\young(
\zero\one\two,%
\two\zero\one,%
\one\two,%
\zero,%
\two
)
$}
&
{\small $
\Yvcentermath1
\young(
\one\one\two,%
\zero\zero\one,%
\one\one,%
\two,%
\one
)
$}
& $1/2$
& $-1$
& $1$
& $-3$
& $2$
& $14\  (2)$
& $1$
\\
\hline
\end{tabular}
\end{center}
\end{table}

\begin{table}[]
\caption{Data for the set ${\cal B}_N^{(3)}$.}
\label{B-3}
\begin{center}
\begin{tabular}{cllccccccc}
\hline
  $|\nu|$ (mod 3)
& $C_3^{(a)}(\nu)$
& $H_3^{(a)}(\nu)$
& $S_\nu^1$
& $S_\nu^2$
& $c\mbox{-}h_\nu^{(0)}$
& $c\mbox{-}h_\nu^{(1)}$
& $c\mbox{-}h_\nu^{(2)}$
& $n(\nu)$ (mod 3)
& $A_\nu$
\\
\hline
  $0\  (0)$
& $\O$
& $\O$
& $0$
& $0$
& $0$
& $0$
& $0$
& $0\  (0)$
& $1$
\\
\hline
  $3\  (0)$
&
{\small $
\Yvcentermath1
\young(
\zero\one,%
\two
)
$}
&
{\small $
\Yvcentermath1
\young(
\zero\one,%
\one
)
$}
& $0$
& $0$
& $0$
& $-1$
& $1$
& $1\  (1)$
& $-1$
\\
\hline
  $6\  (0)$
&
{\small $
\Yvcentermath1
\young(
\zero\one\two,%
\two\zero,%
\one
)
$}
&
{\small $
\Yvcentermath1
\young(
\two\zero\one,%
\zero\one,%
\one
)
$}
& $0$
& $0$
& $0$
& $-1$
& $1$
& $4\  (1)$
& $-1$
\\
\hline
  $9\  (0)$
&
{\small $
\Yvcentermath1
\young(
\zero\one\two,%
\two\zero\one,%
\one\two,%
\zero
)
$}
&
{\small $
\Yvcentermath1
\young(
\zero\one\two,%
\two\zero\one,%
\zero\one,%
\one
)
$}
& $0$
& $0$
& $0$
& $-1$
& $1$
& $10\  (1)$
& $-1$
\\
\hline
\end{tabular}
\end{center}
\end{table}

\begin{table}[]
\caption{Data for the set ${\cal C}_N^{(3)}$.}
\label{C-3}
\begin{center}
\begin{tabular}{cllccccccc}
\hline
  $|\nu|$ (mod 3)
& $C_3^{(a)}(\nu)$
& $H_3^{(a)}(\nu)$
& $S_\nu^1$
& $S_\nu^2$
& $c\mbox{-}h_\nu^{(0)}$
& $c\mbox{-}h_\nu^{(1)}$
& $c\mbox{-}h_\nu^{(2)}$
& $n(\nu)$ (mod 3)
& $A_\nu$
\\
\hline
  $6\  (0)$
&
{\small $
\Yvcentermath1
\young(
\zero\one,%
\two\zero,%
\one,%
\zero
)
$}
&
{\small $
\Yvcentermath1
\young(
\two\two,%
\one\one,%
\two,%
\one
)
$}
& $0$
& $3/2$
& $3$
& $-1$
& $-2$
& $7\  (1)$
& -
\\
\hline
  $9\  (0)$
&
{\small $
\Yvcentermath1
\young(
\zero\one\two,%
\two\zero,%
\one\two,%
\zero,%
\two
)
$}
&
{\small $
\Yvcentermath1
\young(
\one\one\one,%
\two\two,%
\one\one,%
\two,%
\one
)
$}
& $3/2$
& $-3/2$
& $3$
& $-4$
& $1$
& $13\  (1)$
& -
\\
\hline
  $12\  (0)$
&
{\small $
\Yvcentermath1
\young(
\zero\one\two,%
\two\zero\one,%
\one\two,%
\zero\one,%
\two,%
\one
)
$}
&
{\small $
\Yvcentermath1
\young(
\two\two\two,%
\one\one\one,%
\two\two,%
\one\one,%
\two,%
\one
)
$}
& $-3/2$
& $0$
& $3$
& $-1$
& $-2$
& $22\  (1)$
& -
\\
\hline
\end{tabular}
\end{center}
\end{table}



\begin{table}[]
\caption{Data of $K$-regular partitions
for $K=3$ and $N=3$.
The rescaled momentum $\tilde{P}_c$
and excited energy $\tilde{E}_c$
are defined by 
$\tilde{P}_c:=[N/(2\pi)]P_c$
and
$\tilde{E}_c:=[(4/J)(N/(2\pi))^2]E_c$,
respectively.}
\label{K-reglarK3N3}
\begin{center}
\begin{tabular}{cccc}
\hline
 $K$-regular partition (type)
&$c_1,c_2,c_3,c_4,c_5,c_6$
&$\tilde{P}_c$
&$\tilde{E}_c$
\\
\hline
 $\O$ {\scriptsize $\left(\O\right)$}
&0,0,0,0,0,0
&0
&0
\\
\hline
${\Yvcentermath1 \young(
\zero
)}$
{\tiny
${\Yvcentermath1 \left(\yng(1)\right)}$},
${\Yvcentermath1 \young(
\zero,%
\two
)}$
{\tiny
${\Yvcentermath1 \left(\yng(1,1)\right)}$},
${\Yvcentermath1 \young(
\zero,%
\two,%
\one
)}$
{\scriptsize
${\Yvcentermath1 \left(\O\right)}$},
${\Yvcentermath1 \young(
\zero\one
)}$
{\tiny
${\Yvcentermath1 \left(\yng(2)\right)}$}
&1,0,0,0,0,0
&1
&2
\\
${\Yvcentermath1 \young(
\zero\one,%
\two
)}$
{\tiny
${\Yvcentermath1 \left(\yng(2,1)\right)}$},
${\Yvcentermath1 \young(
\zero\one,%
\two,%
\one
)}$
{\tiny
${\Yvcentermath1 \left(\yng(2,1,1)\right)}$},
${\Yvcentermath1 \young(
\zero\one\two,%
\two
)}$
{\tiny
${\Yvcentermath1 \left(\yng(3,1)\right)}$},
${\Yvcentermath1 \young(
\zero\one\two,%
\two,%
\one
)}$
{\tiny
${\Yvcentermath1 \left(\yng(3,1,1)\right)}$},
&
&
&
\\
\hline
${\Yvcentermath1 \young(
\zero\one,%
\two\zero
)}$
{\tiny
${\Yvcentermath1 \left(\yng(2,2)\right)}$},
${\Yvcentermath1 \young(
\zero\one,%
\two\zero,%
\one
)}$
{\tiny ${\Yvcentermath1 \left(\yng(2,2,1)\right)}$},
${\Yvcentermath1 \young(
\zero\one,%
\two\zero,%
\one\two
)}$
{\scriptsize $\left(\O\right)$},
${\Yvcentermath1 \young(
\zero\one\two,%
\two\zero
)}$
{\tiny
${\Yvcentermath1 \left(\yng(3,2)\right)}$}
&1,1,0,0,0,0
&2
&2
\\
${\Yvcentermath1 \young(
\zero\one\two,%
\two\zero,%
\one
)}$
{\tiny
${\Yvcentermath1 \left(\yng(3,2,1)\right)}$},
${\Yvcentermath1 \young(
\zero\one\two,%
\two\zero,%
\one\two
)}$
{\tiny
${\Yvcentermath1 \left(\yng(3,2,2)\right)}$},
${\Yvcentermath1 \young(
\zero\one\two,%
\two\zero\one,%
\one
)}$
{\tiny
${\Yvcentermath1 \left(\yng(3,3,1)\right)}$},
${\Yvcentermath1 \young(
\zero\one\two,%
\two\zero\one,%
\one\two
)}$
{\tiny
${\Yvcentermath1 \left(\yng(3,3,2)\right)}$},
&
&
&
\\
\hline
${\Yvcentermath1 \young(
\zero\one\two\zero,%
\two\zero
)}$
{\tiny
${\Yvcentermath1 \left(\yng(4,2)\right)}$},
${\Yvcentermath1 \young(
\zero\one\two\zero,%
\two\zero,%
\one
)}$
{\tiny ${\Yvcentermath1 \left(\yng(4,2,1)\right)}$},
${\Yvcentermath1 \young(
\zero\one\two\zero,%
\two\zero\one,%
\one
)}$
{\tiny
${\Yvcentermath1 \left(\yng(4,3,1)\right)}$},
&1,1,0,1,0,0
&3
&4
\\
${\Yvcentermath1 \young(
\zero\one\two\zero,%
\two\zero,%
\one\two
)}$
{\tiny ${\Yvcentermath1 \left(\yng(4,2,2)\right)}$},
${\Yvcentermath1 \young(
\zero\one\two\zero,%
\two\zero\one,%
\one\two
)}$
{\tiny ${\Yvcentermath1 \left(\yng(4,3,2)\right)}$},
${\Yvcentermath1 \young(
\zero\one\two\zero,%
\two\zero\one\two,%
\one\two
)}$
{\tiny ${\Yvcentermath1 \left(\yng(4,4,2)\right)}$},
&
&
&
\\
${\Yvcentermath1 \young(
\zero\one\two\zero\one,%
\two\zero\one,%
\one
)}$
{\tiny ${\Yvcentermath1 \left(\yng(5,3,1)\right)}$},
${\Yvcentermath1 \young(
\zero\one\two\zero\one,%
\two\zero\one,%
\one\two
)}$
{\tiny ${\Yvcentermath1 \left(\yng(5,3,2)\right)}$},
&
&
&
\\
${\Yvcentermath1 \young(
\zero\one\two\zero\one,%
\two\zero\one\two,%
\one\two
)}$
{\tiny ${\Yvcentermath1 \left(\yng(5,4,2)\right)}$},
${\Yvcentermath1 \young(
\zero\one\two\zero\one\two,%
\two\zero\one\two,%
\one\two
)}$
{\tiny ${\Yvcentermath1 \left(\yng(6,4,2)\right)}$}
&
&
&
\\
\hline
\end{tabular}
\end{center}
\end{table}



\begin{table}[]
\caption{Data of motifs for $K=3$ and $N=3$.
The rescaled momentum $\tilde{P}_d$
and excited energy $\tilde{E}_d$
are defined by 
$\tilde{P}_d:=[N/(2\pi)]P_d$
and
$\tilde{E}_d:=[(4/J)(N/(2\pi))^2](E_d-E_{d^{(0)}})$,
respectively.}
\label{motif-K3N3}
\begin{center}
\begin{tabular}{lccccc}
\hline
 Motif $d$
&Ribbon diagram
&$SU(3)$ content
&Degeneracy
&$\tilde{P}_d$
&$\tilde{E}_d$
\\
\hline
 $0110=(11)=d^{(0)}$
&{\scriptsize
  ${\Yvcentermath1\young(\hfil,\hfil,\hfil)}$}$\,=\theta^{(0)}$
&$1$
&1
&0
&0
\\
\hline
 $0100=(1)(\,)$
&{\scriptsize
  ${\Yvcentermath1\young(:\hfil,\hfil\hfil)}$}
&{\scriptsize ${\Yvcentermath1\yng(2,1)}$}
&8
&1
&2
\\
\hline
 $0010=(\,)(1)$
&{\scriptsize
  ${\Yvcentermath1\young(\hfil\hfil,\hfil)}$}
&{\scriptsize ${\Yvcentermath1 \yng(2,1)}$}
&8
&2
&2
\\
\hline
 $0000=(\,)(\,)$
&{\scriptsize
  ${\Yvcentermath1\young(\hfil\hfil\hfil)}$}
&{\scriptsize ${\Yvcentermath1\yng(3)}$}
&10
&0
&4
\\
\hline
\end{tabular}
\end{center}
\end{table}

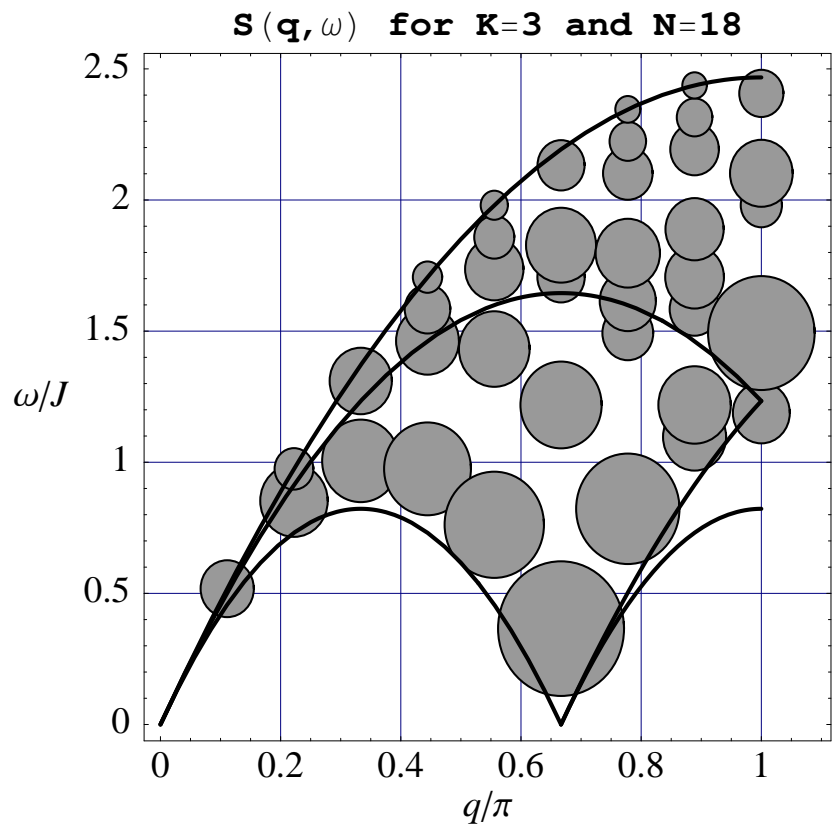
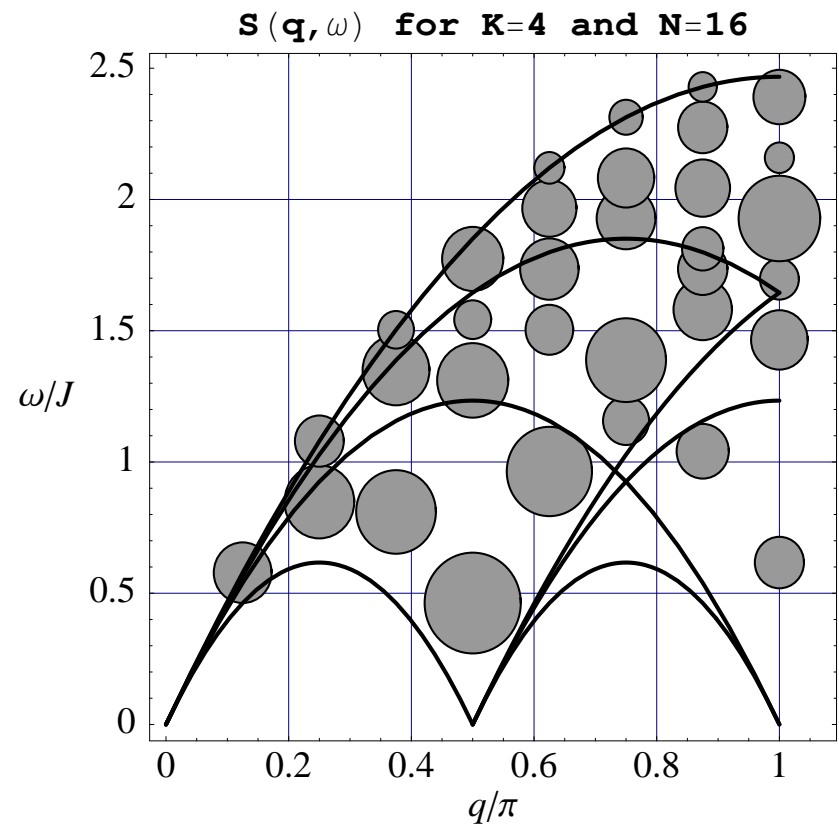
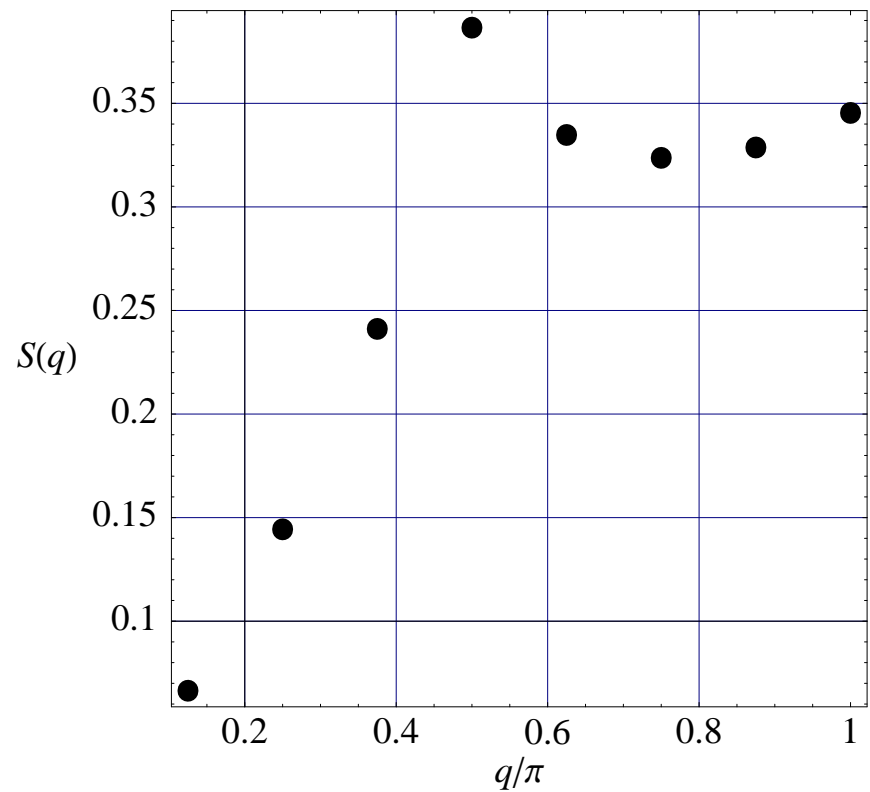
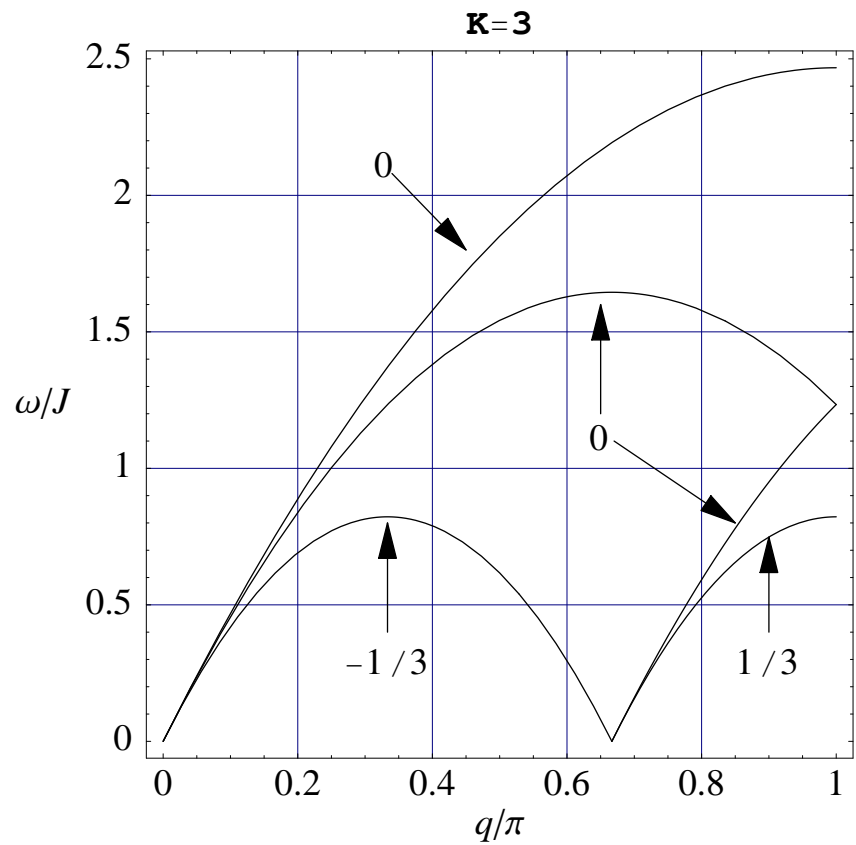
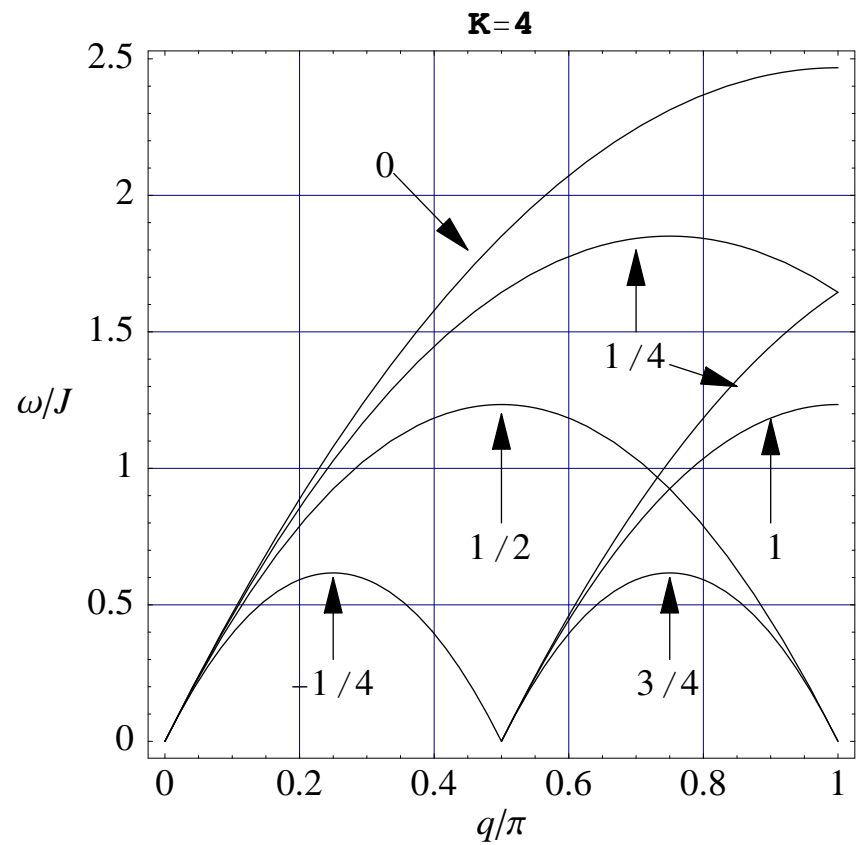
\end{document}